\documentclass[twocolumn, aps, pre, floatfix]{revtex4-2}
\usepackage{amsmath}
\usepackage{float}
\usepackage{graphicx}
\usepackage{amsfonts}
\usepackage{xcolor}
\usepackage{bm}
\usepackage{tabularx}
\usepackage[normalem]{ulem}
\usepackage{afterpage}
\usepackage{soul}
\usepackage{subfigure}
\usepackage{hyperref}
\usepackage[section]{placeins}
\hypersetup{
    colorlinks=true,
    allcolors=.,
}

\newcommand{\micron}{{\mu\mathrm{m}}}

\newcommand{\timeseries}[2]{
    \subfigure[]{\includegraphics[width=0.75\linewidth]{#1} \label{fig: #2 timeseries}}
}
\newcommand{\sampletrajs}[2]{
    \subfigure[]{\includegraphics[width=0.24\linewidth]{#1} \label{fig: #2 sampletrajs}}
}
\newcommand{\streamline}[2]{
    \subfigure[]{\includegraphics[width=0.20\linewidth]{#1} \label{fig: #2 streamline}}
}
\newcommand{\brucknerstream}[2]{
    \subfigure[]{\includegraphics[width=0.21\linewidth]{#1} \label{fig: #2 brucknerstream}}
}
\newcommand{\rowperturbation}[2]{
    \subfigure[]{\includegraphics[width=0.99\linewidth]{#1} \label{fig: #2 perturbation}}
}
\newcommand{\bootview}[5]{
    \begin{figure}[htb]
      \begin{minipage}{.26\linewidth}
        \includegraphics[width=1\linewidth]{#1}
        \makebox[.5\linewidth]{\small (a)}
        
        \medskip
    
        \includegraphics[width=1\linewidth]{#2}
        \makebox[.5\linewidth]{\small (b)}
      \end{minipage}
      \hfill
      \begin{minipage}{.63\linewidth}
        \centering
        \includegraphics[width=0.9\linewidth]{#3}
        \makebox[.5\linewidth]{\small (c)}
      \end{minipage}
      \caption{#4}
      \label{#5}
    \end{figure}
}

\begin{document}
\title{Inferring nonlinear dynamics of cell migration}
\author{Pedrom Zadeh}
 \affiliation{William H. Miller III Department of Physics \& Astronomy, Johns Hopkins University, Baltimore, MD.}
\author{Brian A. Camley}
 \affiliation{William H. Miller III Department of Physics \& Astronomy and Thomas C. Jenkins Department of Biophysics, Johns Hopkins University, Baltimore, MD.}

\begin{abstract}
The motility of eukaryotic cells is strongly influenced by their environment, with confined cells often developing qualitatively different motility patterns from those migrating on simple two-dimensional substrates. Recent experiments, coupled with data-driven methods to extract a cell's equation of motion, showed that cancerous MDA-MB-231 cells persistently hop in a limit cycle when placed on two-state adhesive micropatterns (two large squares connected by a narrow bridge), while they remain stationary on average in rectangular confinements. In contrast, healthy MCF10A cells migrating on the two-state micropattern are bistable, i.e., they settle into either basin on average with only noise-induced hops between the two states. We can capture all these behaviors with a single computational phase field model of a crawling cell, under the assumption that contact with non-adhesive substrate inhibits the cell front. Our model predicts that larger and softer cells are more likely to persistently hop, while smaller and stiffer cells are more likely to be bistable. Other key factors controlling cell migration are the frequency of protrusions and their magnitude of noise. Our results show that relatively simple assumptions about how cells sense their geometry can explain a wide variety of different cell behaviors, and show the power of data-driven approaches to characterize both experiment and simulation.
\end{abstract}
\maketitle

\section{Introduction}
Living cells navigate intricate spatial constraints and complex environments and respond to them by changing their motility and behavior \cite{charras2014physical,paul2017cancer}. This can range from developing biased migration in response to asymmetric micropatterns \cite{Mahmud2009, caballero-1, caballero2014protrusion} to altering mechanisms of migration \cite{Camley-Nain,doyle2009one,stroka2014water} or developing completely different migration patterns such as rotations and oscillations \cite{Camley-PRM-models,camley2013periodic,ipina2023secreted,d2021cell}. How can we characterize all of these complex types of migration and determine how they arise from the influence of confinement? One promising recent approach is the use of data-driven models \cite{SELMECZI2005912,Bruckner2019,li2011dicty,amselem2012stochastic,selmeczi2008cell,bruckner2023learning}, in which equations of motion are learned directly from experimental data. In particular, we focus on a recent study on the migration of cancerous MDA-MB-231 and healthy MCF10A cells \cite{Bruckner2019}, which summarized experimental observations into a data-driven representation of the dynamics of the cell position $x$ and velocity $v$. Using this representation, cells were characterized with the language of nonlinear dynamics. For instance, cancerous MDA-MB-231 cells were well described by a limit cycle in $x$ and $v$ when placed on a ``two-state geometry'', consisting of two square basins connected by a narrow bridge (Fig. \ref{fig: geometries}). This limit cycle characterizes cells persistently hopping from basin to basin. This is strongly dependent on the presence of the narrow constriction: when MDA-MB-231 were placed on a rectangular substrate, they remained stationary on average. By contrast, MCF10A cells were bistable on the two-state micropattern, settling into either basin on average and only executing noise-induced hops \cite{Bruckner2019}.

Why does the two-state geometry break the symmetric migration of cancerous cells and cause them to limit-cycle? Why do MDA-MB-231 and MCF10A cells behave differently? To what extent does the experimental measurement of these nonlinear dynamics constrain more complicated models of cell motility? We attempt to construct a single model of a cell that can capture the rich landscape of behaviors observed by \cite{Bruckner2019}. We use the phase field framework to describe our cell motion, with a simple assumption that a cell's polarity is inhibited when the cell leaves the micropattern. We apply the data-driven approach of \cite{Bruckner2019} to determine the equation of motion for the cell's velocity and position. We show that our model can successfully recapitulate limit cycle, bistable, and stationary behaviors depending on cell attributes and geometry of the micropattern. Our model suggests that larger and softer cells tend to limit-cycle on two-state micropatterns, while smaller and stiffer ones are more likely to be bistable. We also study the role of other perturbations of our cell polarity model, and characterize the variability in the predictions of the data-driven model due to finite sampling. Our results suggest that the qualitative difference between bistable and limit cycle behavior can be achieved by relatively small quantitative changes in mechanical, physical, or polarity properties of the cell, including cell size and cell tension. These results highlight the power of data-driven approaches to detect fine changes in cell motility.

\section{Model}
\label{sec: model}

We model a single cell within the phase field framework \cite{Nonomura,Doostmohammadi,Cao,Kulawiak,Camley-PRM-models,palmieri2015multiple}, which describes objects with arbitrary and continuously deforming shapes. Because the experiments of \cite{Bruckner2019} confine the cell on two-dimensional adhesive micropatterns, we simplify our model to two dimensions and adopt a ``top view'' of the cell (see Fig. \ref{fig: lc timeseries}). The cell shape is set by a field $\phi({\bm r}, t)$, which smoothly varies from zero outside the cell and one inside it, implicitly defining the boundary at $\phi=1/2$. The evolution of the cell boundary is governed by energy minimization and advection of the interface \cite{Doostmohammadi, Camley-PF-eq2, Biben-PF-eq1},
\begin{equation}
    \label{eq: phase-field-model}
   \frac{\partial \phi({\bm r},t)}{\partial t} + {\bm v}({\bm r},t) \cdot \nabla \phi=-\mathcal{M}\ \frac{\delta \mathcal{F}}{\delta \phi},
\end{equation}
where $\mathcal{M}$ is a transport coefficient, and ${\bm v}({\bm r},t)$ is the velocity field of the cell driving its interface. The free energy is $\mathcal{F}=\mathcal{F}_{\mathrm{CH}} + \mathcal{F}_{\mathrm{area}}$. The Cahn-Hilliard energy is \cite{Doostmohammadi}
\begin{equation}
    \label{eq: cahn-hilliard-eq}
   \mathcal{F}_{{\rm CH}}=\int d{\bm r} \frac{\gamma}{\lambda} \left[16\phi^2(1-\phi)^2 + \lambda^2 \left(\nabla \phi\right)^2 \right],
\end{equation}
which is minimal at $\phi = 0,1$ and penalizes deformations of the interface. $\lambda$ has units of length and sets the phase field interfacial thickness, while in two dimensions, $\gamma$ sets the line tension of the cell with units of energy/length. The term $\mathcal{F}_\textrm{area}$ models cell incompressibility by penalizing deviations of cell area from its target $\pi R_0^2$ \cite{Doostmohammadi},
\begin{equation}
    \label{eq: area-conservation-eq}
   \mathcal{F}_{\mathrm{area}}=\kappa \left[ 1-\frac{1}{\pi R_0^2} \int d{\bm r} \phi^2 \right]^2.
\end{equation}
In the absence of cell motility, these energy terms equilibrate the cell to a circle with radius $R_0$.

We introduce cell motility through the velocity field $\bm v(\bm r, t)$. Modeling the cell as an overdamped system, we locally balance forces per unit area and write \cite{Doostmohammadi}
\begin{equation}
    \label{eq: phase-field-velocity-eq}
    \eta {\bm v}=\frac{\delta \mathcal{F}}{\delta \phi} \nabla \phi + {\bm f}_{\mathrm{motility}}.
\end{equation}
The left hand side represents the friction force per unit area, while the first term on the right denotes the force density arising from energy minimization. The last term is the active motility force per unit area, which depends on cell polarity and drives the cell front.

\subsection{Micropattern geometries}
\begin{figure}[htb]
    \centering
    \includegraphics[width=1\linewidth]{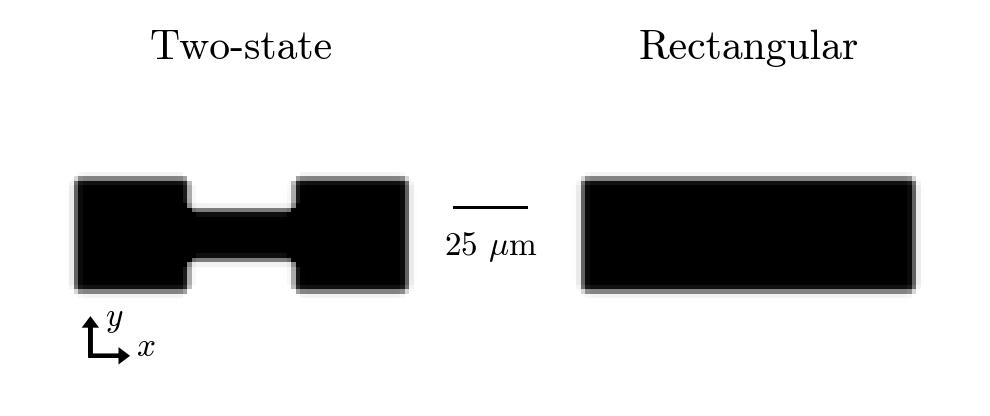}
    \caption{A representation of the two substrate geometries by a phase field $\chi(\bm r)$, which is 0 (black) in the interior and 1 (white) in the exterior. The two-state geometry has two basins measuring $38 \micron \times 38 \micron$ that are connected by a $10 \micron \times 35 \micron$ bridge. The rectangular substrate measures $111\micron \times 38 \micron$.}
    \label{fig: geometries}
\end{figure}

The experiments of \cite{Bruckner2019} use an adhesive fibronectin micropattern to effectively confine the cells---cells can only adhere to the substrate on the micropattern. We incorporate these micropatterns in our phase field framework using a static field $\chi(\bm r)$. We set $\chi$ to be zero in the interior and one in the exterior of the micropattern, transitioning smoothly at the boundary (Fig. \ref{fig: geometries}). We assume that the cell interacts with the substrate through its polarity, which controls cell motility. This neglects any adhesive energy to the micropattern \cite{pedrom, Cao,albert2014dynamics}.

\subsection{Defining cell polarity and motility}
\label{sec: cell_polarity}

Migrating cells have an underlying chemical polarization, indicating the areas of the cell that are likely to protrude (``front-like'') and those likely to contract (``rear-like'') \cite{rappel2017mechanisms}. This can include asymmetric distribution of Rho GTPases, with Rac1 activity driving the cell front through lamellipodial extensions, and RhoA promoting myosin contractility in the rear \cite{Ridley}. Rather than explicitly modeling the dynamics of one or more Rho GTPases \cite{Kulawiak, Camley-PRM-models, Merchant, merchant2020rho, Mori, Holmes-plos}, we summarize cell polarity with a single spatiotemporal field $\mathbb{P}({\bm r}, t)$ whose sign determines whether the cell boundary protrudes ($\mathbb{P}>0$) or contracts ($\mathbb{P}<0$). This leads us to write the motility force as 
\begin{equation}
    \label{eq: motility_force}
    \bm{f}_{\mathrm{motility}}=\alpha \mathbb{P}(\bm{r}, t) \phi^2(1-\phi)^2(1-\chi)\bm{\hat{n}}.
\end{equation}
Here, $\alpha$ is a constant, {$\bm{\hat{n}}\equiv-\nabla\phi/|\nabla \phi|$ is the unit vector normal to the cell interface}, and the phase field terms ensure that motility forces exist only where the cell interface (i.e. $\phi^2(1-\phi)^2 > 0$) is in contact with the micropattern (i.e. $1-\chi >0$)\cite{Ridley, Cao, Aronson}.

\begin{figure}[htb]
    \centering
    \includegraphics[width=0.7\linewidth]{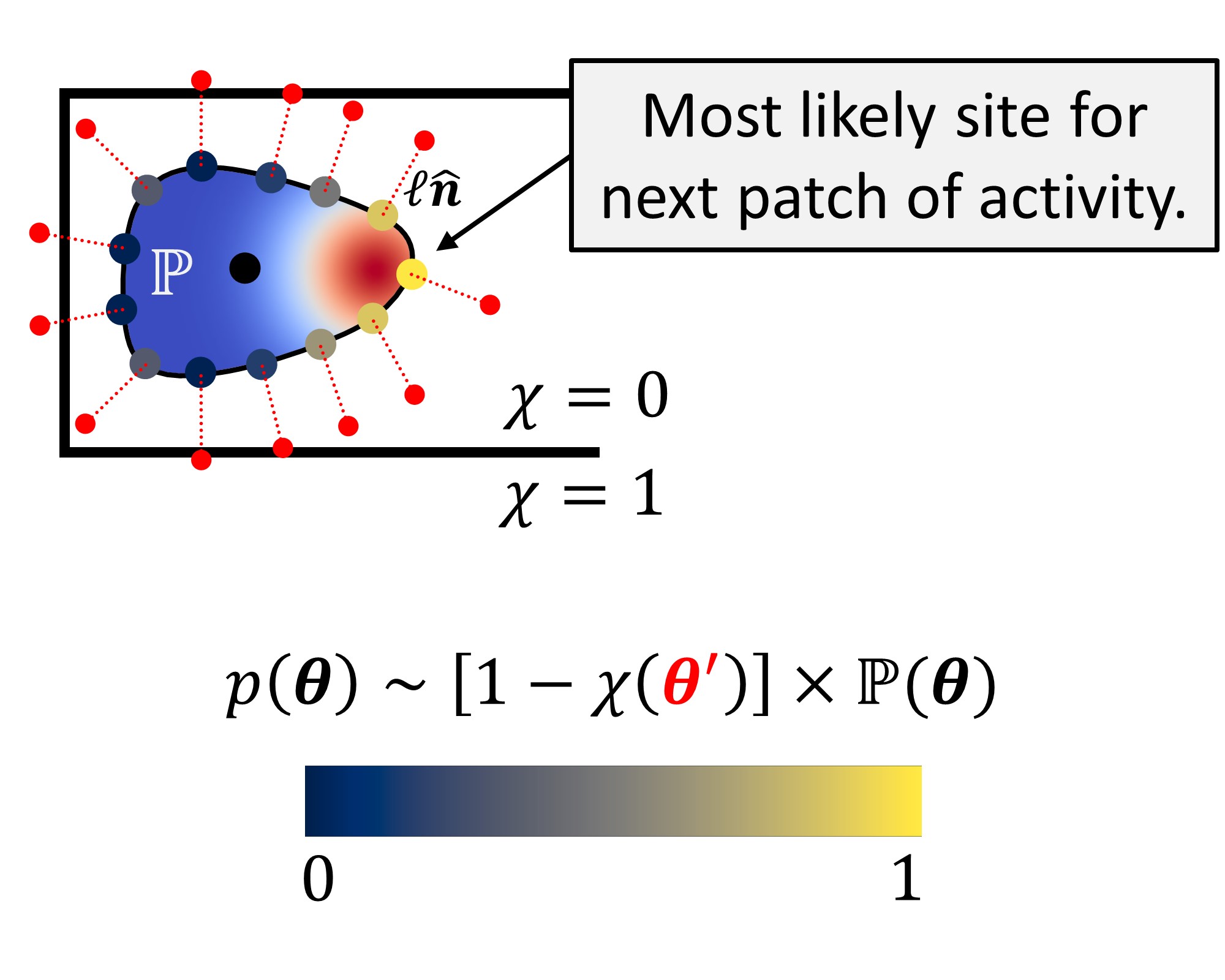}
    \caption{A cartoon of how polarity at contour points and their proximity to the micropattern edge combine to define the probability $p(\bm \theta)$. The presence of the micropattern is measured at $\bm \theta'$, which is obtained by projecting each contour point normally outward a distance $\ell$ (red dots). In this sketch, the highest probability is associated with the contour point that measures the highest polarity and amount of micropattern ahead of it.}
    \label{fig: polarity sketch}
\end{figure}

There is no single established approach for modeling the complex dynamics of cell polarity \cite{Camley-Topical-Review}. Here, we evolve $\mathbb{P}(\bm r, t)$ using stochastic fluctuations in polarity at random locations on the cell boundary, inhibition of polarity in the absence of the micropattern, inhibition of polarity when the cell's tension surpasses a certain threshold, and a basal decay of polarity. Mathematically, we write
\begin{align}
    \label{eq: dpol_dt}
    \begin{split}
        \frac{\partial \mathbb{P}(\bm r, t)}{\partial t} &= 
        \beta(t) f\left(|\bm r - \bm \theta(t)|; \sigma\right)\delta(t \text{ mod }\tau_f) \phi\\
        &- \frac{1}{\tau_\chi} \chi \phi\\
        &- \frac{1}{\tau_s}\left(\frac{s}{s_0}\right)^2H(s-s_0)\\
        &- \frac{1}{\tau}\mathbb{P}(\bm r, t).
    \end{split}
\end{align}
The first term introduces radially-symmetric patches $f\left(|\bm r - \bm \theta(t)|; \sigma\right)=e^{-|\bm r - \bm \theta(t)|^2/2\sigma^2}$ that activate polarity at locations $\bm \theta$ on the cell boundary. The amplitude of the patch is a random variable $\beta \sim \mathrm{Gaussian}(\mu_\beta, \sigma_\beta)$. We choose $\mu_\beta > 0$ so these patches locally increase the polarity on average, but $\beta(t)$ can experience large fluctuations, including negative values. Lastly, we write $\delta(t \text{ mod }\tau_f) \phi$, where $\delta(t)$ is the Dirac delta, which lets us introduce these noisy activity patches at a controlled frequency $\tau_f^{-1}$ and ensure they are confined within the cell interior. To complete this description, we must specify a location $\bm \theta$ for the polarity patch every $\tau_f$. The polarity only drives protrusion and retraction at the interface, so we choose $\bm \theta$ to be a contour point along the boundary of the cell ($\phi=1/2$). We assume cells choose these locations with two important factors: 1) a generic positive feedback, where existing polarity promotes new polarity, akin to other polarity models \cite{rappel2017mechanisms,edelstein2013simple}, and 2) feedback from successful adhesion, where new cell-substrate adhesion promotes polarity \cite{xia2008directional}. We choose the locations of polarity patches according to these principles by sampling $\bm \theta$ from a probability $p(\bm{\theta})$ every $\tau_f$, where
\begin{equation}
    \label{eq: cntr_pmf}
    p(\bm{\theta})\sim \left[1-\chi(\bm\theta')\right] \times \mathbb{P}(\bm \theta),
\end{equation}
and $\bm{\theta'} = \bm{\theta}+\ell \bm{\hat{n}}(\bm{\theta})$ is a point a distance $\ell$ normally outward from the contour (Fig. \ref{fig: polarity sketch}). In the $\left[1-\chi(\bm\theta')\right]$ term, the cell is essentially measuring to discover if the adhesive micropattern is present (indicated by regions where $1-\chi > 0$) at the point $\bm{\theta}'$, and if so, enhancing polarity at $\bm{\theta}$. 
The idea behind this ``filopodial probability'' is that small stochastic protrusions (filopodia) that are not explicitly represented in our model of the cell boundary test the region near the cell edge, and upon contact with the adhesive micropattern, mature and reinforce polarity \cite{caballero2014protrusion, Ridley}. Lastly, our assumption that the probability to pick a contour point $p(\bm{\theta})$ is proportional to the polarity at that contour point $ \mathbb{P}(\bm \theta)$ is a simple positive feedback mechanism, where higher polarity at $\bm{\theta}$ means higher likelihood of sampling it for subsequent addition of an activity patch. Taken together, sampling from the probability of Eq. \eqref{eq: cntr_pmf} generates more activity patches at points that have a reasonable amount of substrate in front of them and currently higher polarity values (Fig. \ref{fig: polarity sketch}).

The second term in Eq. \eqref{eq: dpol_dt} says that cells decrease their polarity in regions where they lose contact with the micropattern (places where $\chi\phi\neq0$). This term, along with the filopodial probability in Eq. \ref{eq: cntr_pmf}, are the only two mechanisms by which the cell interacts with the micropattern. {This inhibition term reflects a signaling interaction between the cell and the adhesive substrate. Contact between the cell and an adhesive substrate can up-regulate and localize Rac activity via focal adhesions and FAK \cite{xia2008directional,chang2006fak}. We treat the absence of this activation as an effective inhibition -- so we would expect the strength of the $\chi\phi$ term to decrease if the adhesivity of the substrate decreases or FAK is inhibited.}

The third term in Eq. \eqref{eq: dpol_dt} uniformly inhibits polarity when the cell perimeter $s$ exceeds a threshold $s_0$, using the Heaviside function $H(s-s_0)$. This assumption is motivated by experiments where membrane tension was observed acting as a global, long-range inhibitor of actin polymerization in neutrophils \cite{Houk2012} (though also see \cite{shi2018cell,de2023cell}). 

The last term in Eq. \eqref{eq: dpol_dt} decays the polarity over a timescale $\tau$, setting the characteristic time over which the cell can maintain a given polarity. 

{We note that Eq.  \eqref{eq: dpol_dt} does not have a term advecting the polarity to match the cell boundary motion. This equation thus assumes that the cell's polarity does not move significantly with respect to the substrate. We often implicitly think of polarity as localized to the basal membrane, making this an assumption that the membrane is relatively strongly linked to the substrate. This is a useful simplifying assumption, often used in models coupling cell shape and polarity \cite{shao2010computational,camley2017crawling,thuroff2019bridging}. Generalizations are possible, but require mechanical modeling of the cell interior \cite{camley2013periodic,shao2012coupling} or assumptions relating interior flow to boundary flow \cite{wolgemuth2011redundant}.}

\subsection{Parameter setting}
{The numerical values listed in Table \ref{table: params} define our default simulation cell, which is calibrated to behave similarly to cancerous MDA-MB-231 cells in \cite{Bruckner2019}. We take the values of $\lambda$, $\kappa$, $\eta$, and $\mathcal{M}$ from our earlier work \cite{pedrom}, where we calibrated to a different experimental study of MDCK cells migrating on narrow channels \cite{Jain}.  As in our earlier work \cite{pedrom}, we define many of our parameters relative to an arbitrary line tension scale $\gamma_0$---the results are not sensitive to the true value of the line tension, only how these different parameters relate to $\gamma_0$.}

\begin{center}
  \begin{table}[htb]
    \centering
    \begin{tabularx} {0.4\textwidth}{| >{\raggedright\arraybackslash}X | >{\raggedright\arraybackslash}X | >{\raggedright\arraybackslash}X |}
      \hline
      Parameter & Value & Units \\
      \hline
      $\gamma$ & 0.8 & $\gamma_0$ \\
      $\lambda$ & 4.8 & $\micron$ \\
      $\kappa$ & 6000 & $\gamma_0\ \micron$ \\
      $R_0$ & 18 & $\micron$ \\
      $\eta$ & 0.67 & $\gamma_0\ \rm min\ \micron^{-3}$ \\
      $\mathcal{M}$ & 0.75 & $\micron\ (\gamma_0\ \rm min)^{-1}$ \\
      $\mu_\beta$ & 7.5 & unitless \\
      $\sigma_\beta$ & 37.5 & unitless \\
      $\sigma$ & 6.7 & $\micron$ \\
      $s_0$ & $2\pi (1.5R_0)$ & $\micron$ \\      
      $\tau_f$ & 2.2 & $\rm s$ \\
      $\tau_\chi$ & 9.6 & $\rm s$ \\
      $\tau_s$ & 8 & $\rm min$ \\
      $\tau$ & 4 & $\rm min$ \\
      $\ell$ & 6 & $\micron$ \\
      $\alpha$ & 1.4 & $\gamma_0\ \micron^{-2}$ \\
      \hline
    \end{tabularx}
    \caption{A list of simulation parameters along with the values that define our default cell, which is calibrated to behave similarly to MDA-MB-231 cells in \cite{Bruckner2019}. See Appendix \ref{appendix: numerical integration} for parameters relevant to the numerical integration scheme.}
    \label{table: params}
  \end{table}
\end{center}

{Many parameters in the model are set fairly roughly to represent large penalty terms. For instance, the constraint penalizing deviations from the preferred area, $\kappa$, is a large value, which  ensures the area of the default cell varies less than $0.5\%$ from its target -- we do not think of this as a relevant biological parameter but primarily a simple way to set a constraint. Similarly, the timescale over which cell polarity decays to zero in regions where the cell is near substrate edges, $\tau_\chi$, is set so that when the cell boundary leaves the micropattern, the polarity rapidly becomes negative, leading to the membrane contracting away from the micropattern edge, preventing the cell from leaving the pattern entirely. The perimeter-induced degradation in polarity prevents the cell from becoming too elongated. Since we do observe that cells in \cite{Bruckner2019} can be quite elongated, we cannot set the timescale of this process $\tau_s$ too short, or it will prevent cell stretching. We choose $\tau_s$ to be on the order of the global decay $\tau$. We set the magnitude of the motility force $\alpha$ so that our simulated cells achieve speeds and accelerations that roughly match the experiments. The rest of the parameters in Table \ref{table: params} will be discussed in \S\ref{sec: results-intro} as we consider their variations and effect on cell behavior.}

{The phenomenological parameters describing cell polarity dynamics like $\tau_f$, the frequency of polarity patches, and the magnitude and variability $\mu_\beta$ and $\sigma_\beta$ of patch polarity, are largely set to recapitulate the core limit cycle behavior for the default cell. Matching these values to other experiments is not straightforward, but the value of $\tau_f = 2.2$ s is at least plausible -- the Rho GTPase cdc42, a key element of mammalian cell polarity, is pulsatile with a timescale of $\sim$ a few seconds \cite{yang2016locally}. However, the lifetime of protrusions may be quite different than this value ($\sim$ minutes, on our rough scale of $\tau$), both in experiment (see e.g. \cite{marshall2024excitable}) and in our simulation, given the feedback mechanisms we include. Future work to more precisely calibrate the model would require measurements of protrusion statistics on MDA-MB-231 in the absence of a geometric signal.}

{We also note that, though some of the values shown in Table \ref{table: params} are not round numbers, the model is not too carefully tuned -- these numbers are essentially historical artifacts of the original fitting process, and were set as some percentage of an original value. Other awkward values arise because of scaling between simulation parameters and physical units. We address the sensitivity of the model in Section \ref{sec: perturbing-activity-patch} below.}

\subsection{Inferring motility patterns from simulations}
\label{sec: footprint inference}

We begin each simulation by either positioning the cell randomly in one of the two basins of the two-state substrate or uniformly distributing it across the rectangular micropattern. Upon initialization, we numerically evolve the phase field equations of motion for 48 hours (see Appendix \ref{appendix: numerical integration}). \textit{Unlike} the experiments, which recorded the position of the cell nucleus, we track the $x$ position of the cell center of mass. While nuclear dynamics could be modeled with another phase field \cite{Camley-PRM-models}, we neglect it here for simplicity and claim that cell center of mass is a fair approximation of cell nucleus for most of the morphologies we observe. 

We adapt the approach of Br\"uckner et al. \cite{Bruckner2019} to our simulations by representing the dynamics of the cell center-of-mass position and velocity with the stochastic equation of motion,
\begin{equation}
\dot{v}=F(x, v)+\sigma(x, v)\eta(t),
\end{equation}
where $\eta(t)$ is a Gaussian white noise with $\langle \eta(t) \rangle=0$ and $\langle \eta(t)\eta(t')\rangle=\delta(t-t')$. Here, we focus only on $F(x,v)$, which denotes the average center-of-mass acceleration of the cell when it has position $x$ and velocity $v$. Since cell motility is stochastic, we perform $N=4800$ simulations of 48 hours for every cell-micropattern pair. Mirroring the approach of \cite{Bruckner2019}, we compute velocities and accelerations as numerical derivatives, construct the stochastic acceleration field $\dot{v}(x, v)$ from each simulation, and define the deterministic footprint by computing the average acceleration $\dot{v}$ conditional on the value of $x$ and $v$, i.e.  $F(x, v)\equiv\langle \dot{v}|x, v\rangle$ (refer to Appendix \ref{appendix: acc. footprint} for details on this inference scheme).

Similar to \cite{Bruckner2019}, we interpret $F(x, v)$ as the blueprint that defines the deterministic migration of the cell on a given micropattern geometry. One way to visualize the motility pattern encoded by $F$ is to trace its streamlines---trajectories that outline the instantaneous direction of motion in phase-space. For instance, we can begin with a point in the $xv$ phase-space, say $(x_0, v_0)$, and integrate it one timestep forward to ($x_0+v_0 dt,\ v_0 + dt F(x_0, v_0)$). Repeating this procedure will eventually reveal the phase-space trajectory, i.e., the streamline, dictated by $F(x,v)$ when starting at $(x_0, v_0)$. A streamline will terminate once it reaches a point at which both acceleration and speed are zero, i.e., the point $(x_f, v_f)$ such that $F(x_f, v_f)=0$ and $v_f=0$.

We generally see streamline plots that have one of three typical forms: 1) a single limit cycle, where streamlines converge onto a closed loop, 2) ``stationary'' behavior, where streamlines all converge to $v = 0$ but do not have a strong dependence on $x$, and 3) bistability, where two sets of disconnected streamlines converge to zero speed at either basin $x=\pm x_{\mathrm{basin}}$. When a streamline terminates, we mark its end point with an $\bm \times$ and interpret it as the point to which the cell would relax to if its equation of motion were to evolve without any noise. When multiple terminal points occur, we group trajectories with common endpoints and represent each group with distinct colors. We then draw the corresponding separatrices with solid lines to delineate the boundaries between regions of phase space that are attracted to different terminal points. 

The blueprints $F(x, v)$ are inferred from a finite number of observations, leading to some uncertainty in the values of $F(x,v)$. This can lead to notable uncertainty in some terminal points and separatrices. We use a bootstrapping technique to show the scale of expected variations in our streamline plots, terminal points, and separatrices in Appendix \ref{appendix: separatrix variations}.

\section{Results}
\label{sec: results-intro}

\subsection{Model predicts limit-cycles on two-state and stationary behavior on rectangular micropatterns}
\label{sec: lc-stationary-states}

Can our proposed model of the cell simultaneously capture limit-cycles on two-state and stationary behavior on rectangular geometries, as observed for cancerous MDA-MB-231 cells by \cite{Bruckner2019}?

Figure \ref{fig: lc timeseries} plots one hopping cycle of the cell as it navigates the two-state geometry in our simulations (see Movie M\ref{mov: limit-cycle} for a longer trajectory). The cell adopts a singly-polarized morphology with a clear cell front that is rich in polarity and elongates to travel across the bridge between the two square micropatterns. We plot trajectories of the cell center of mass from a few simulations in Fig. \ref{fig: lc sampletrajs}. Our default cell, calibrated to MDA-MB-231 cells, is persistently hopping between basins, as MDA-MB-231 cells in \cite{Bruckner2019} do. We then infer the deterministic acceleration footprint $F(x, v)$, which is a concise way of summarizing the migration of the cell observed across 4800 simulations. Figure \ref{fig: lc streamline} plots this footprint in the $xv$ space using a colormap. When we compute the streamlines (as described in Section \ref{sec: footprint inference}), we find that the motion of the cell described by $F(x,v)$ forms a closed loop in the $xv$ space. This mean that our model, at its default parameters, creates a limit-cycle behavior, where cells hop persistently between basins. This finding is consistent with the limit cycle behavior reported in \cite{Bruckner2019}, which we have reprinted in Fig. \ref{fig: lc brucknerstream}.

Our default cell behaves entirely differently on the rectangular geometry. Figure \ref{fig: ss timeseries} plots a few snapshots of the cell as it migrates on the rectangular substrate, where we observe the cell adopting a bipolar shape and undergoing a ``tug-of-war'' with competition between two protrusions, similar to that of MDA-MB-231 cells in experiments \cite{Bruckner2019}. The center of mass trajectories in Figure \ref{fig: ss sampletrajs} show that the cell is not persistently migrating but has small fluctuations and occasional larger steps. Note that unlike migration in the two-state geometry, cells can remain near $x=0$ on the rectangular micropattern for long periods of time, not just transiently while hopping. Mirroring the above analysis, we infer $F(x, v)$ from simulations and plot it in Figure \ref{fig: ss streamline}. This time, tracing the streamlines of $F(x,v)$ reveals trajectories that bring the cell quickly to the $v = 0$ line, with termination points distributed along $x$, suggesting that our model on the rectangular geometry, at its default parameters, generates an $F(x,v)$ where cells quickly relax to become stationary with a position near its initial position. We call this sort of streamline plot ``stationary'' throughout the paper. This finding is consistent with the similar stationary behavior of cancerous MDA-MB-231 cells observed in \cite{Bruckner2019}, which we have reprinted in \ref{fig: ss brucknerstream}.

As a side note, we should mention that the specific locations of the terminal points in Fig. \ref{fig: ss streamline} are somewhat uninformative. For this stationary case, the cell will have near-zero $F(x,v)$ near $v = 0$ for a broad range of values of $x$, small sampling error can change the positions of the $\times$-labeled termination points. In later plots that show this ``stationary'' dynamics, we do not label putative termination points.

\begin{figure*}[htbp]
    \centering
    \textbf{Case 1: Default cell limit-cycles on two-state geometry}
    \timeseries{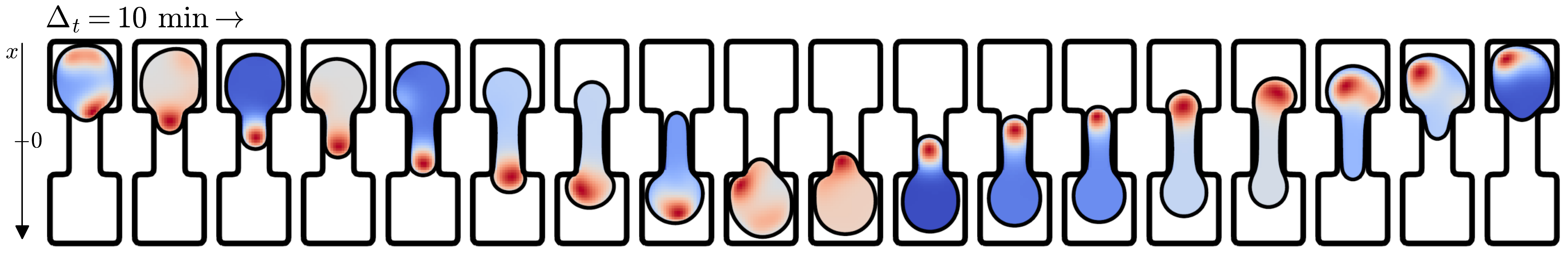}{lc}
    \sampletrajs{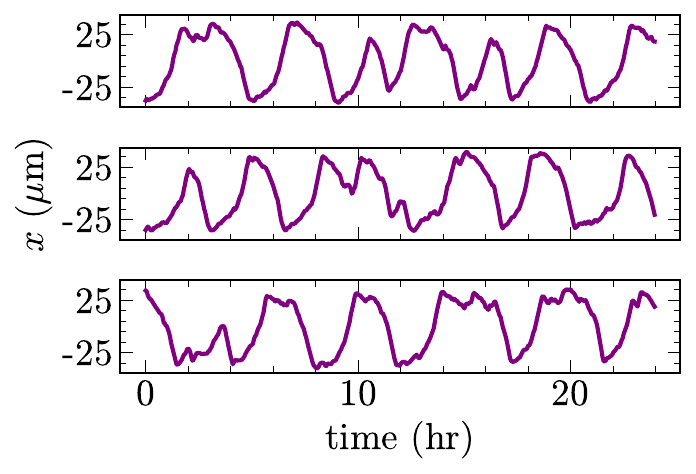}{lc} \qquad
    \streamline{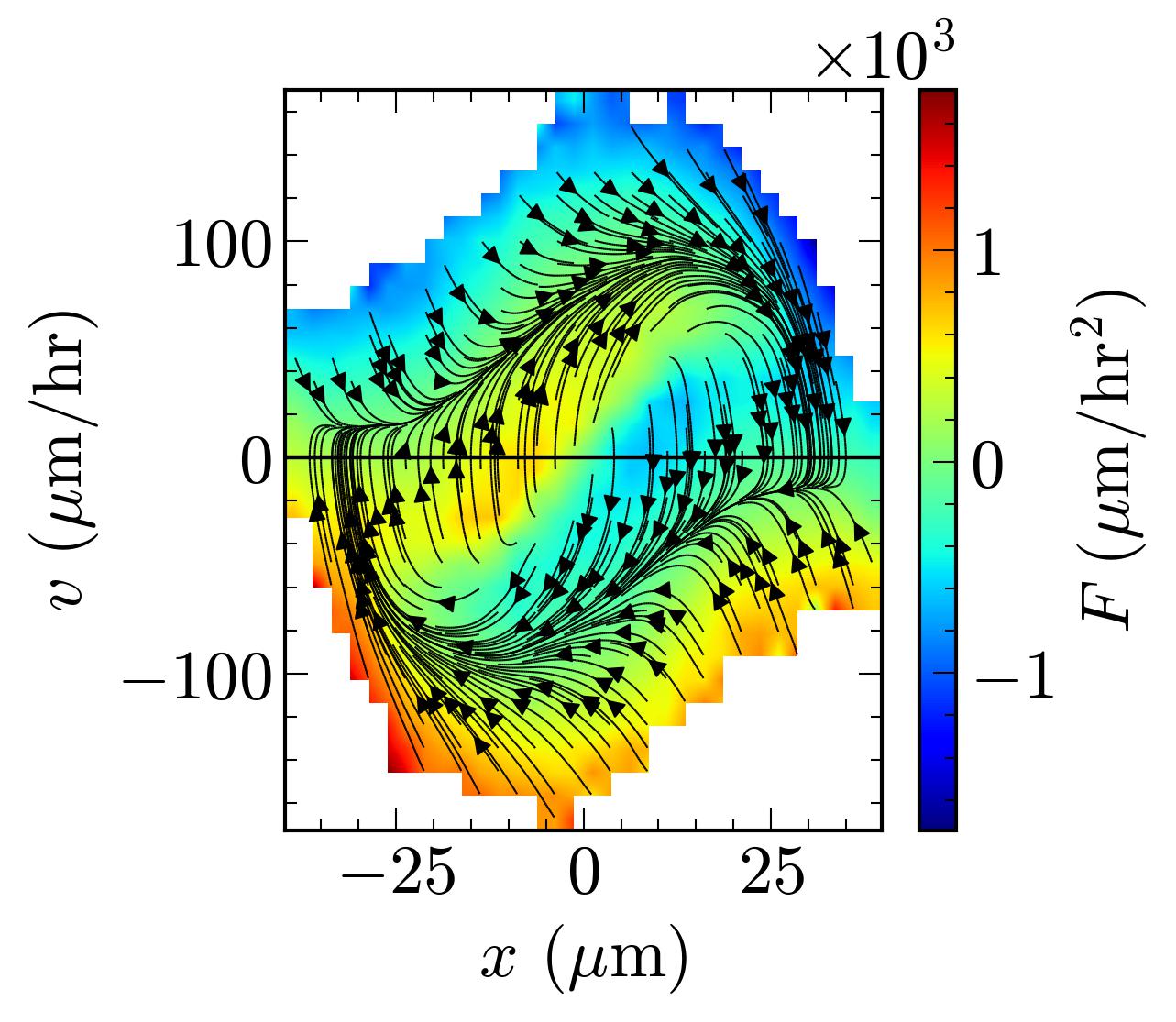}{lc} \qquad
    \brucknerstream{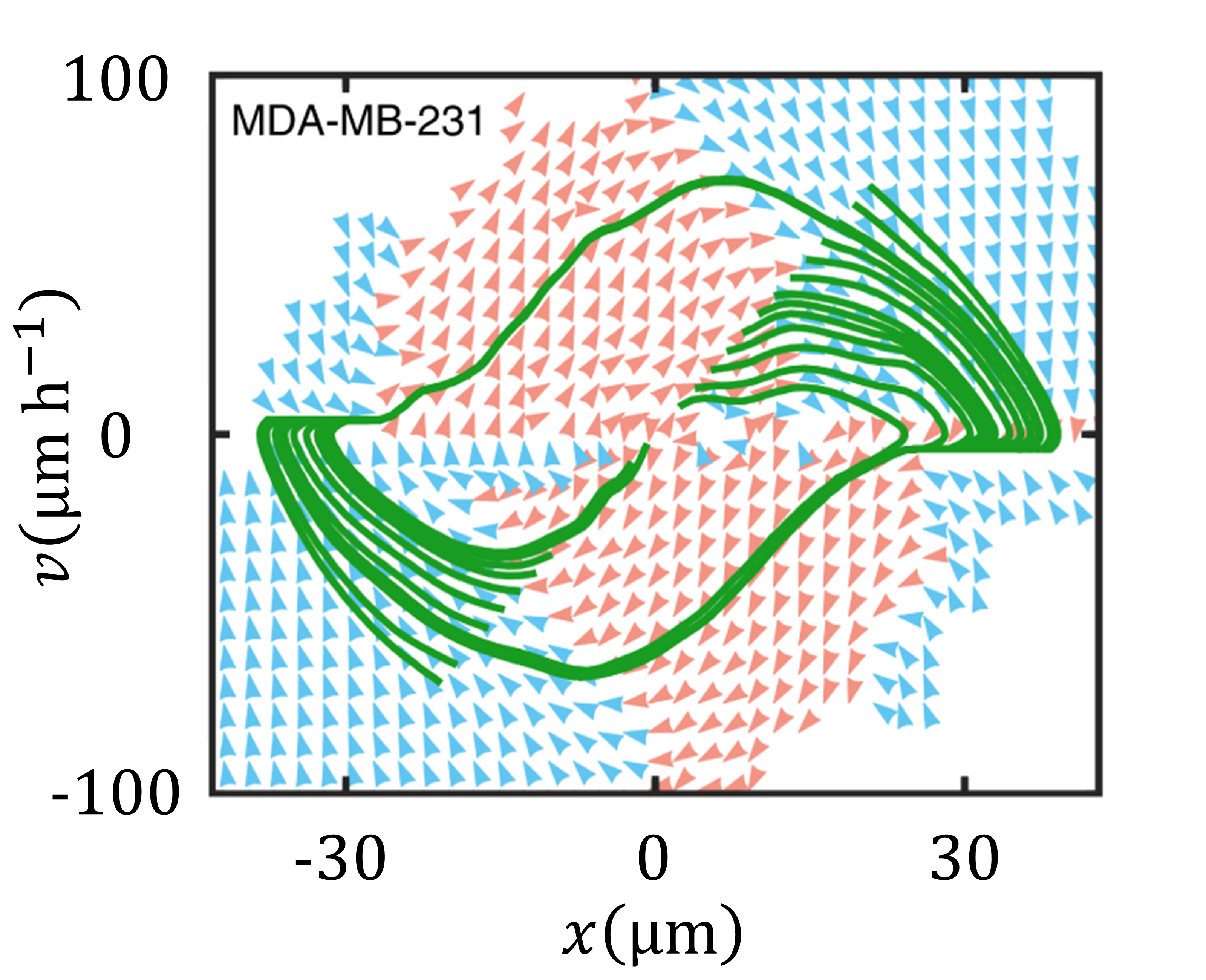}{lc}

    \textbf{Case 2: Default cell is stationary on rectangular geometry}
    \timeseries{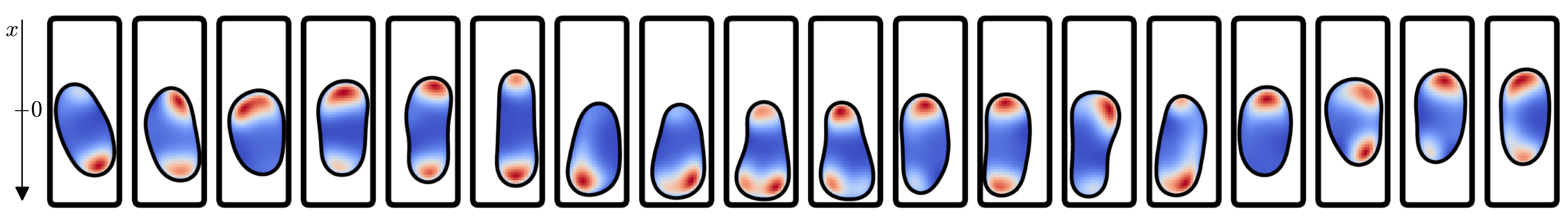}{ss}
    \sampletrajs{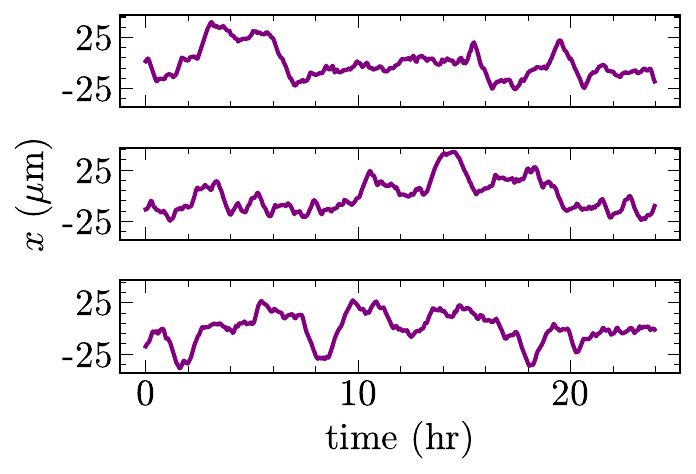}{ss} \qquad
    \streamline{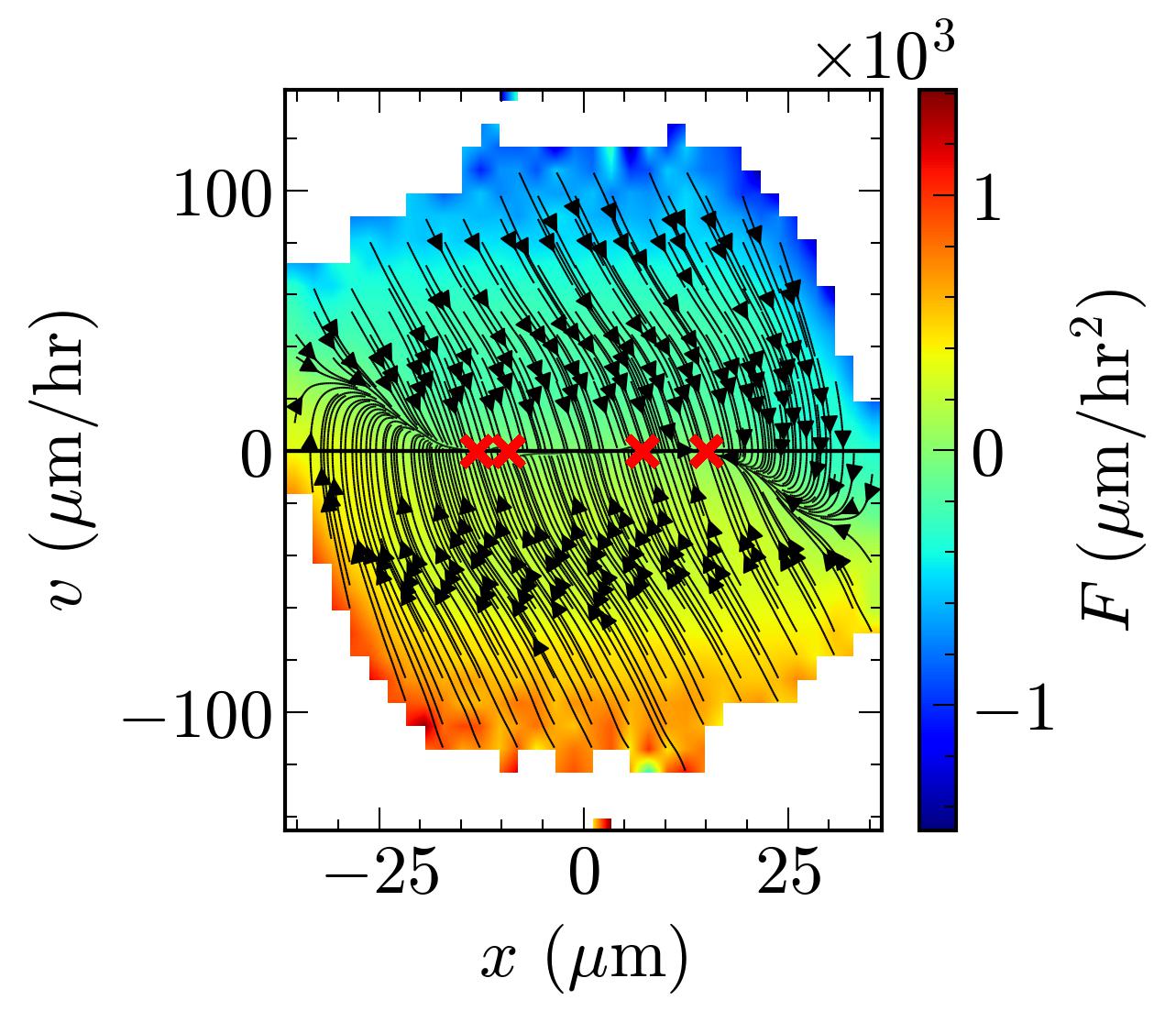}{ss} \qquad
    \brucknerstream{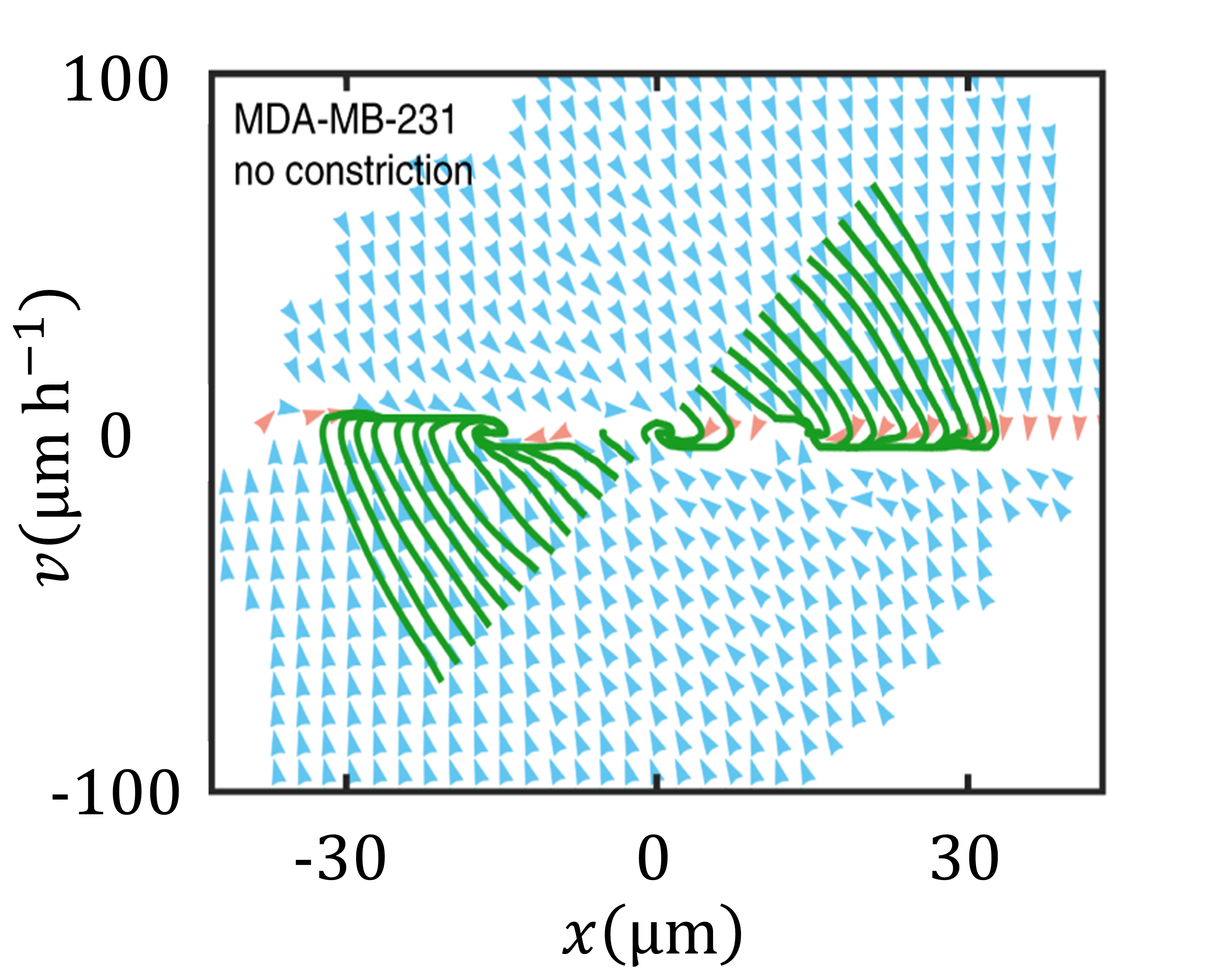}{ss}

    \textbf{Case 3: Decreasing cell size ($R_0=15\micron$) causes bistability on two-state geometry}
    \timeseries{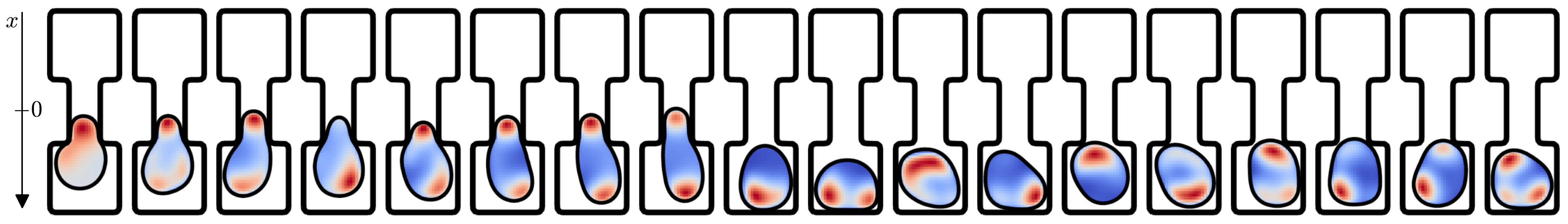}{bs}
    \sampletrajs{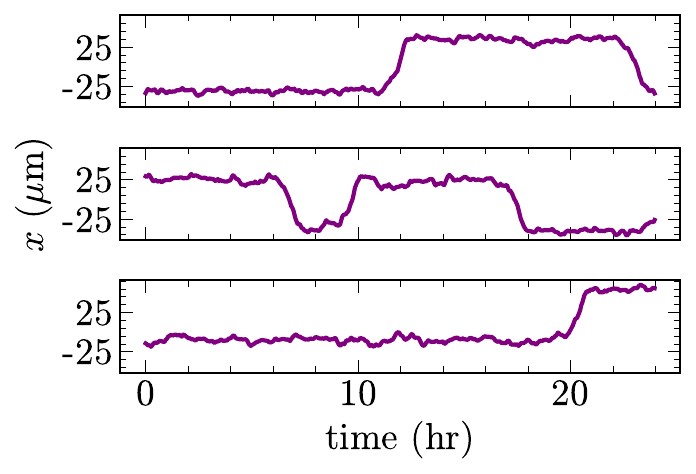}{bs} \qquad
    \streamline{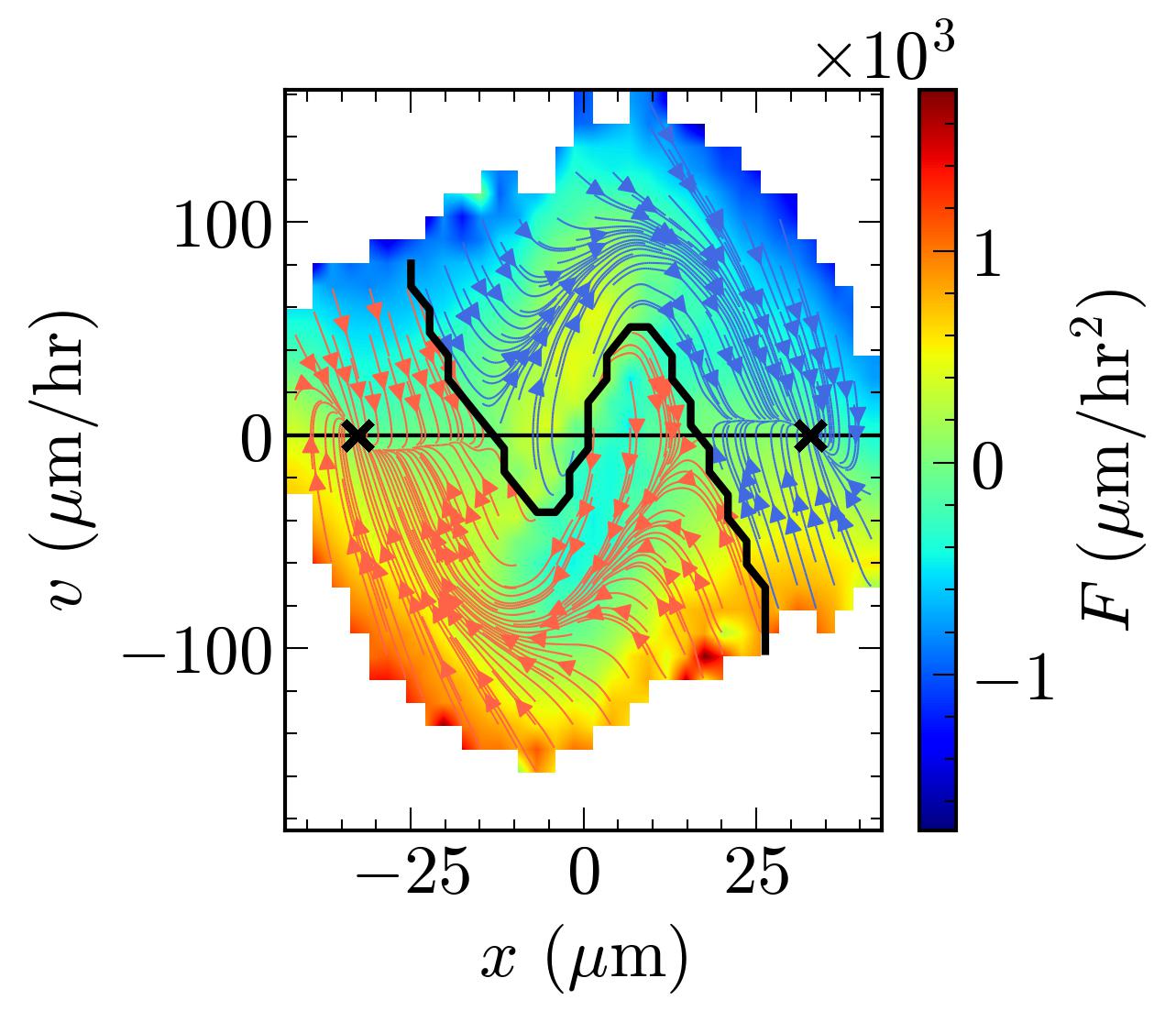}{bs} \qquad
    \brucknerstream{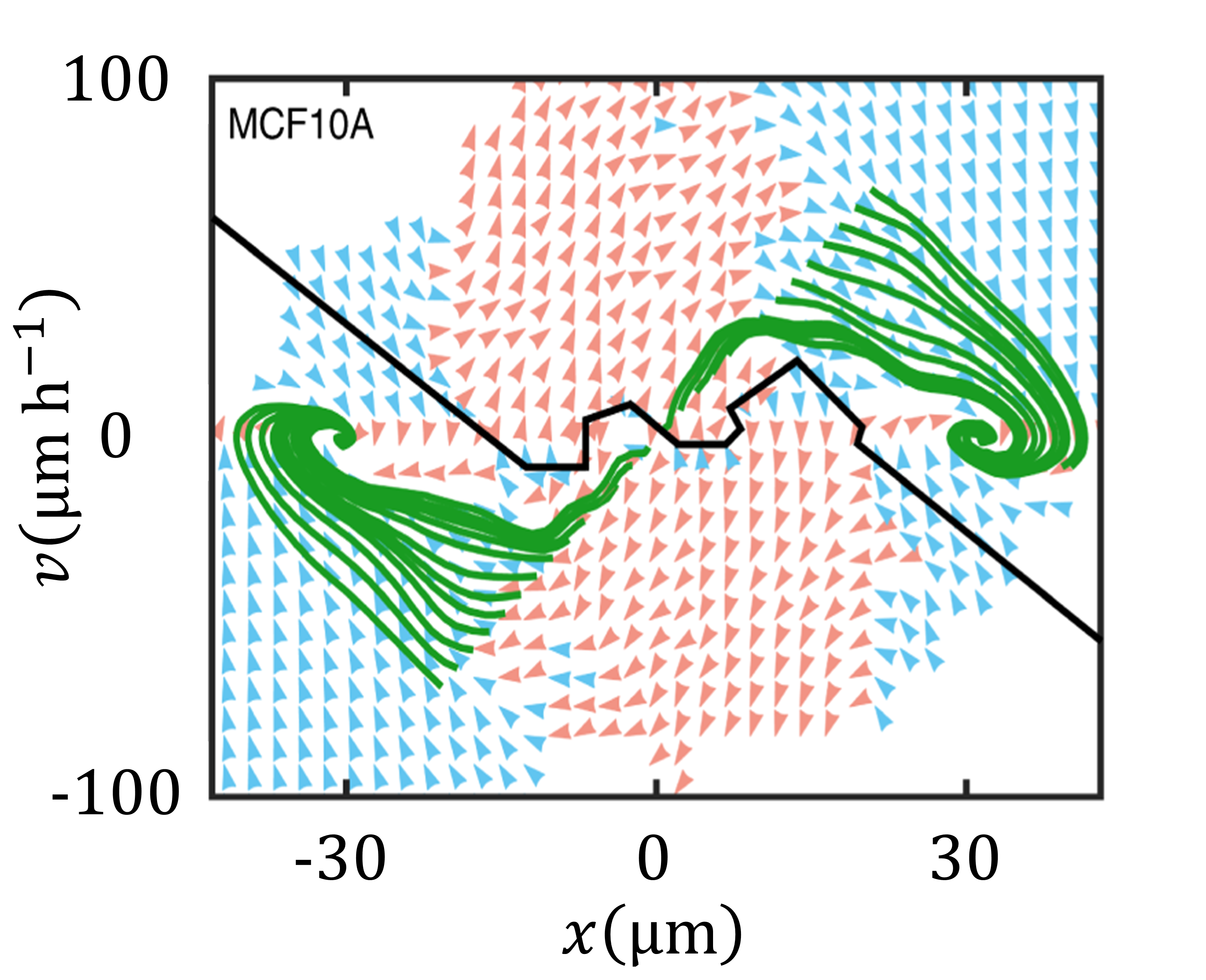}{bs}
    
    \caption{Overview of limit-cycle (\subref{fig: lc timeseries}-\subref{fig: lc brucknerstream}), stationary (\subref{fig: ss timeseries}-\subref{fig: ss brucknerstream}), and bistable (\subref{fig: bs timeseries}-\subref{fig: bs brucknerstream}) behavior. Each case shows sample snapshots {separated by time $\Delta_t = 10 \textrm{min}$} of typical morphologies and polarity (red: high, blue: low) (\subref{fig: lc timeseries}, \subref{fig: ss timeseries}, \subref{fig: bs timeseries}), outlines center-of-mass trajectories (\subref{fig: lc sampletrajs}, \subref{fig: ss sampletrajs}, \subref{fig: bs sampletrajs}), plots $F(x, v)$ inferred from simulations (\subref{fig: lc streamline}, \subref{fig: ss streamline}, \subref{fig: bs streamline}), and reprints the experimental footprints of \cite{Bruckner2019} with permission for comparison (\subref{fig: lc brucknerstream}, \subref{fig: ss brucknerstream}, \subref{fig: bs brucknerstream}). Streamlines are colored by equilibrium point; $\times$ indicates equilibrium points where streamlines terminate. Solid black lines are separatrices---crossing this line changes the long-run behavior of the cell. See Section \ref{sec: footprint inference} for details.}
    \label{fig: trimodal trajectories & streamlines}
\end{figure*}

\subsubsection{Geometry sensing and the origin of limit-cycles}

Let's consider migration in the two-state geometry. When the cell is inside a basin and large enough to fill it up, most of its interface lies near the edges of the confinement except the part facing the bridge. Since we degrade polarity in regions where the cell escapes the micropattern (Eq. \eqref{eq: dpol_dt}), and suppress the generation of activity in regions that are near to the end of the micropattern (filopodial probability of Eq. \eqref{eq: cntr_pmf}), sustained polarity can only develop in regions facing the bridge. This region further experiences a targeted increase in average cell polarity due to the positive feedback assumption of Eq. \eqref{eq: cntr_pmf}. The tendency to polarize toward free micropattern space and positive feedback work cooperatively to yield a persistent, singly-polarized cell that migrates through the bridge and into the other basin. Once on the other side, the cell reverses in polarity as the bridge now faces the opposite direction and repeats the cycle above. The result is a persistently polarized cell that limit-cycles on the two-state geometry. Interestingly, we find that the filopodial probability of Eq. \eqref{eq: cntr_pmf} is dispensable---the assumption that leaving the micropattern decreases cell polarity is sufficient for reproducing limit-cycles in the two-state micropattern (see Section \ref{sec: model-knockouts} for details).

\subsubsection{Origin of stationary behavior on rectangular substrates}

Let's now consider the evolution of cell polarity on the rectangular substrate. Suppose polarity is initially constant in space. Since the cell is confined at its top and bottom, sustained polarity can only develop at the sides, yielding elongation in $\pm x$. Given the stochastic evolution of cell polarity and our assumption of positive feedback, which favors a one-side-wins-all configuration, we would initially expect the cell to be singly-polarized and have one clear front, just like it is on the two-state substrate. Instead, we observe the average cell shape is bipolar (see Fig. \ref{fig: ss timeseries} and Movie M\ref{mov: stationary}). This bipolarity arises because of a coupling between the cell polarity and the evolution of cell shape. When we numerically solve the polarity model of Eqs. (\ref{eq: dpol_dt}) and (\ref{eq: cntr_pmf}) on a \textit{static} phase field (see Appendix \ref{appendix: evolving field required}), we consistently arrive at singly-polarized cells. What is different about a cell whose boundary dynamically moves? The polarity field $\mathbb{P}(\mathbf{r},t)$ is confined within the body of the cell, and while it responds to motion of the cell boundary, it does not do so instantaneously. When a high-amplitude patch of activity hits the contour ${\bm \theta}_k$, it increases the polarity in that region and leads to a protrusion of the cell membrane. This protrusion moves ${\bm \theta}_k$ normally outward and away from its previous location. Because $\mathbb{P}$ does not instantly migrate with the cell boundary, the polarity measured at the same contour point shortly after protrusion is not nearly as high as before---i.e. polarity can be implicitly {\it suppressed} by protrusion. {We hypothesize that this negative feedback, combined with the stochasticity of the protrusion dynamics, can prevent our positive feedback terms in Eq. (\ref{eq: cntr_pmf}) from leading to a singly-polarized cell. Earlier work has also found that simple Rho GTPase polarity models can have coexisting peaks over relevant timescales, though there is a tendency of peak merger, but the timescale of coexistence is strongly parameter-dependent \cite{chiou2018principles}. It may thus be possible to reproduce this coexistence by tuning parameters even in a stationary cell. Nevertheless, our discovery that cell shape dynamics can qualitatively change polarization dynamics is part of a broader realization that complex behaviors can emerge from the coupling of cell polarization to cell shape \cite{Nwogbaga2023, Camley-PF-eq2, shape-polarity}. }

\subsection{Cell size modulates bistability on two-state substrates}
\label{sec: cell-size-bistability}

Thus far, we have demonstrated our model's ability to recapitulate limit cycles on two-state and stationary behavior on rectangular substrates. Now, we ask which properties of the cell might dictate whether it shows limit cycles or bistability on the two-state micropattern. These are potential candidates for being key differences between MDA-MB-231 cells and MCF-10A cells. In this section, we focus on perturbing cell size, which we control through the target radius $R_0$, leaving all other parameters the same as those in Table \ref{table: params}.

Figure \ref{fig: bs timeseries} presents a few simulation snapshots of a smaller cell with $R_0=15\micron$ on the two-state geometry. According to these snapshots, the smaller cell experiences difficulty hopping between the basins and appears more likely to repolarize back into either state of the micropattern. Note that while the snapshots we highlighted don't include a hopping event, smaller cells can indeed hop between the basins of the micropattern, and when they do, their morphology is highly similar to that of the default cell in Fig. \ref{fig: lc timeseries} (see Movie M\ref{mov: bistable} for a simulation that includes hopping). The lack of persistent hopping in smaller cells is readily apparent from the center-of-mass trajectories of Figure \ref{fig: bs sampletrajs}, where we observe smaller cells to remain in a given basin for much longer times with infrequent hopping to the other side. When we infer $F(x, v)$ and trace its streamlines, we find, in contrast with the limit-cycle dynamics of Fig. \ref{fig: lc streamline}, trajectories that terminate either at the left or right basin. We color trajectories that terminate in the left basin red and those that terminate in the right basin as blue. The two stable steady states of $F(x,v)$ show that our model predicts smaller cells to be bistable (Fig. \ref{fig: bs streamline}), with streamlines similar to the bistability reported in \cite{Bruckner2019} (reprinted in Fig. \ref{fig: bs brucknerstream}).

If we decrease the cell size significantly below $R_0 = 15 \micron$, we find that the cells essentially migrate freely within the basins---very small cells rarely interact with the edges of the micropattern and almost never enter the bridge or successfully hop. Lastly, making cells larger than our default values does not affect their migration pattern: cells larger than the default $R_0=18\micron$ still limit-cycle with morphologies and streamlines that are highly similar to the default cell shown in Figures \ref{fig: lc timeseries}-\subref{fig: lc streamline}.

Why are smaller cells bistable in our model?  When the cell is relatively small, its body does not overlap with the edges of the basin as much, so polarity is not as strongly suppressed near the micropattern edges. This leads to increased chances of developing a cell front anywhere along the cell periphery, not just in regions pointing at the bridge. This in turn increases the rate of repolarizations. Since smaller cells have lower chances of consistently polarizing into the bridge, they do not hop persistently, but rather stochastically, and are bistable. On the other hand, larger cells fill up the basin nearly entirely, leading to a consistent dampening of polarity along the cell periphery except around the regions pointing toward the bridge. This promotes persistent elongation into the bridge and thus limit cycles. The underlying mechanism is just that cells with larger spread area are forced to have more contact with the micropattern edges---so we expect that larger cells are more likely than small ones to limit-cycle in any model that drives cells to repolarize away from the edges of the micropattern. We expect this result would be fairly robust to changes in our detailed model assumptions.

Although the experiments of \cite{Bruckner2019} do not monitor cell size quantitatively, their example movies show that MDA-MB-231 cells are typically spread over a larger area compared to MCF10A cells. There is also evidence that cancer cells have increased lamellipodial activity compared to their healthy counterparts \cite{yamaguchi}. Additionally, the absence of the tumor-suppressing micro RNA family miR-200 has been shown to result in increased sizes of gastric cancer cells \cite{Yu2022}. Collectively, these findings support the notion that cancer cells tend to have larger spread areas, rendering our model's nomination of cell size as a predictor of motility behavior in two-state confinement plausible.

\subsection{Cell tension could modulate bistability on two-state substrates}
\label{sec: cell-tension-bistability}

\begin{figure}[htb]
    \centering
    \subfigure[]{\includegraphics[width=0.8\linewidth]{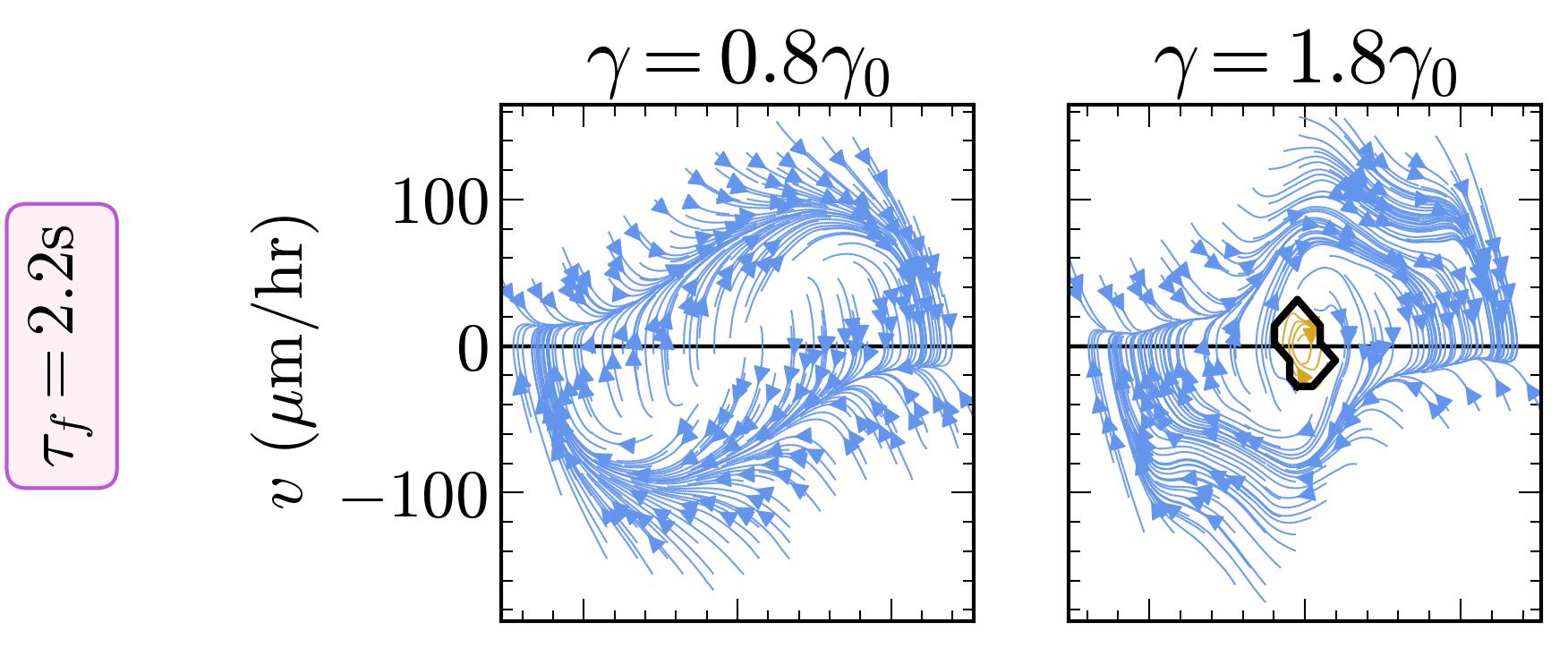} \label{fig: tension lc}}
    \subfigure[]{\includegraphics[width=0.8\linewidth]{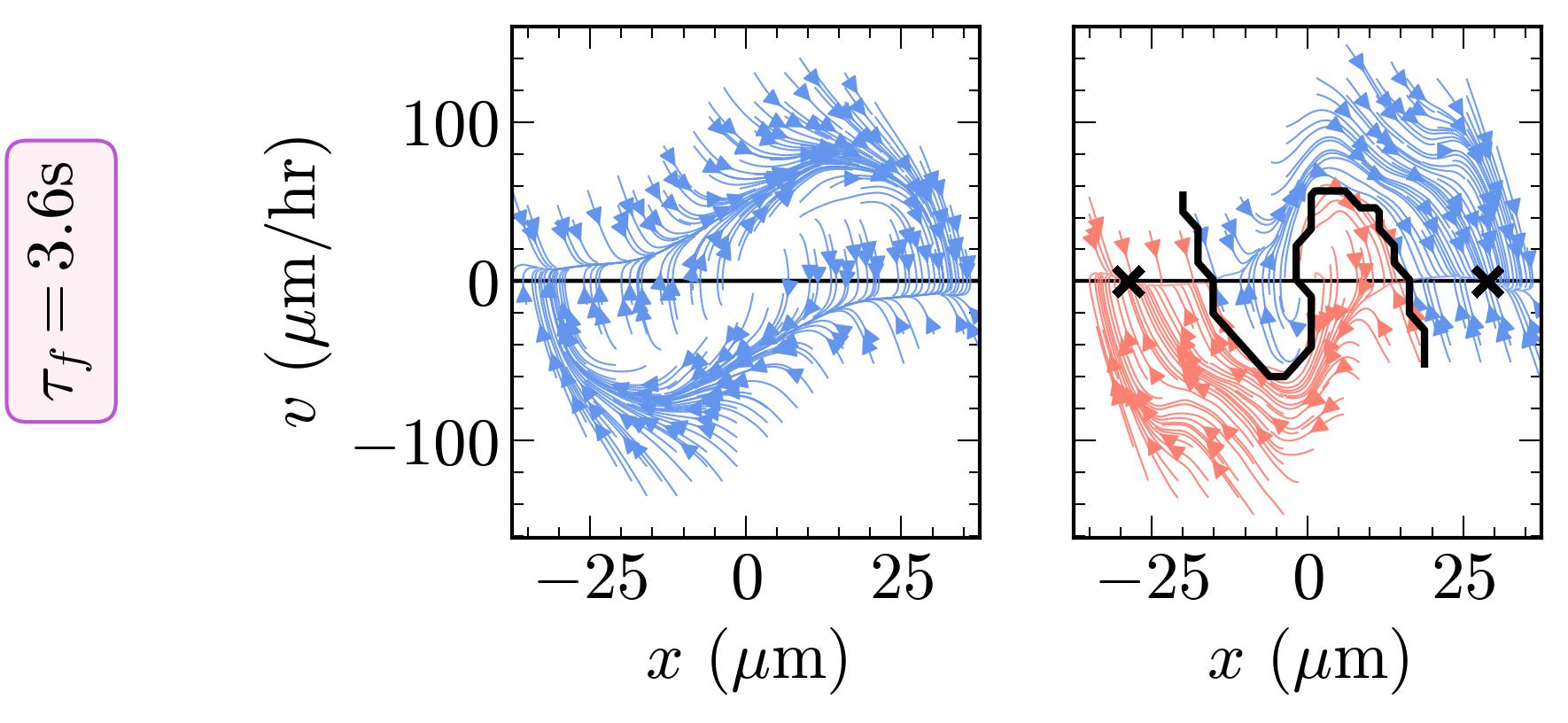} \label{fig: tension bs}}
    \caption{Motility patterns of soft ($\gamma=0.8\gamma_0$) and stiff ($\gamma=1.8\gamma_0$) cells on the two-state micropattern. \subref{fig: tension lc} Under the default simulation parameters, stiff cells mostly exhibit limit-cycles, but \subref{fig: tension bs} they can transition to bistable behavior when accompanied by a reduction in the intensity of cell polarity. Streamlines are colored by equilibrium point; $\times$ indicates equilibrium points where streamlines terminate. Solid black lines are separatrices---crossing this line changes the long-run behavior of the cell. See Section \ref{sec: footprint inference} for details.}
    \label{fig: varying_cell_tension}
\end{figure}

To hop between the basins, cells must elongate into the narrow bridge and tolerate a great deal of shape deformations (see the snapshots in Fig. \ref{fig: lc timeseries}). When $\gamma$, which sets the cell membrane tension, is low, i.e., the cell is soft, deformations are not very costly and can occur more easily. At high $\gamma$, the cell is stiff and shape deformations are costly. Therefore, we would expect soft cells to hop readily and stiff cells to resist hopping. To test this hypothesis, we first perturb our default cell by increasing its membrane tension to $\gamma=1.8\gamma_0$ and then simulate its migration on the two-state micropattern. Contrary to our intuition, we observe that the stiff cell still limit-cycles (Fig. \ref{fig: tension lc}), albeit with a small subset of the phase-space that originates and gets stuck oscillating inside the bridge (yellow region near $x = 0, v = 0$).
When we look at the migration of stiff cells ($\gamma=1.8\gamma_0$) on two-state substrates, we see an unexpected behavior: stiff cells round up when passing through the bridge, which causes parts of the cell body to temporarily escape the micropattern (see Movie M\ref{mov: stiff-lc}). Further increasing stiffness does not dampen limit-cycles---we simply observe a greater fraction of cells escaping the micropattern. Why does this happen? While higher tension naturally tends to make the cell round up, the active forces will still push the cell across the bridge. In the lower-$\gamma$ simulations (Fig. \ref{fig: lc timeseries}), the cell's boundary contracts to within the micropattern because the polarity quickly becomes negative when the cell leaves the micropattern, leading to an inward-pointing force. However, once the cell boundary manages to leave the micropattern region, e.g. because the membrane tension resisting increasing the cell perimeter in the inward contraction is too large to overcome, no force is locally generated to bring it back (Eq. \ref{eq: motility_force}). As a result, the cell will round up and migrate across the bridge. We are not confident if this behavior is likely to be observed in experiment; it certainly would be strongly altered if we explicitly modeled an adhesive energy between the cell and the micropattern or altered our assumptions about where the cell could generate force. 

Based on our analysis above, we think we can avoid the suspect behavior of the cell leaving the micropattern if we decrease the strength of active motility in some way. We can dampen the intensity of cell polarity by increasing the timescale $\tau_f$ over which new stochastic activity is added. We repeat the above numerical experiments with $\tau_f=3.6$s (compared to the default value $\tau_f = 2.2 s$). Under this alternate parameter regime, where cell polarity is less intense, we observe stiff cells ($\gamma = 1.8 \gamma_0$) to be bistable (Fig. \ref{fig: tension bs}) and softer cells ($\gamma = 0.8 \gamma_0$) to limit-cycle.

Our analysis shows that cell tension---or more generally, factors controlling cell deformability---can control whether the cell shows a bistable or limit cycle behavior. In \cite{Bruckner2019}, cancerous cells limit-cycled and healthy cells were bistable. Evidence suggests that cell stiffness is a potential biomarker for cell malignancy with cancer cells tending to be softer \cite{stiffness-cancer-biomarker, guck2005optical, lincoln2004deformability, remmerbach2009oral}. Taken together, this lends confidence to our model's nomination of cell stiffness as a potential predictor of bistability on the two-state micropattern.

\subsection{Model recapitulates ``negative friction'' on two-state and normal friction on rectangular geometries}
\label{sec: nonlinear dynamics}

\begin{figure}[htb]
    \centering
    \includegraphics[width=0.9\linewidth]{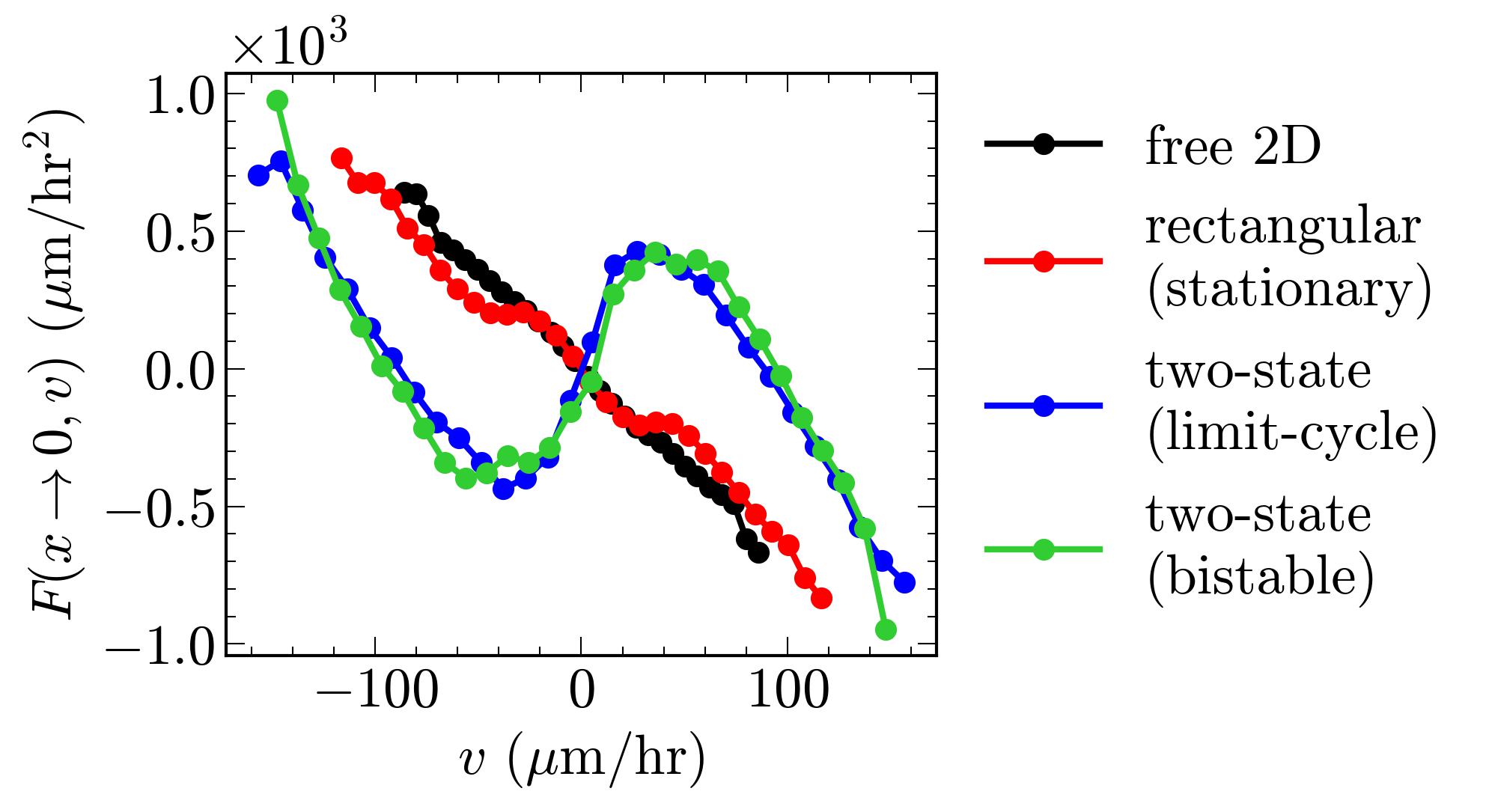}
    \includegraphics[width=0.9\linewidth]{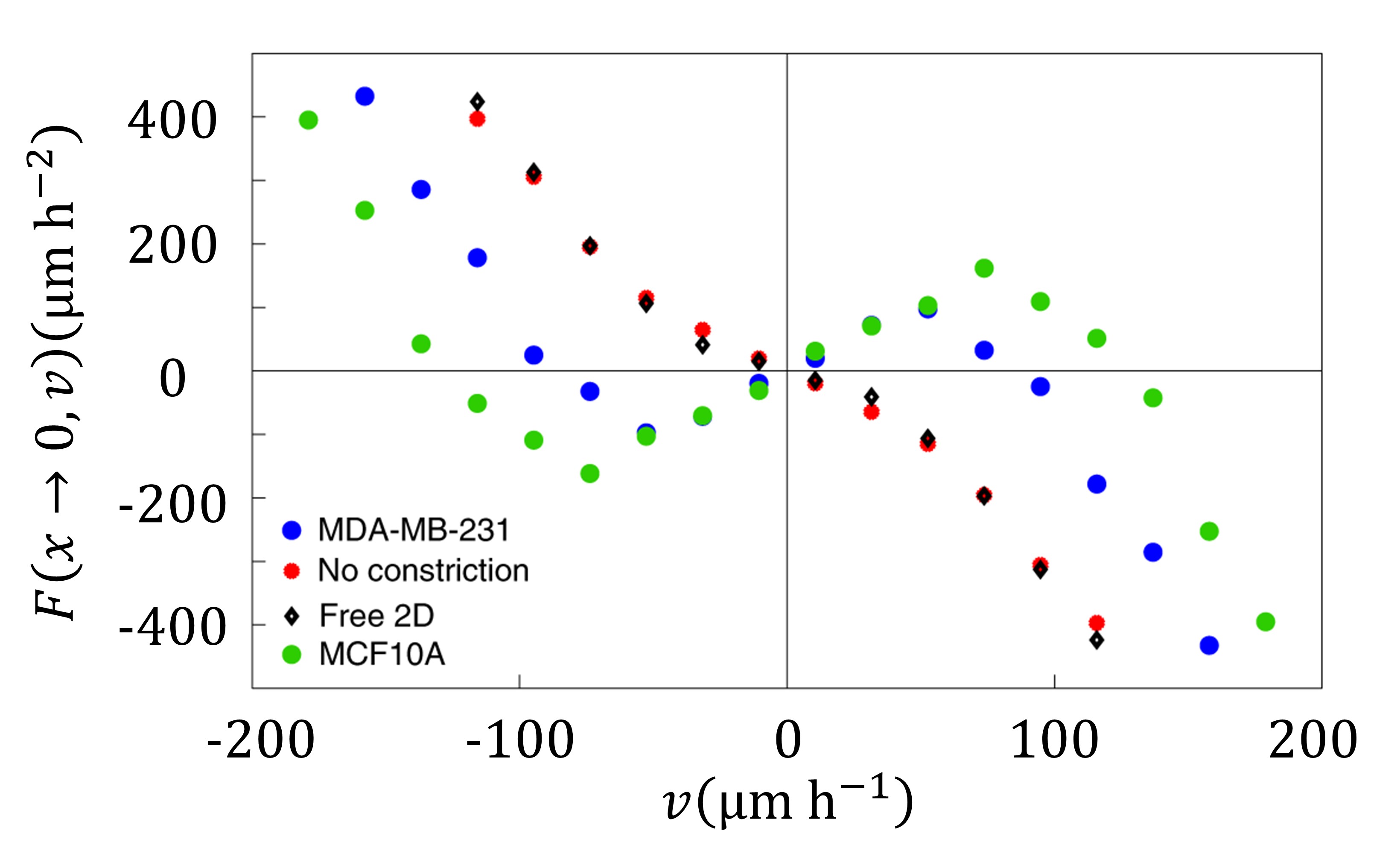}
    \caption{(Top) Simulation: Center-of-mass acceleration near the micropattern center exhibits a nonlinear (nearly linear) velocity dependence in two-state (rectangular) geometries. (Bottom) Experimental acceleration-velocity relationships for cell nucleus from \cite{Bruckner2019}, reprinted with permission.}
    \label{fig: nonlinear plots}
\end{figure}

Following \cite{Bruckner2019}, we take a cut through the acceleration footprint near $x=0$, which denotes the symmetry axis of the micropattern, and ask how the center-of-mass acceleration at $x = 0$ varies as a function of velocity, $F(x\to0, v)$. For typical cell migration on a 2D substrate, which can often be captured by an Ornstein-Uhlenbeck-type model $\dot{\bm v}=-\xi \bm v + \eta(t)$ \cite{Huang, Gail-Boone, SELMECZI2005912, bruckner2024Review}, we would expect that the average acceleration $F(x \to 0,v)$ would be linear, i.e., $F(0,v) = -\xi v$, akin to a normal friction. Instead, our model predicts a nonlinear relationship for cells migrating on the two-state micropattern, with cells with small velocities tending to accelerate. This ``negative friction'' regime is consistent with the experimental measurements of \cite{Bruckner2019} (Fig. \ref{fig: nonlinear plots}). $F(x\rightarrow0, v)$ is the average acceleration of cells whose center of mass is in the middle of the bridge. When the center of mass is in the bridge middle, the cell is in a dumbbell-like shape: its front is about to enter the new basin and the rear is about to retract (see Fig. \ref{fig: lc timeseries}). Both of these features mean that the cell will likely accelerate, as once the rear enters the bridge, it is not longer blocking forward motion, and once the front enters the new basin, it can generate protrusive force over a broader range. We thus think that the negative friction seen here requires the constriction of the cell as it enters the bridge.

This negative friction is entirely due to the narrow constriction in the two-state micropattern. For cells migrating on rectangular micropatterns, we observe a nearly linear acceleration-velocity relationship, which is also consistent with the experiments of \cite{Bruckner2019}  (Fig. \ref{fig: nonlinear plots}). {We also see a similar near-linear behavior in $F(x\to0,v)$ in unconstrained cells on two-dimensional substrates fully coated with fibronectin (``free 2D'' curve of Fig. \ref{fig: nonlinear plots}). Our observation that the rectangular and two-dimensional geometries have nearly-identical, nearly-linear curves is also consistent with the experimental results of \cite{Bruckner2019}. A movie of a cell in an unconfined geometry is shown in Movie 5.}

\subsection{Perturbing the polarization patch model}
\label{sec: perturbing-activity-patch}

\begin{figure*}[htb]
    \centering
    \begin{minipage}[t]{0.46\textwidth}
    \centering
    Two-state Micropattern
    \end{minipage}
    \begin{minipage}[t]{0.53\textwidth}
    \centering
    \hspace{0.23\textwidth} Rectangular Micropattern
    \end{minipage}
    
    \rowperturbation{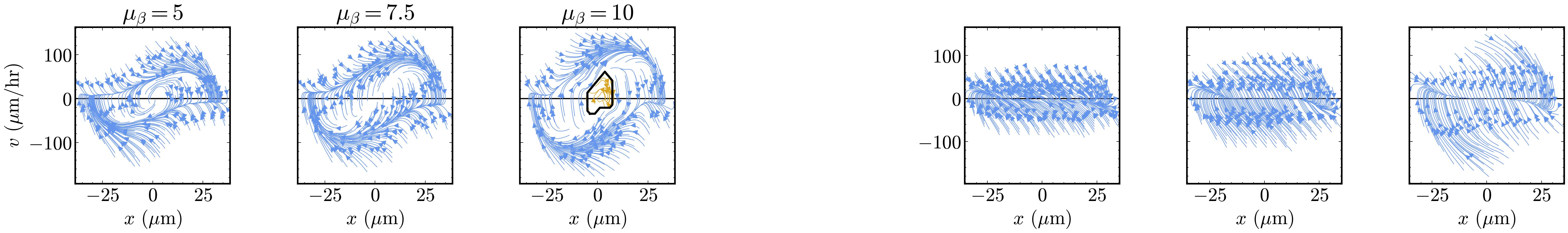}{mub}
    \rowperturbation{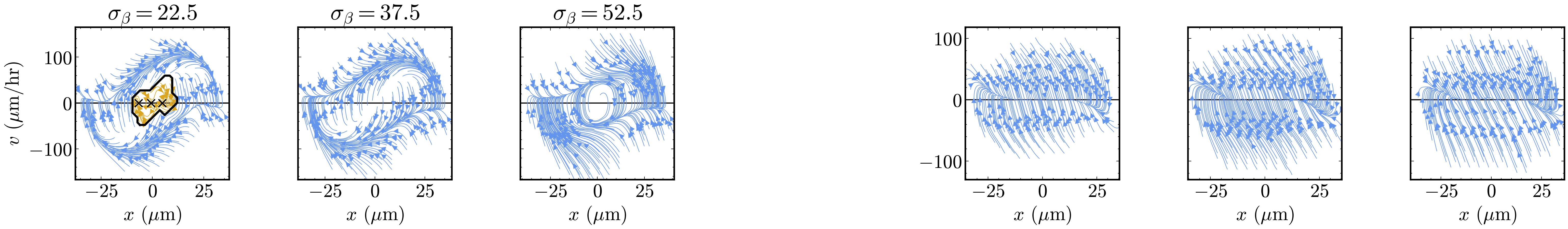}{sigmab}
    \rowperturbation{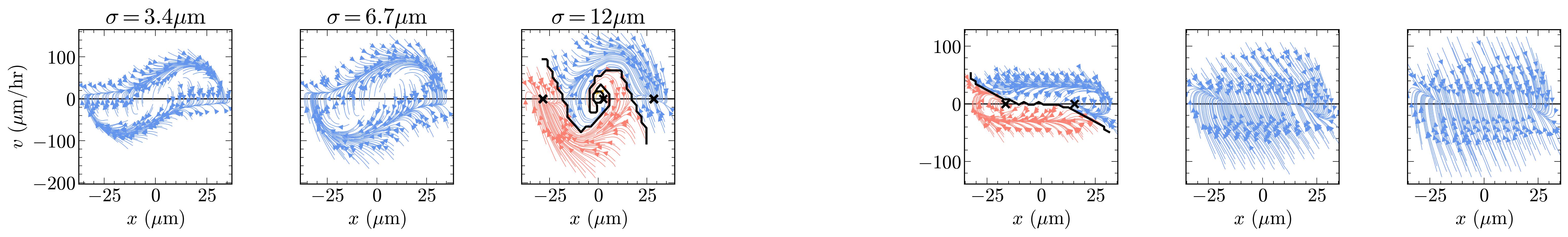}{sigma}
    \rowperturbation{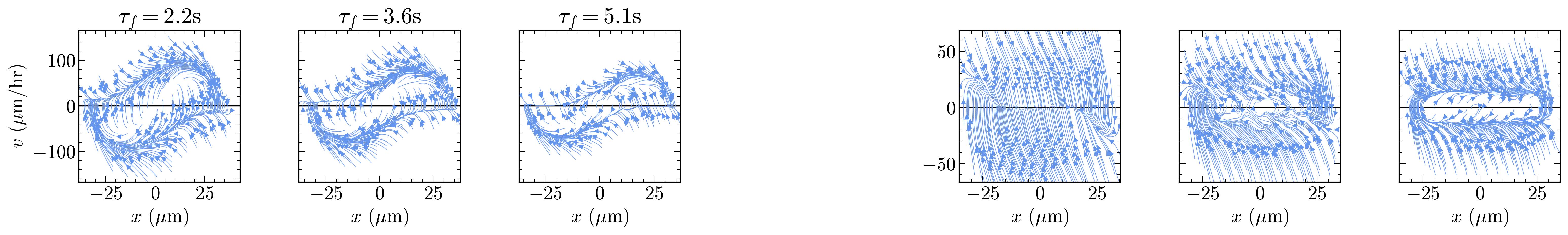}{tauf}
    \caption{An overview of how various perturbations in the spatially-correlated activity patch we add to cell polarity affect the deterministic migration of the default cell on the two-state (left column) and rectangular (right column) micropatterns. In each column, the middle figure illustrates the behavior of the default cell, except in the last row, where the left figure ($\tau_f = 2.2$ s) is the default cell. The perturbations considered are in \subref{fig: mub perturbation} the mean and \subref{fig: sigmab perturbation} standard deviation of the magnitude of the noisy activity patch, \subref{fig: sigma perturbation} the spatial extent of the correlated activity patch, and \subref{fig: tauf perturbation} the frequency of adding noisy activity to cell polarity. Streamlines are colored by equilibrium point; $\times$ indicates equilibrium points where streamlines terminate. Solid black lines are separatrices---crossing this line changes the long-run behavior of the cell. See Section \ref{sec: footprint inference} for details. {Results are obtained from 2400 simulations.}}
    \label{fig: varying-activity-patch}
\end{figure*}

A key assumption of our model is characterizing the stochastic behavior of cell polarity by generating patches of activity every $\tau_f$ seconds, with a profile of $\beta(t) f\left(|\bm r - \bm \theta(t)|; \sigma\right)$, with amplitude $\beta$ and location $\bm \theta$ chosen randomly. In this section, we perturb this activity module away from its default state and gauge how sensitive our predictions are to its various components. Figure \ref{fig: varying-activity-patch} outlines all the variations considered. As discussed below, our predictions are robust against certain aspects of the activity module, while other parameters are tightly constrained to recapitulate experimental observations.

\subsubsection{Perturbing \texorpdfstring{$\beta$},, the magnitude of noisy activity}
\label{subsub: vary-beta}

$\beta(t) \sim \mathrm{Gaussian}(\mu_\beta, \sigma_\beta)$ is a random variable that sets the magnitude of the stochastic patches we add to cell polarity. We find that our model predicts limit-cycles on two-state and stationary behavior on rectangular micropatterns relatively robustly against variations in both $\mu_\beta$ (Fig. \ref{fig: mub perturbation}) and $\sigma_\beta$ (Fig. \ref{fig: sigmab perturbation}). However, there are a few interesting features that can appear within the larger limit-cycle dynamics. For the two cases where $\mu_\beta=10$ and $\sigma_\beta=22.5$, a new behavior emerges---in addition to the limit-cycles---that is confined to the interior of the bridge. This suggests that cells that happen to enter the bridge of the two-state micropattern with near-zero velocity may become trapped there, and either perform a small limit-cycle or reach a stationary state. These behaviors at the bridge are ones we are slightly skeptical about observing in experiments akin to  \cite{Bruckner2019}, in part because our model does not have an explicit representation of the nucleus, while \cite{Bruckner2019} use the cell's nucleus to parameterize $x$. We would expect the nucleus to not be stably trapped exactly in the middle of the micropattern. 

Lastly, we note that for $\sigma_\beta=52.5$, the limit-cycles do not reach from the left basin to the right basin, but instead the streamlines loop around a narrower region. This corresponds to an oscillation of the cell center of mass around the bridge in our plot of $F(x,v)$. When we inspect our movies of cell trajectories, we see that in this case, cells quickly re-enter the bridge once they have entered a basin (see Movie M\ref{mov: noisy-lc}).

We varied $\mu_\beta$ and $\sigma_\beta$ over a range set by monitoring the behavior of the cell. For values lower than $\mu_\beta = 5$ and $\sigma_\beta = 22.5$, there isn't a sufficient buildup of polarity, and we do not observe the cells to hop. Similarly, for higher values, we see many cell body ruptures as polarity becomes large enough to split one cell into many. The perturbations in this section show that our predictions are not overly sensitive to the magnitude of the activity patch so long as $\beta(t)$ is within a reasonable range.

\subsubsection{Perturbing \texorpdfstring{$\sigma$},, the spatial extent of correlated activity}
\label{subsub: vary-cov-mat}

When we add a patch of activity to the cell, it is spread over a region $f\left(|\bm r - \bm \theta(t)|; \sigma\right)=e^{-|\bm r - \bm \theta(t)|^2/2\sigma^2}$---the parameter $\sigma$ sets the spatial extent over which one stochastic event affects the polarity.

Figure \ref{fig: sigma perturbation} summarizes the motility of cells with different values of $\sigma$. At the low value of $\sigma=3.4\micron$, we still observe the cells to limit-cycle on the two-state micropattern. However, a new behavior is observed on the rectangular geometry---cells are now bistable, settling deterministically into the left and right sides of the micropattern with two distinct terminal points. We do not have a strong intuition for why cells prefer the sides of the rectangular micropattern when their activity patches are tightly localized. At the large value of $\sigma=12\micron$, we recover the original stationary behavior on the rectangular geometry, but we observe cells to be mostly bistable on the two-state substrate (note the small fraction of lattice sites terminating near $(x=0, v=0)$ in Fig. \ref{fig: sigma perturbation}). While we do not know with certainty why disperse patches lead to nearly bistable behavior in our model, we think it has to do with how spatially-extended patches activate a broader set of contour points. This, in turn, may increase the likelihood of activity occurring elsewhere on the cell boundary, thereby reducing the probability of persistent singly-polarized morphology. We conclude that $\sigma$ significantly impacts the migration of the cell in our model, leading to a tight range of values within which we can recapitulate experimental observations.

\subsubsection{Perturbing \texorpdfstring{$\tau_f$},, the frequency of noisy activity}
\label{subsub: vary-tau-f}

The last parameter in our stochastic patch model is $\tau_f$, the frequency with which we add noisy patches to cell polarity. We find that our model robustly predicts limit-cycles on two-state micropatterns and stationary behavior on rectangular micropatterns when changes in this timescale are small (Fig. \ref{fig: tauf perturbation}). When we nearly double $\tau_f$ from its default value, we observe limit-cycles not only on the two-state substrate but also on the rectangular micropattern. This may occur because as $\tau_f$ increases, the time between stochastic protrusion events becomes large enough for the cell to develop a robust singly-polarized state, which then leads to a cell migrating persistently along the rectangular micropattern. This result shows that even the seemingly simple and intuitive result that cells on the rectangular pattern have near-stationary dynamics can be constraining to a detailed model.

{Balance is required between some pairs of parameters to generate successful migration}
{We also study the variation of pairs of model parameters simultaneously. We find that there is a balance required between terms that tend to promote cell polarization, and those that degrade it. For instance, if the generation of protrusion pulses is made significantly slower, increasing $\tau_f$, cells will not be able to generate enough force to actually cross from one basin to the other, and streamlines will be disconnected (Fig. \ref{fig:two_taus}.) However, if the polarity degradation time $\tau$ is also increased, the limit cycle behavior can be recovered. Similarly, decreasing $\tau_f$ can be compensated by increasing the amplitude of polarity patches $\mu_\beta$ (Fig. \ref{fig:tauf_mub}).}

\subsection{Can we simplify our model assumptions?}
\label{sec: model-knockouts}

Having confirmed our model's ability to recapitulate the findings of \cite{Bruckner2019}, we now investigate whether a simplified version of it can achieve the same success. We do this by knocking out various aspects of the model and observing how each affects the migration dynamics of our default cell on the two-state and rectangular micropatterns.

\begin{figure}[htb]
    \centering
    \includegraphics[width=\linewidth]{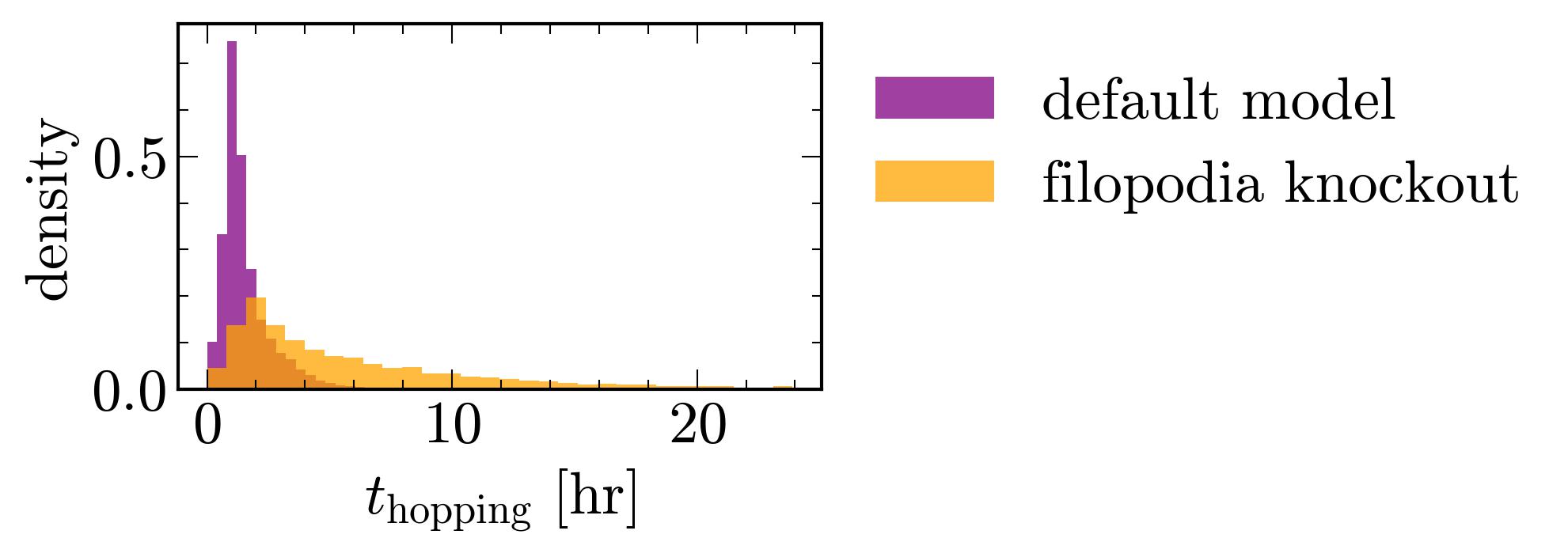}
    \caption{Distribution of ``hopping times'' predicted by the default model \textit{vs.} one with the filopodial mechanism absent.}
    \label{fig: hopping times}
\end{figure}

We begin by asking if the two mechanisms by which the cell integrates information about substrate geometry, i.e., the filopodial probability in Eq. \ref{eq: cntr_pmf}, and the coupling to $\chi$ in Eq. \ref{eq: dpol_dt}, are both needed. We knock out the filopodial sensing mechanism by simplifying the probability of Eq. \eqref{eq: cntr_pmf} to $p(\bm \theta)\sim \mathbb{P}(\bm \theta)$. Upon this knockout, we observe no significant change in the motility pattern or the velocity dynamics of the default cell on either the two-state or rectangular micropatterns (Fig. \ref{fig: filopodia_knockout_global}). However, we observe a notable change in the distribution of ``hopping times'', defined as the duration of time spent in a basin of the two-state geometry before hopping (Fig. \ref{fig: hopping times}). While our default model predicts a tightly-peaked distribution, consistent with the findings of \cite{Bruckner2019}, this perturbation leads to a broader density. The broadening is expected; without the filopodial probability, cells have reduced ability to aim their protrusive activity toward the bridge, which naturally delays the hopping process and extends the hopping time. Considering the hopping time distribution, we opt to retain the filopodial probability to ensure a more faithful reproduction of the observations of \cite{Bruckner2019}.

\begin{figure}[htb]
    \centering
    \subfigure[]{\includegraphics[width=\linewidth]{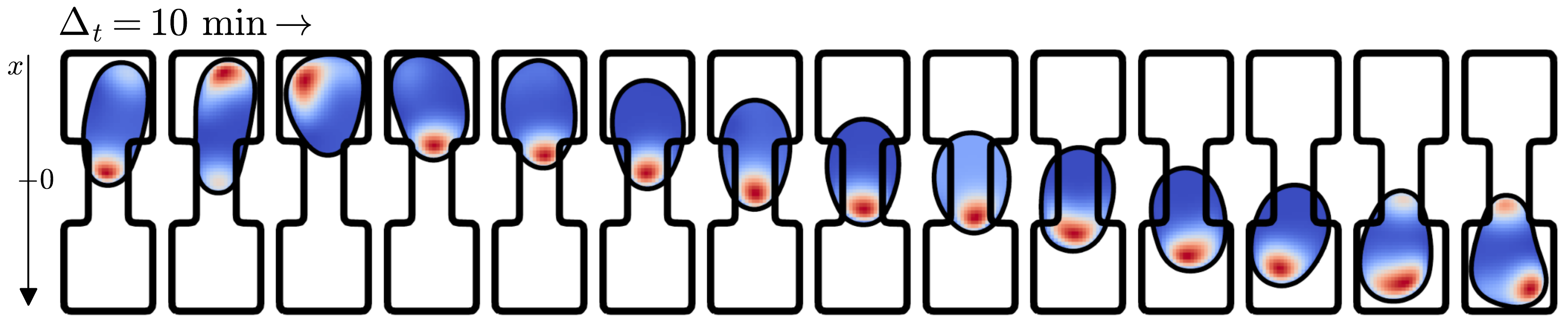} \label{fig: tauXknx timeseries}}
    \hfill
    \subfigure[]{\includegraphics[width=\linewidth]{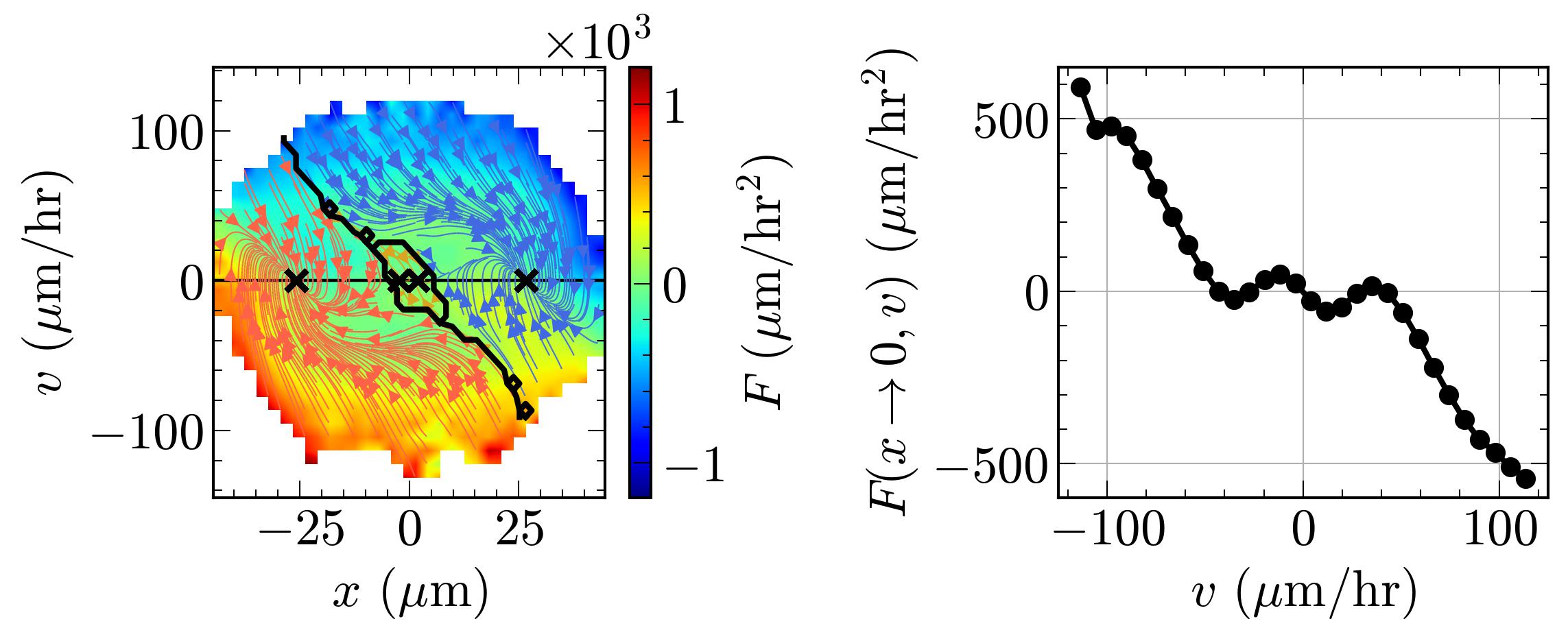} \label{fig: tauXknx streamline}}
    \caption{\subref{fig: tauXknx timeseries} Sample simulation snapshots of our default cell migrating on the two-state micropattern, along with \subref{fig: tauXknx streamline} a summary of its motility dynamics when cell-confinement coupling is turned off, i.e., $\tau_\chi\rightarrow\infty$.}
    \label{fig: substrate_knockout}
\end{figure}

Next we knock out the coupling of cell polarity to the substrate by taking $\tau_\chi$ to infinity (Fig. \ref{fig: tauXknx timeseries}). In this case, the cell is still largely confined within the substrate due to the filopodial sensing mechanism disfavoring any polarization that would cause the cell to escape. While motility in the rectangular substrate remains unaffected (Fig. \ref{fig: tauX_knx_rect}), migration in the two-state micropattern develops multiple equilibrium points: while the two points at either basin are robust to sampling uncertainties, the putative equilibrium points near the origin are not (see Fig. \ref{fig: tauXknx err}). We thus interpret the streamlines of Fig. \ref{fig: tauXknx streamline} as effectively suggesting a bistable behavior. Additionally, the acceleration at $x \to 0$ no longer displays the pronounced negative friction ($F(x\to0,v) \sim + v$) that is characteristic of migration in the two-state micropattern (compare Fig. \ref{fig: tauXknx streamline} to the blue and green curves in Fig. \ref{fig: nonlinear plots}). Both the filopodial probability and the cell-micropattern coupling sense spatial cues, altering polarity in response to the presence of the micropattern, yet only the latter disrupts the motility dynamics of the cell. Why is that? The key difference between each sensory input is likely how they affects cell polarity. The filopodial sensing mechanism changes the distribution of polarity activation patches. Because the average effect of these activation patches is positive, redistributing these activity patches will not generally create a negative polarity at any point in the cell. By contrast, the micropattern coupling term $-\tau_\chi^{-1} \chi \phi$ can make the polarity negative and lead to the membrane retracting away from the micropattern edge. When the cell is about to enter the bridge and the cell-micropattern coupling is present, the constant decrease in polarity pinches the membrane near the corners of the constriction and facilitates the formation of tubular protrusive fronts while holding back the cell body (Fig. \ref{fig: lc timeseries}). Without this coupling, the cell does not contract its membrane at the pinch points of the constriction; it fails to grow tubular fronts and adopts a morphology that is unfazed by the micropattern constriction (see Fig. \ref{fig: tauXknx timeseries} and Movie M\ref{mov: substrate_knox}). Considering this morphological change, the absence of a clear negative friction dynamic with this knockout is not surprising given we think that squeezing through the bridge is necessary for exhibiting nonlinear dynamics (see our argument in Section \ref{sec: nonlinear dynamics}).

Lastly, we knock out the perimeter-induced degradation of cell polarity by taking $s_0\rightarrow\infty$ and observe no significant change in the motility pattern or the velocity dynamics of the default cell on either micropatterned geometry (Fig. \ref{fig: perimeter_knockout_global}). Although we could safely eliminate this assumption and still recapitulate the experimental observations, we choose to keep it because it leads to a cell morphology that is closer to that of MDA-MB-231 cells in \cite{Bruckner2019}. Without it, there is a higher chance for cells to become bipolar midway through the bridge or for one bipolar cell to break apart into two (see Movie M\ref{mov: perim_knox}). Given the clear experimental rationale for this sort of term \cite{Houk2012}, we keep the perimeter-dependent global inhibition in our default model.

{We could also consider other potential model variants. For instance, it would be natural to consider eliminating our micropattern-sensing terms in Eq. \ref{eq: dpol_dt} and replacing them with, e.g. an energy of adhesion to the micropattern \cite{pedrom,Cao,albert2014dynamics}. This would then create an energy barrier for the cell leaving the micropattern. This energy barrier would successfully confine the cell and prevent it from escaping the micropattern, but this variant of the model would not suffice to repolarize the cell -- there would be no inhibition of polarity when the cell hit the micropattern edge, preventing it from turning around. Within our model, this would lead to an absence of limit cycles. We believe the role of cell-substrate adhesion to have both an important mechanical effect as well as a signaling (repolarization) effect. Adhesive energy, in combination with alternate models of polarity, might be sufficient to create limit cycles, though this is speculative. Within the context of our current work, we do not treat this adhesive energy, since it is not necessary to recapitulate our core results.}

\begin{figure*}[htbp]
    \centering
    \includegraphics[width=1\linewidth]{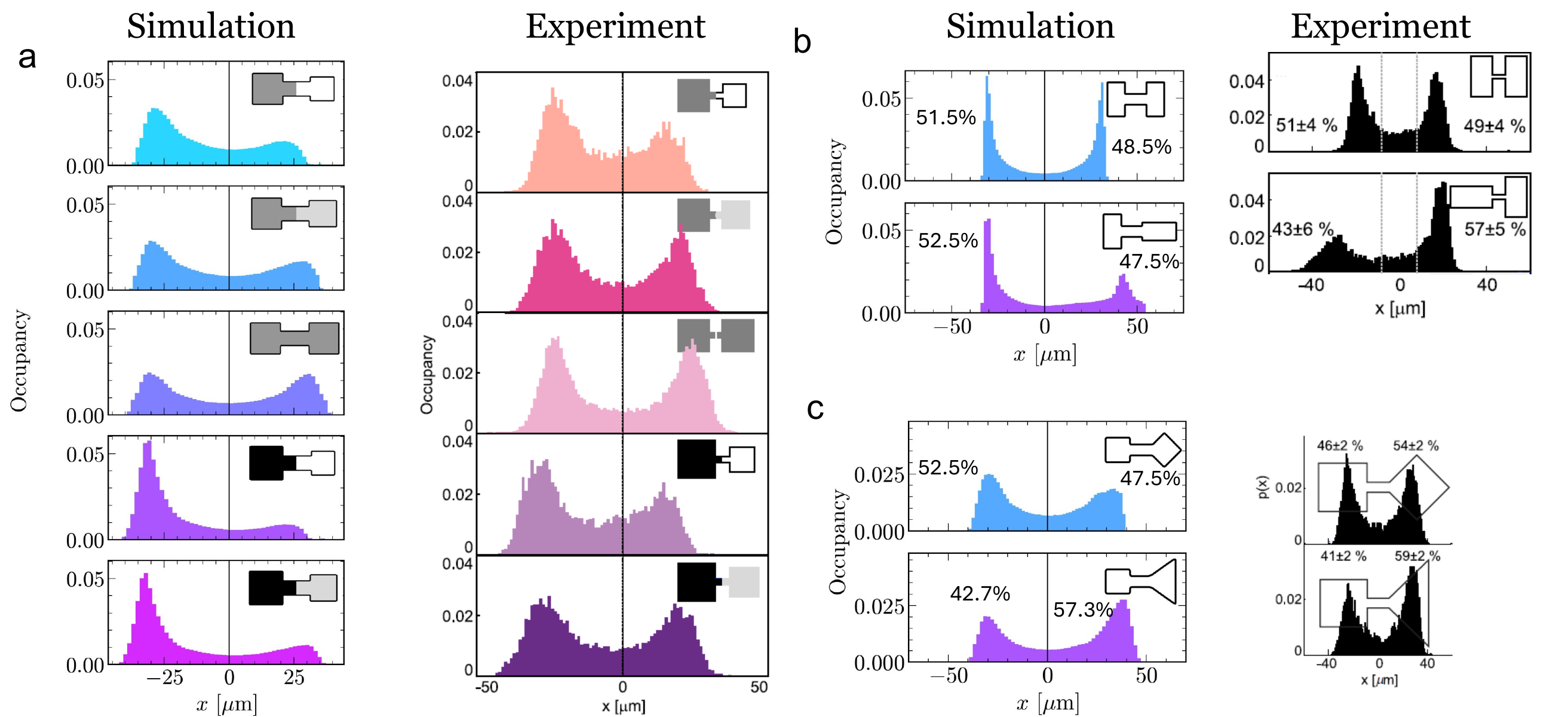}
    \caption{{Simulation {($N=2400$)} and experiment of cell center of mass distributions in geometries of different shape. Experiment is edited from \cite{Fink2020}. {As in \cite{Fink2020}, areas of basins in the asymmetric shapes may differ slightly. Our reported percentages are the fraction of the time the cell spends in one basin, weighted by the area of the basin, $P_\textrm{left} = \frac{\left(\textrm{time in left}\right)/A_\textrm{left}}{\left(\textrm{time in left}\right)/A_\textrm{left} + \left(\textrm{time in right}\right)/A_\textrm{right}}$ as in \cite{Fink2020}.}}}
    \label{fig:generalize}
\end{figure*}

{
\subsection{Generalization to other micropattern geometries}
In a follow-up to \cite{Bruckner2019}, Fink et al. studied the effect of varying micropattern shape and size, quantifying the distributions of occupancies of MDA-MB-231 cells in each micropatterned ``basin'' \cite{Fink2020}. We originally tuned our model to reproduce the qualitative features of MDA-MB-231 observed in \cite{Bruckner2019}. How successful is our model at predicting these experiments? We show simulations of cell motility in analogous geometries in Fig. \ref{fig:generalize}. We have not quite matched the experimental geometries -- as with our earlier simulations, we find that our cells struggle to move through the absolute narrowest bridges without breaking, so we have chosen bridges to be wide enough for cell to travel through easily. We find in Fig. \ref{fig:generalize} that our model correctly predicts the side of the micropattern which the cell spends the most time on in eight out of the nine geometries tested. This includes both relatively trivial cases -- we find that the cell will spend more time in the larger-area basin as the basin size is increased (Fig. \ref{fig:generalize}a) -- but also less intuitive results. For instance, we correctly predict that cells will spend more time in the vertically-oriented rectangular basin, and less in the horizontally-oriented rectangular basin (Fig. \ref{fig:generalize}b). This deviation is smaller than found in the experiment, though it is larger than the presumably statistical asymmetry for the symmetric rectangle-rectangle geometry. We also see that the histograms for the vertical-rectangular vs horizontal-rectangular case are notably asymmetric, as found in the experiment. Cells also spend more time in the triangular basin than a square basin (Fig. \ref{fig:generalize}c) when normalized by area. The only distinct failure is the square-diamond geometry (Fig. \ref{fig:generalize}c), in which we predict more localization to the square, but experiments observe slightly more localization to the diamond.}

{Overall, we view this as a relatively good performance in generalization. We have a few possible hypotheses for the failure in the square-diamond micropattern. First, this discrepancy with experiment may reflect simply that we have only roughly fit our parameters -- we originally sought to qualitatively reproduce the key features of \cite{Bruckner2019}, and did not tune most of our parameters beyond that necessary to reproduce the key features. We note in particular that in this square-diamond micropattern, cells can develop three independent fronts (see Fig. 4A of \cite{Fink2020}). Our model does not commonly do this -- perhaps unsurprisingly, since as we discussed earlier, the coexistence of multiple fronts is an emergent property of our model and one that may require some degree of parameter tuning, based on previous work \cite{chiou2018principles}. We suspect this behavior is not  constrained by the requirement that our model reproduces the results of \cite{Bruckner2019}. Another possibility is that the asymmetry in the square-diamond pattern just arises from statistical fluctuations in the experiment.  The experimental square-diamond asymmetry is one of the smallest effects observed in \cite{Fink2020} (asymmetry of $\sim 2$ times the standard error). Given the multiple comparisons made here and in \cite{Fink2020} it is not too surprising that {\it one} of these many comparisons would vary by this magnitude.}

{While \cite{Fink2020} did not show streamline plots, we can easily compute $F(x,v)$ and its streamlines (Fig. \ref{fig:streams_different_geometries}); these predictions could be compared with data generated by \cite{Fink2020}. However, we do not see immediate qualitative changes in the streamline plots between the symmetric and asymmetric streamline plots -- the changes are primarily quantitative.}

\section{Discussion}
\label{sec: discussion}

Br\"uckner \textit{et. al.} revealed the distinct migratory dynamics of cancerous MDA-MB-231 and healthy MCF10A cells crawling on two-state and rectangular micropatterned substrates by inferring the cell's equation of motion from observations. Here, we have proposed a single model of the cell and showcased its successful recapitulation of observations in \cite{Bruckner2019}. Our model correctly predicts limit cycle, stationary, and bistable behavior depending on cell properties and the geometry of the micropattern. It also successfully predicts the nonlinear (nearly linear) dynamics of cells on the two-state (rectangular) substrate, consistent with \cite{Bruckner2019} (Fig. \ref{fig: nonlinear plots}).

The behavior of MDA-MB-231 cells on rectangular and two-state micropatterns is radically different, with cells switching from a weakly-polarized fluctuating state to a highly persistent oscillation (Figs. \ref{fig: lc sampletrajs} and \ref{fig: ss sampletrajs}). We have been able to recapitulate this shape sensing with a very minimal assumption---that the cell’s polarity is locally inhibited by absence of cell-fibronectin contact. This assumption seems well-supported by existing literature showing strong effects of fibronectin on Rho GTPase polarity \cite{xia2008directional}. Our simple assumption may provide a mechanistic view of what could drive the geometrically-adaptable polarity hypothesized by \cite{Bruckner-prx}.  

Within our model, we can switch cells between showing limit-cycle dynamics and bistability either by changing physical parameters like cell size or tension, or by changing polarity dynamics like the size of protrusion fluctuations. Our results indicate that these distinct motility patterns can reflect quantitative changes in the cell, rather than requiring fundamentally different assumptions in how we model MCF10A and MDA-MB-231 cells. Consistent with this idea, it was later found that subpopulations of cancerous cells, derived from their inherent cell-to-cell variability, can also exhibit distinct motility behaviors \cite{cancer-ccv-bruckner}. These findings complement our predictions that dramatic, qualitative changes in cell behavior can possibly be explained by simple, quantitative changes in the cell. Expression of the tumor-suppressing micro RNA miR-200c in MDA-MB-231 cells has recently been shown to induce a switch in their motility behavior on the two-state micropattern, with cells expressing miR-200c becoming bistable \cite{Kohler2023}. While miR-200c has many targets and its impact on cell migration is complex, it does modify both cell stiffness and size. However, the dynamics here are somewhat complex -- its expression can lead to decreased cell stiffness \cite{Ljepoja2019} and in different assays miR-200c induction increased cell spread \cite{Ljepoja2019} but miR-200 family knockout increased cell size and decreased polarity \cite{Yu2022}. If miR-200c induction decreased spread in \cite{Kohler2023}, that could lead to bistability, according to our model. However, perturbations of polarity can also lead to switches between bistability and limit cycles within our model.

{To what extent can the parameter variations we have studied be realized in experiment? Past experiments have studied drug perturbations including inhibiting the Rho/ROCK pathway, preventing actin polymerization with latrunculin, and inhibiting myosin contractility with blebbistatin \cite{Bruckner-prx}, finding that these can disrupt the nonlinear dynamics as visualized in $F(x\to0,v)$. Within our model, we would expect the application of blebbistatin -- decreasing myosin contractility -- to decrease our value of $\gamma$, which we see can induce significant perturbations to nonlinear dynamics (Fig. \ref{fig: tension bs}. Latrunculin, disrupting actin polymerization, would definitely alter our protrusion strength $\alpha$. We have not seen particularly interesting effects from changing $\alpha$ other than changing the scale of cell speeds, etc. However, F-actin also has roles in cell polarity and can feed back on Rho GTPases \cite{miao2019wave} -- so latrunculin treatment may also alter our polarity parameters like $\tau_f$, $\mu_\beta$, etc. Another natural approach would be to try eliminating filopodia, as we studied in Fig. \ref{fig: filopodia_knockout_global}. This could be done with, e.g. knockdown of myosin X \cite{bohil2006myosin} or formins. However, we caution against overinterpretation of these drug assays -- off-target effects \cite{nishimura2021formin} and feedback loops can confound the apparent results. Another possibility would be to change the concentration of fibronectin on the micropattern. Within our model, this would change the contrast between the micropattern and the region outside -- essentially lowering our value of $\chi$. We have performed simulations lowering $\chi$, and find that there are relatively few new nonlinear features until $\chi$ is small enough that the cell is not well-confined to the micropattern. This intervention, too, may not be straightforward to interpret. The effect of substrate adhesiveness on cell speed is non-monotonic \cite{gupton2006spatiotemporal,barnhart2011adhesion}, and this non-monotonicity depends on actomyosin contractility \cite{gupton2006spatiotemporal} and may be complicated to model \cite{mai2020hydrodynamic}. In addition, as we are modeling primarily the {\it signaling} effect of cell-substrate adhesion, we are not as confident in predicting a dose-response between fibronectin concentration and cell polarization. We view changes in micropattern shape as the intervention most likely to yield agreement between simulation and experiment without the need to introduce additional mechanisms.}

If the key difference between MCF10A and MDA-MB-231 cells were the cell spread area, which we have shown is sufficient to explain a transition between limit cycle and bistable behaviors, this could be easily tested through both quantification of cell spread areas and variation of the micropattern size. Within our model, simulated cells with similar $A_\textrm{cell}/A_\textrm{basin}$ have similar dynamics, assuming all other parameters are held constant, and cells will transition to a bistable state as the basin size is increased (Fig. \ref{fig:basin_size}). %

In our application of the data-driven approach of \cite{Bruckner2019}, we discovered that uncertainties in the inferred nonlinear dynamics---driven by finite sampling---can be significant. This is despite our large dataset of 4800 simulations of 48 hours in length, providing roughly 24x as much trajectory time as observed experimentally. This means a single streamline plot should be taken with some skepticism. For instance, the bistable dynamics seen in Fig. \ref{fig: sigma perturbation} with $\sigma = 12\micron$ generates a streamline plot that is slightly asymmetric at the center, and has a small domain of attraction at the center. These features, though, are not robust to sampling (see Appendix \ref{appendix: separatrix variations}). We suggest that the bootstrap approach we take in Appendix \ref{appendix: separatrix variations} should be used as an essential robustness check. 

{How does our model relate to other models of two-state motility, and can it be discriminated from them? Recent work from the Broedersz group and collaborators \cite{Bruckner-prx,flommersfeld2024geometry} has hypothesized that cells may be able to polarize while in the constricted region of the micropattern, while being unpolarized in the basins. In \cite{Bruckner-prx}, this assumption was built into the model, while \cite{flommersfeld2024geometry} provided a mechanism based on the alignment of actin filaments. Our model agrees that cells are often polarized while traveling through the constricted region and not as strongly polarized in the basins, though this emerges from our model's coupling between the cell polarity and the micropattern via sensing the micropattern boundary. In this sense, we expect that our model is in the same broad class as the assumptions of \cite{Bruckner-prx}, but has a different specific set of assumptions than that of \cite{flommersfeld2024geometry}. To distinguish between the two, the most natural test would be to attempt to design a micropattern that created confinement but disrupted actin alignment -- possibly by having curved edges on the bridge between basins.}

\section*{Data and code availability}
Code to reproduce this paper has been deposited at Zenodo \href{https://doi.org/10.5281/zenodo.14373111}{https://doi.org/10.5281/zenodo.14373111}

\begin{acknowledgments}
PZ and BAC are supported by NIH R35 GM142847. This work was carried out at the Advanced Research Computing at Hopkins (ARCH) core facility  (rockfish.jhu.edu), which is supported by the National Science Foundation (NSF) grant number OAC 1920103. We thank Kurmanbek Kaiyrbekov and Yongtian Luo for a close reading of a draft of this paper. 
\end{acknowledgments}

\newpage
\appendix
\onecolumngrid
\renewcommand\thefigure{S\arabic{figure}}
\setcounter{figure}{0}

\section{Numerically evolving the phase field equations of motion}
\label{appendix: numerical integration}
The phase field equations of motion are
\begin{equation}
  \left\{
    \begin{aligned}
      & \frac{\partial \phi({\bm r},t)}{\partial t} + {\bm v}({\bm r},t) \cdot \nabla \phi=-\mathcal{M}\ \frac{\delta \mathcal{F}}{\delta \phi},\\
      & \eta {\bm v}=\frac{\delta \mathcal{F}}{\delta \phi} \nabla \phi + {\bm f}_{\mathrm{motility}},
    \end{aligned}
  \right.
  \label{eq: pf_motion}
\end{equation}
where the total free energy of the cell is $\mathcal{F}=\mathcal{F}_{\mathrm{CH}}+\mathcal{F}_{\mathrm{area}}$. Given the definitions in Eqs. (\ref{eq: cahn-hilliard-eq}) and (\ref{eq: area-conservation-eq}), the functional derivatives are
\begin{equation}
  \frac{\delta \mathcal{F}_{\rm CH}}{\delta \phi}=36\frac{\gamma}{\lambda} \phi (2\phi-1) (\phi-1)-2\lambda \gamma \nabla^2 \phi,
\end{equation}
and     
\begin{equation}
    \frac{\delta \mathcal{F}_{\rm area}}{\delta \phi}=2 \kappa \left[ 1-\frac{1}{\pi R_0^2} \int d{\bm r} \phi^2 \right] \left[ -\frac{2}{\pi R_0^2}\phi \right].
\end{equation}
With $\frac{\delta \mathcal{F}}{\delta \phi}$ worked out, we numerically evolve Eq. (\ref{eq: pf_motion}) on a 200$\times$200 grid with a spatial resolution of $\Delta x=1.5\micron$ and simulation timestep of $\Delta t=0.36$s \cite{pedrom}. We also leverage scikit-image's \texttt{measure} function \cite{scikitimage} to find contour points at the 1/2 level and matplotlib's \texttt{streamplot} \cite{matplotlib} to trace out streamlines.

\section{Constructing acceleration footprints from simulated trajectory data}
\label{appendix: acc. footprint}
We record the $x$ position of cell center-of-mass every $\Delta t=3$ min in our simulations, and at the end of each run, we calculate the velocity and acceleration from this time series:
\begin{align}
v(t)&=\frac{x(t+\Delta t) - x(t)}{\Delta t},\\
a(t)&=\frac{x(t+\Delta t)-2x(t)+x(t-\Delta t)}{\Delta t^2}.
\label{eq: va_from_x}
\end{align}
Note that $\Delta t$ here is not the time step of the previous section, but the time resolution with which we measure the cell's center of mass. With the above equations, we compute the tuples $(x, v, a)$ for all center-of-mass trajectories observed across all 4800 simulations, which amounts to roughly 2.3M records. We construct the $xv$ phase-space by discretizing the domain of each coordinate with $n=32$ bins, where each bin has dimensions $dx=(x_{\rm max} - x_{\rm min})/n$ and $dv=(v_{\rm max} - v_{\rm min})/n$. Considering the spatial and velocity domains typical to our simulations, our bins are approximately $3\micron \times 11 \micron$/hr. To populate this phase-space, we group together the set of all tuples $(x, v, a)$ that fall into the $(x, v)$ bin and assign that bin the average acceleration of the set. Repeating this procedure for all the $n\times n$ bins present yields the acceleration footprint $F(x, v)=\langle a(x, v) | x, v\rangle$. 

{We note that the conditional averaging method we use here, adapted from \cite{Bruckner2019}, can have systematic errors due to finite observation time, etc. \cite{bruckner2024Review} These can be corrected if the conditional averaging is done using a basis function approach -- but the methods do not generalize well to the straightforward binning of \cite{Bruckner2019}. We have chosen to use the binning approach so that our results are as comparable as possible with the original experiments of \cite{Bruckner2019}.}

{\section{Varying two parameters simultaneously}}
\label{app:twoparametervariations}

\begin{figure}
    \centering
    \includegraphics[width=1\linewidth]{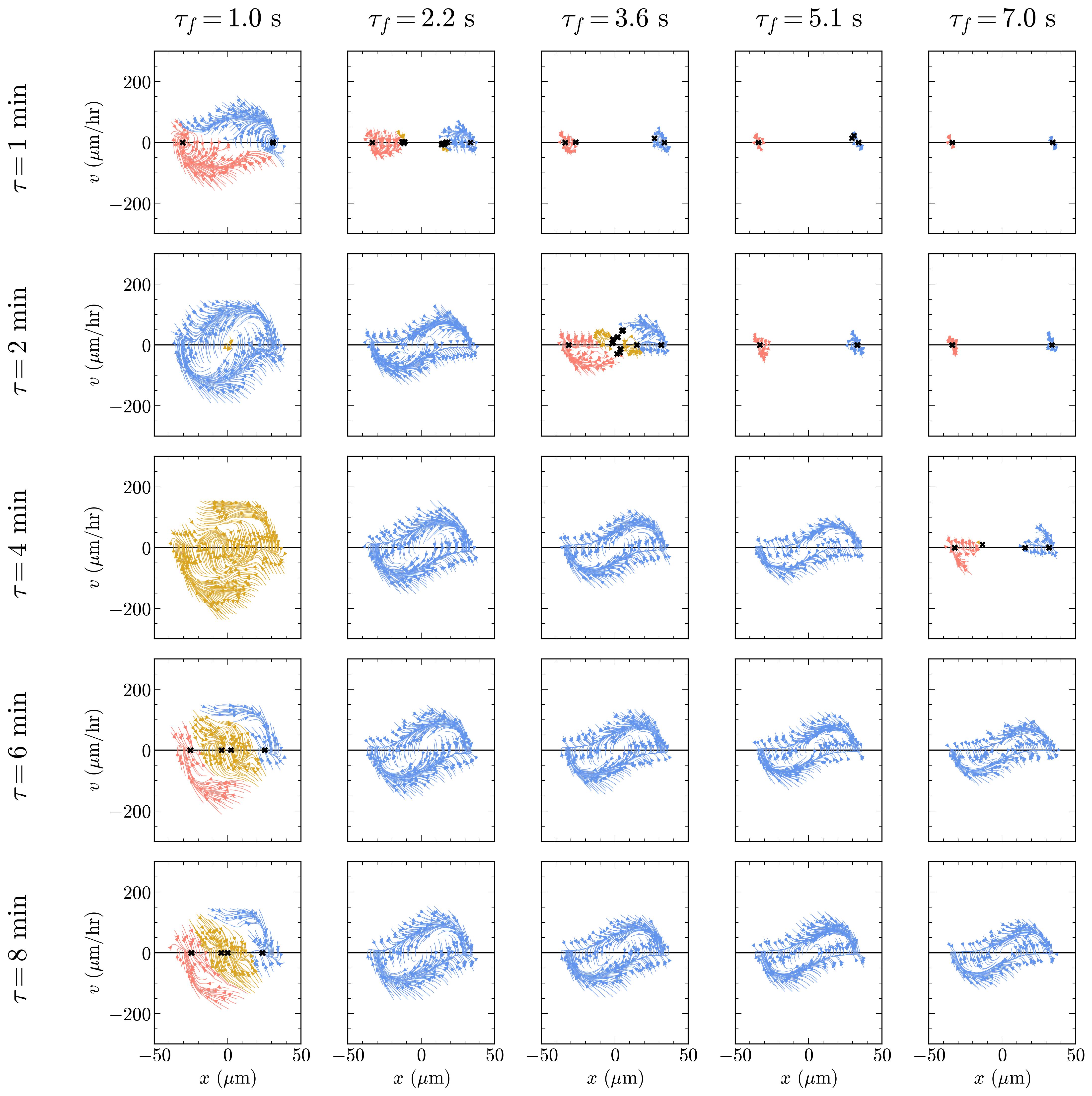}
    \caption{{Simultaneous perturbation of the timescales $\tau$ and $\tau_f$ shows that limit cycle behavior requires that the ratio of these two rates not get too far from our default value. Results are obtained from 2400 simulations.}}
    \label{fig:two_taus}
\end{figure}

\begin{figure}
    \centering
    \includegraphics[width=1\linewidth]{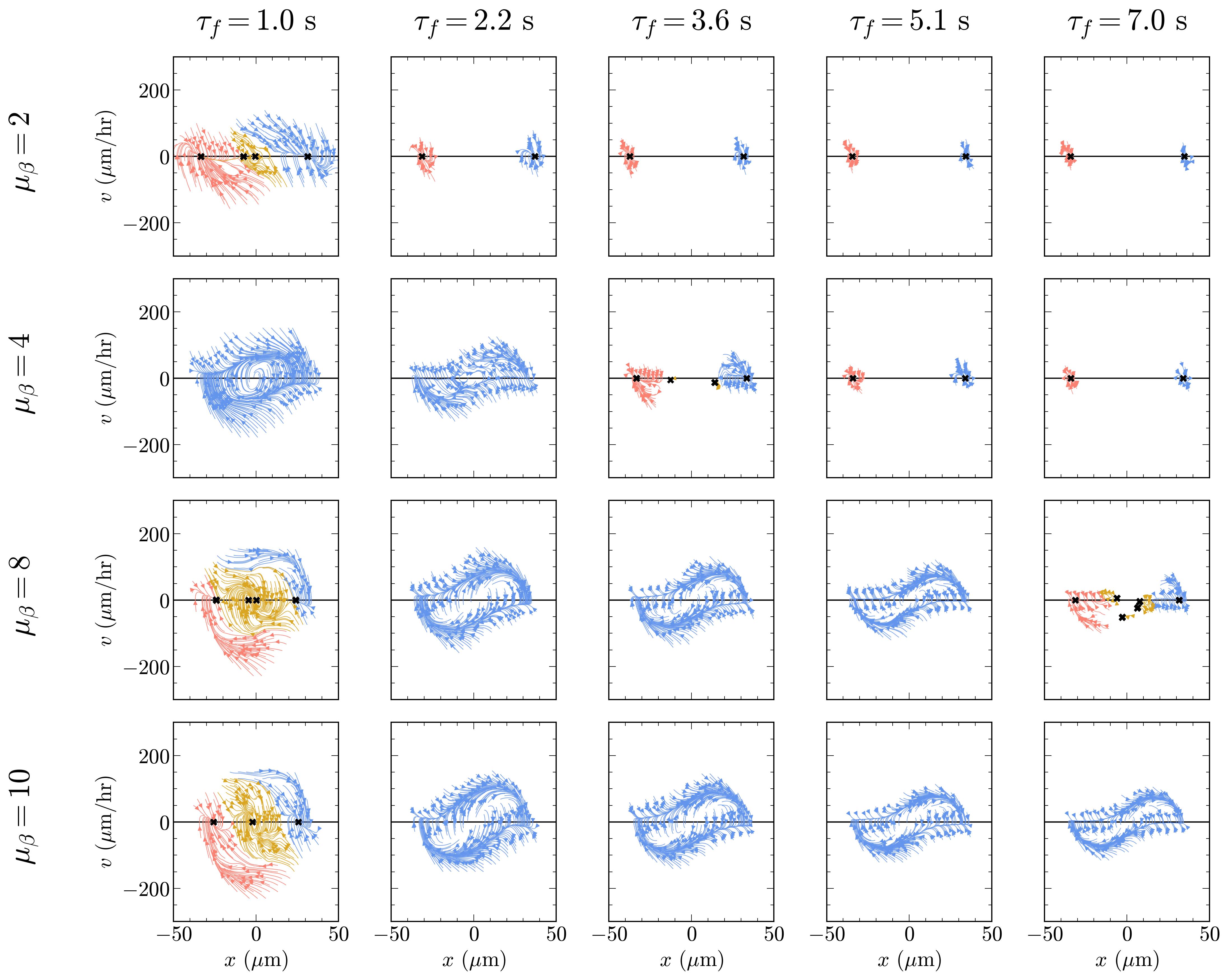}
    \caption{{Simultaneous perturbation of $\tau_f$ and $\mu_\beta$ shows that a balance between the rate of patch generation and the mean polarity increase at each patch is necessary to gain limit cycle behavior. Results are obtained from 2400 simulations.}}
    \label{fig:tauf_mub}
\end{figure}

\section{A dynamically-evolving cell boundary is required for bipolar morphology}
\label{appendix: evolving field required}

\begin{figure}[htb]
    \centering
    \includegraphics[width=0.7\linewidth]{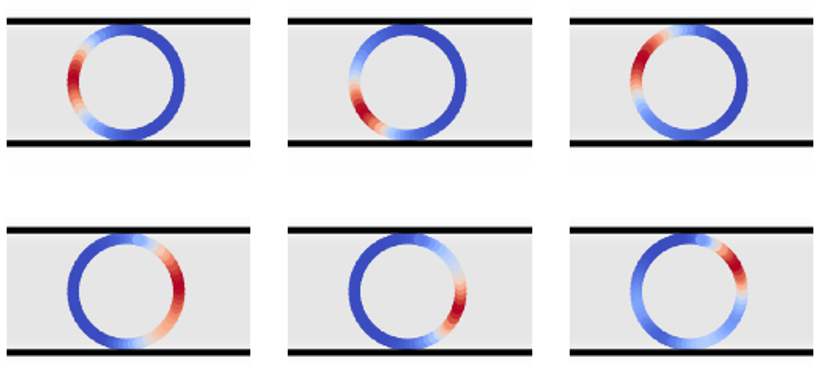}
    \caption{A sample of the steady-state distribution of cell polarity as evolved according to Eqs. \eqref{eq: dpol_dt} and \eqref{eq: cntr_pmf} on a \textit{static} phase field initialized on the rectangular micropattern. Although each sample starts with a random initial distribution and evolves stochastically, all of them equilibrate to a singly-polarized configuration.}
    \label{fig: static pol evol}
\end{figure}

In Section \ref{sec: lc-stationary-states}, we outlined our model's ability to recapitulate limit-cycles on two-state and stationary behavior on rectangular substrates. The emergence of persistently hopping cells is plausible considering our assumption of positive feedback for cell polarity (see Eq. \ref{eq: dpol_dt})---this mechanism amplifies small differences in polarity into one hot spot of activity, thus promoting singly-polarized cells that can migrate persistently. The same principle underlies the migration of the cell on the rectangular micropattern, yet the cell adopts a \textit{bipolar} morphology on average, leading to near zero speeds and the observed stationary states (compare Movies M\ref{mov: limit-cycle} and M\ref{mov: stationary}). To investigate how our assumptions for cell polarity can lead to bipolar cells on the rectangular micropattern, we decided to study the evolution of cell polarity in isolation. This simplified version of the model represented the cell with a \textit{static}, circular phase field initialized in the center of the rectangular geometry. Here, we did not update the cell interface itself, i.e., $\frac{\partial}{\partial t} {\phi}_{\rm static}=0$, but did evolve the polarity according to Eqs. (\ref{eq: dpol_dt}) and (\ref{eq: cntr_pmf}) on the static cell boundary. Surprisingly, we observed the stochastic evolution of cell polarity to always localize into a single hot spot of activity, consistently yielding a singly-polarized cell (Fig. \ref{fig: static pol evol}). Since the only change we made was making the phase field static, we conclude that the emergence of bipolar cell morphologies in our model requires the cell interface to be \textit{dynamic}.

\newpage
\section{Predicted motility dynamics in response to simplifying model knockouts}
\begin{figure}[htb]
    \centering
    \subfigure[]{\includegraphics[width=0.49\linewidth]{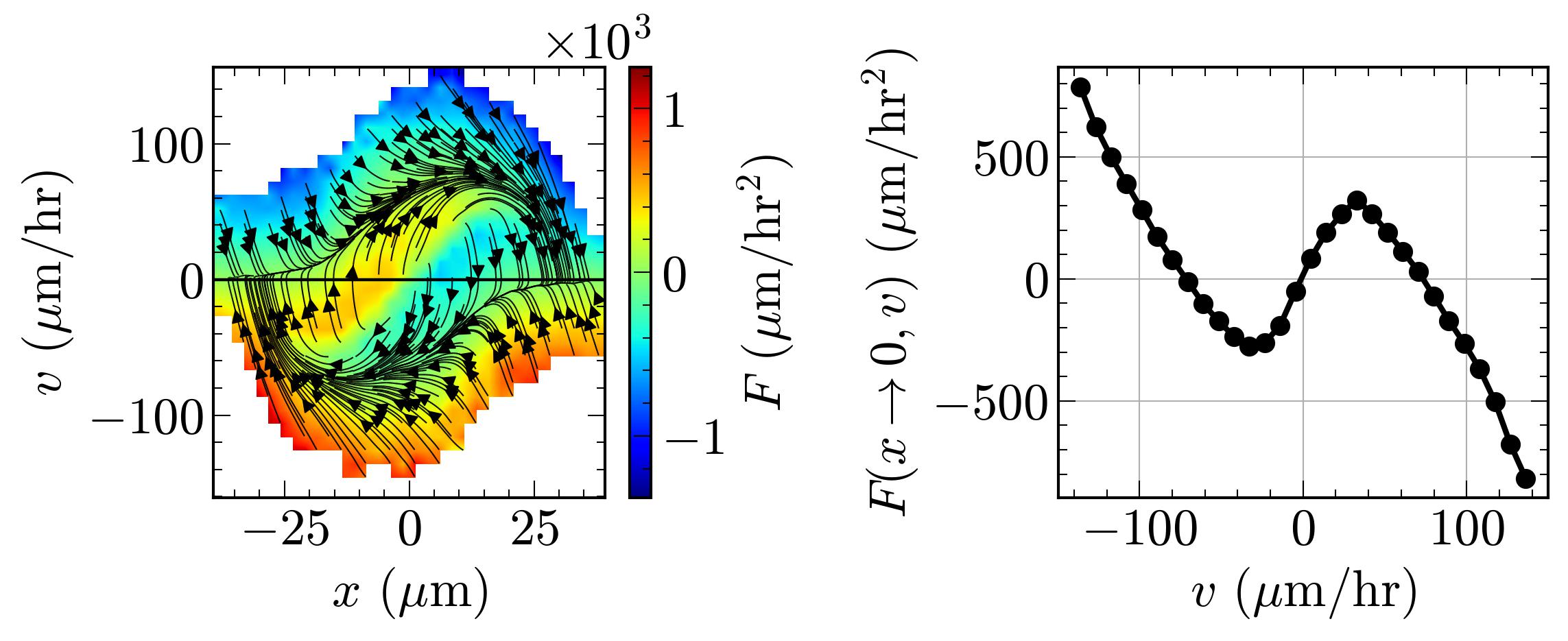} \label{fig: filopodia_knx}}
    \hfill
    \subfigure[]{\includegraphics[width=0.49\linewidth]{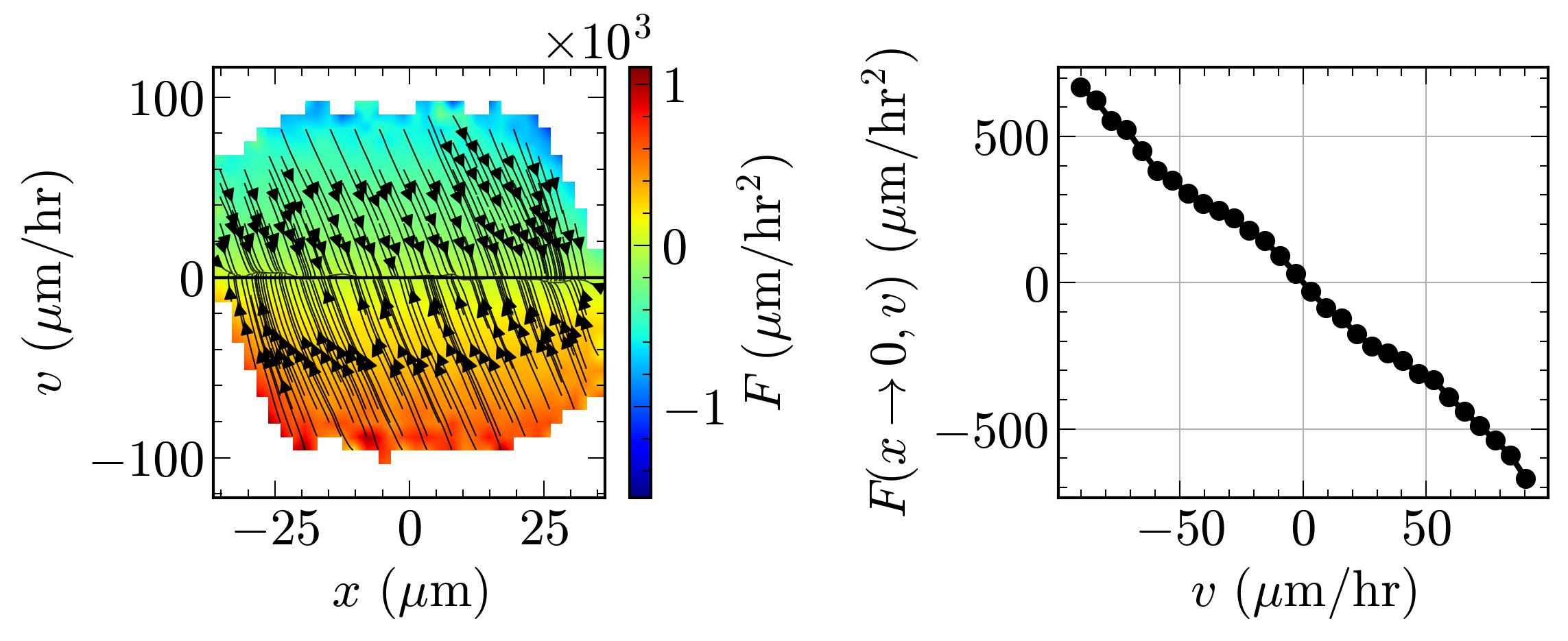} \label{fig: filopodia_knx_rect}}
    \caption{Predicted motility dynamics of our default cell on \subref{fig: filopodia_knx} the two-state and \subref{fig: filopodia_knx_rect} the rectangular micropatterns with the filopodial mechanism absent, i.e., $p(\bm \theta)=\mathbb{P}(\bm \theta)$.}
    \label{fig: filopodia_knockout_global}
\end{figure}

\begin{figure}[htb]
    \centering
    \includegraphics[width=0.5\linewidth]{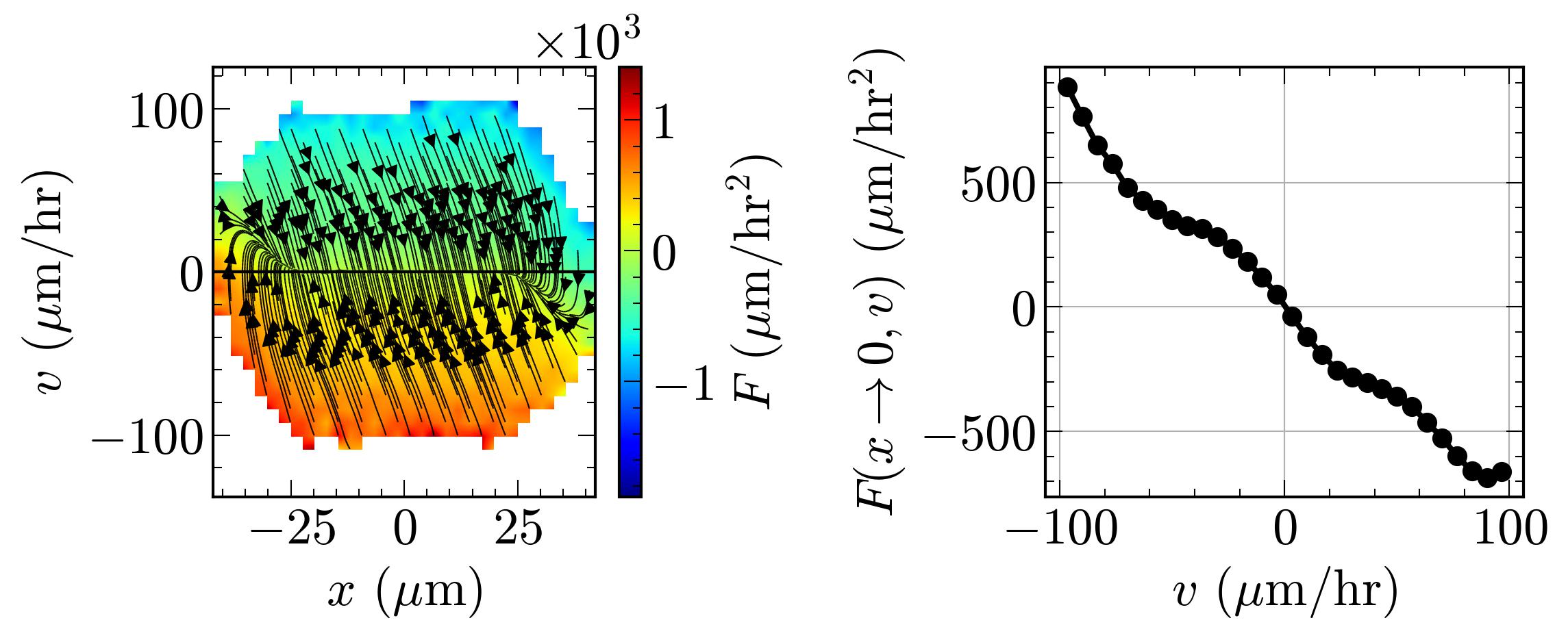}
    \caption{Predicted motility dynamics of our default cell on the rectangular micropattern with the degradation of polarity outside the micropattern absent, i.e., $\tau_X\to\infty$.}
    \label{fig: tauX_knx_rect}
\end{figure}

\begin{figure}[htb]
    \centering
    \subfigure[]{\includegraphics[width=0.49\linewidth]{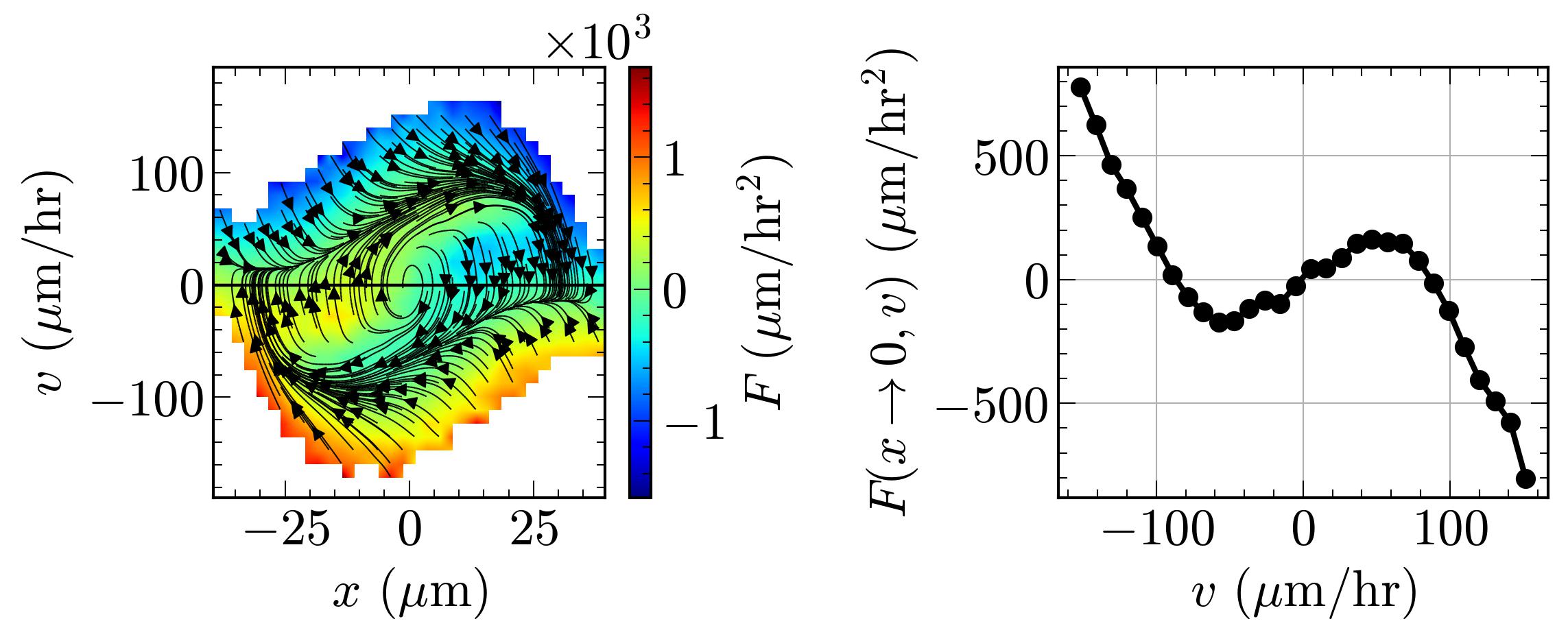} \label{fig: tension_knx}}
    \hfill
    \subfigure[]{\includegraphics[width=0.49\linewidth]{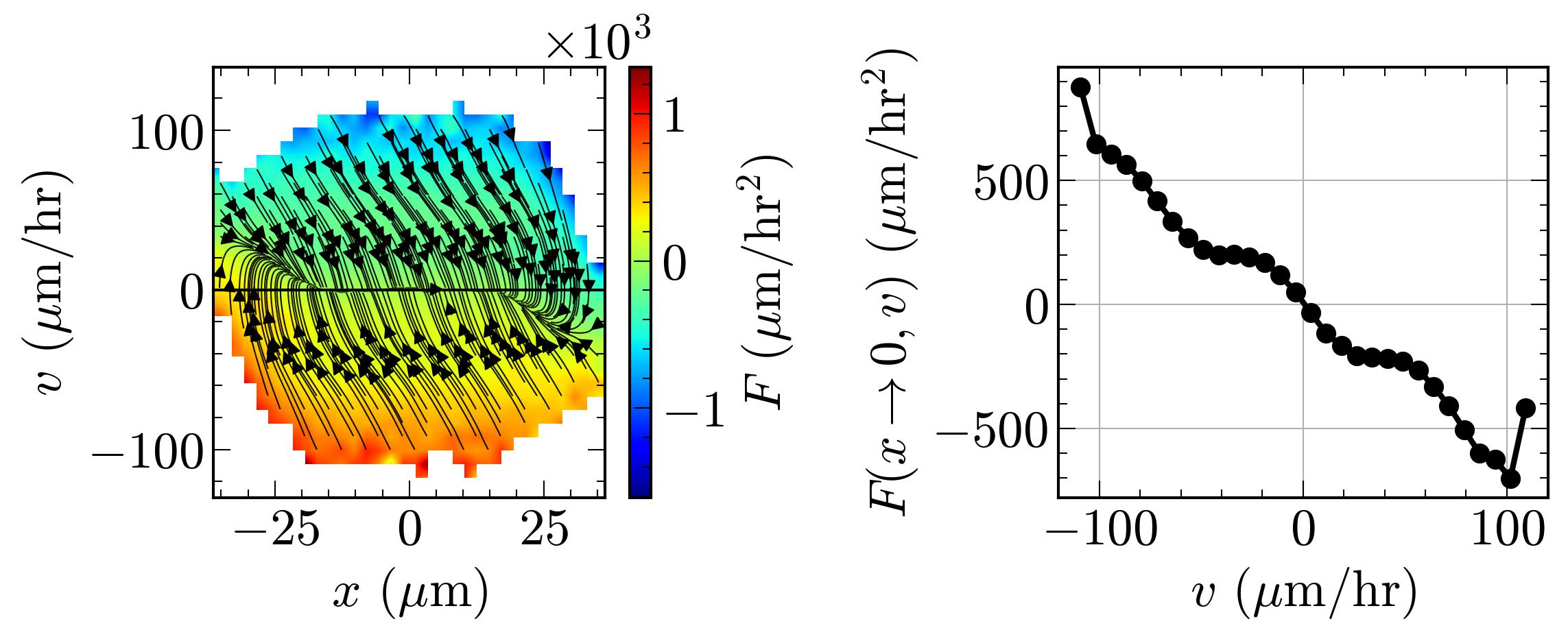} \label{fig: tension_knx_rect}}
    \caption{Predicted motility dynamics of our default cell on \subref{fig: tension_knx} the two-state and \subref{fig: tension_knx_rect} the rectangular micropatterns with the perimeter-induced degradation of polarity absent, i.e., $s_0\to\infty$.}
    \label{fig: perimeter_knockout_global}
\end{figure}

{\section{Streamline plots for different geometries}}

{Here we show streamline plots for alternate geometries, including those showcased in Fig. \ref{fig:generalize}}

\begin{figure}
    \centering
    \includegraphics[width=0.4\linewidth]{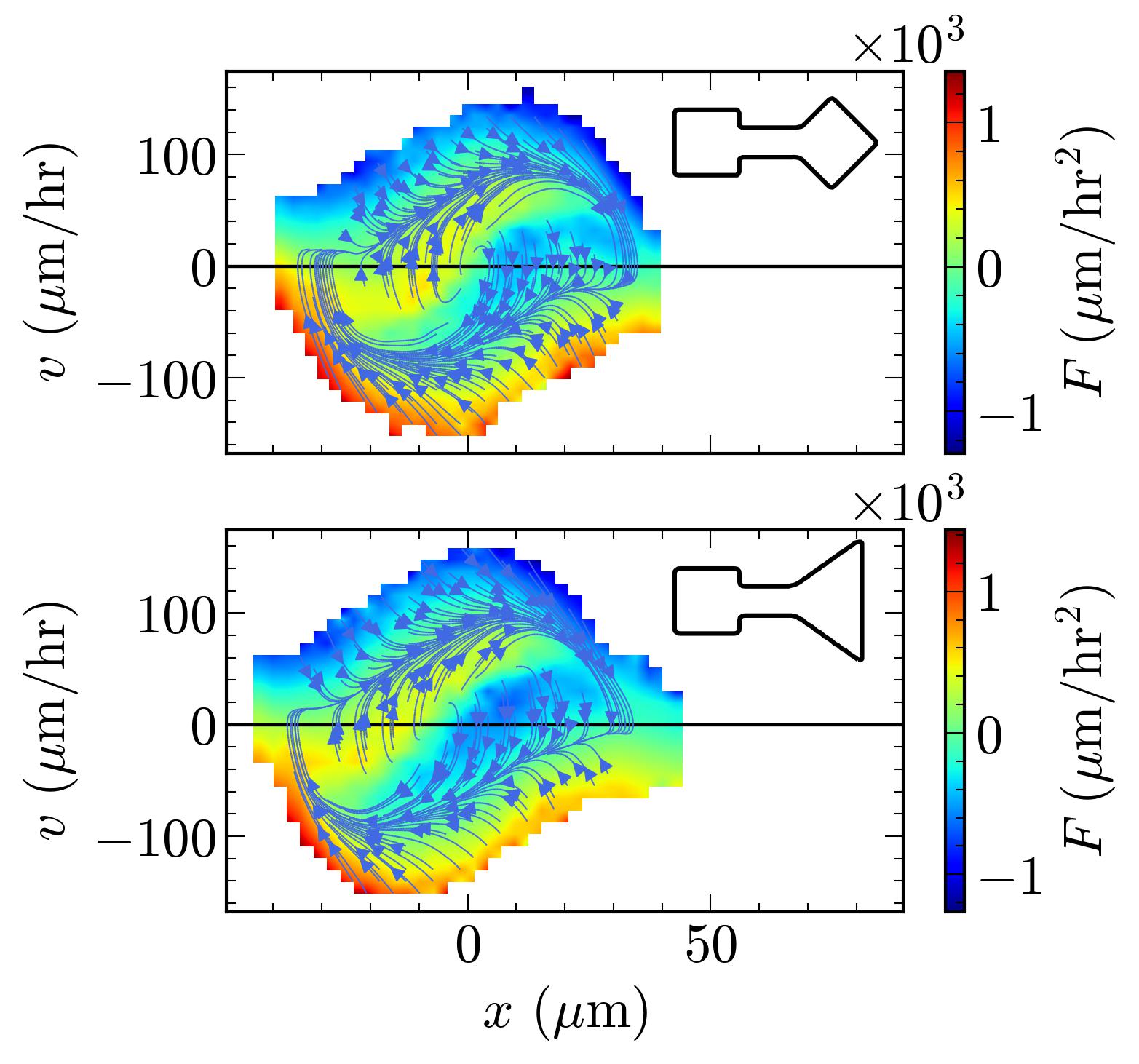}     
    \caption{{Predicted streamlines for two alternate two-state geometries are similar to those studied in \cite{Fink2020}. Results are obtained from 2400 simulations.}}
    \label{fig:streams_different_geometries}
\end{figure}

\begin{figure}
    \centering
    \includegraphics[width=0.5\linewidth]{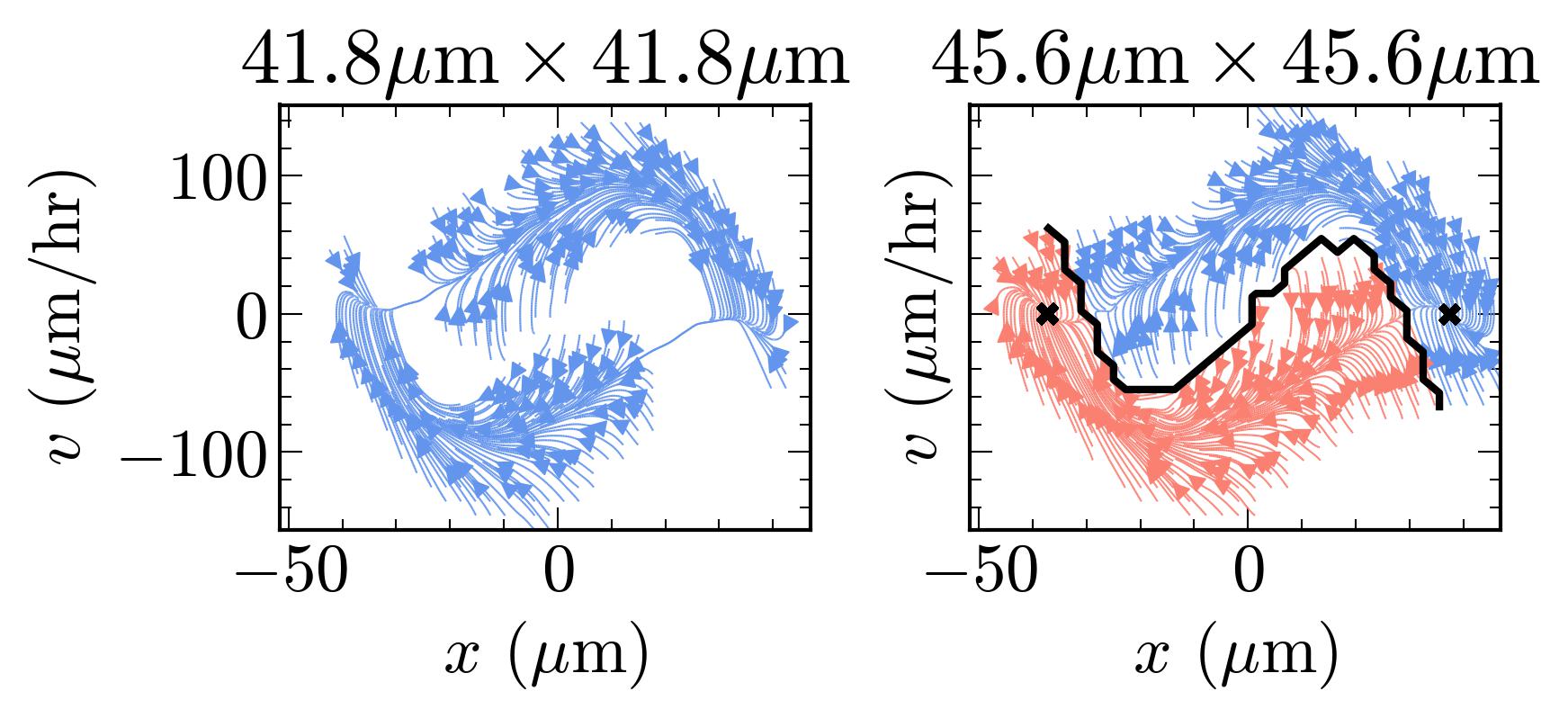}
    \caption{{The default cell switches between limit cycle and bistable as the micropattern size is increased.}}
    \label{fig:basin_size}
\end{figure}

\section{Showcasing the uncertainty of phase-space separatrices}
\label{appendix: separatrix variations}

Our deterministic acceleration footprints $F(x, v)$ have associated uncertainties since we infer them from a finite number of simulations. Here, we focus on footprints that exhibit multiple phases and seek to demonstrate the typical variability we can expect when interpreting the motility patterns encoded by them. Note that we specifically highlight the uncertainty of these acceleration footprints because we characterize them by their separatrices and want to better understand the stability of these boundaries. To assess variability, we employ bootstrapping on our original set of simulations. Each bootstrap involves sampling 4800 times from the set of independent phase field simulations with replacement, allowing us to infer $F_k(x, v)$ for each bootstrapped ensemble $k$. We generate a total of 48 such bootstrapped ensembles, i.e., $k=[1, 2, ..., 48]$. The below figures showcase the variability of their respective acceleration footprint, each consisting of three panels:
\begin{itemize}
    \item[(a)] This is the original $F(x, v)$ inferred from our phase field simulations. A copy of this is in the main text.
    \item[(b)] This is a heatmap showing the number of distinct categories each phase-space lattice site has ever been a member of across the entire set of bootstrapped ensembles. There is a maximum of 3 phases a lattice site can be classified as based on whether it terminates at the left, center, or right of the micropattern. In this view, lattice sites that fluctuate are colored darker and serve as a visual representation of the variability we can expect when delineating the boundary of distinct phases.
    \item[(c)] This is an exemplary view of 9 out of the 48 bootstrapped acceleration maps. It is a useful way to explicitly demonstrate the variability we can expect in separatrix boundaries.  
\end{itemize}

\bootview{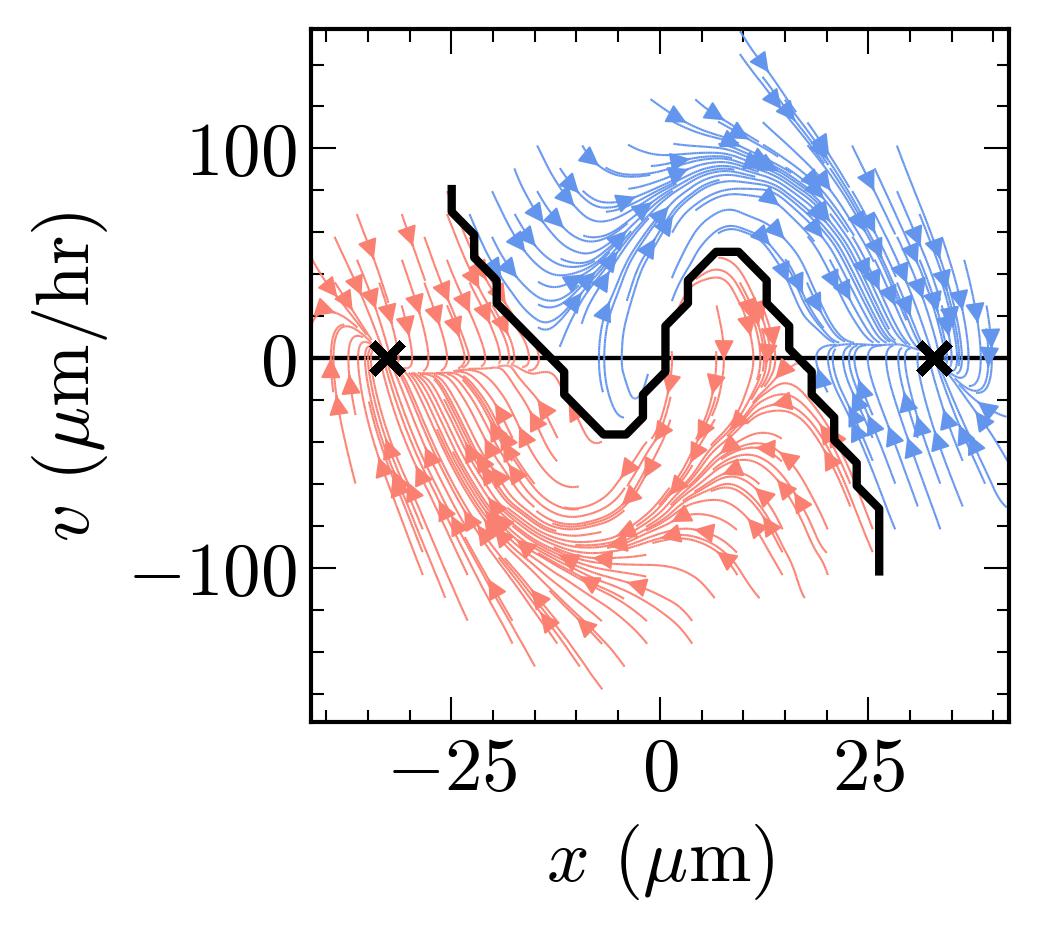}{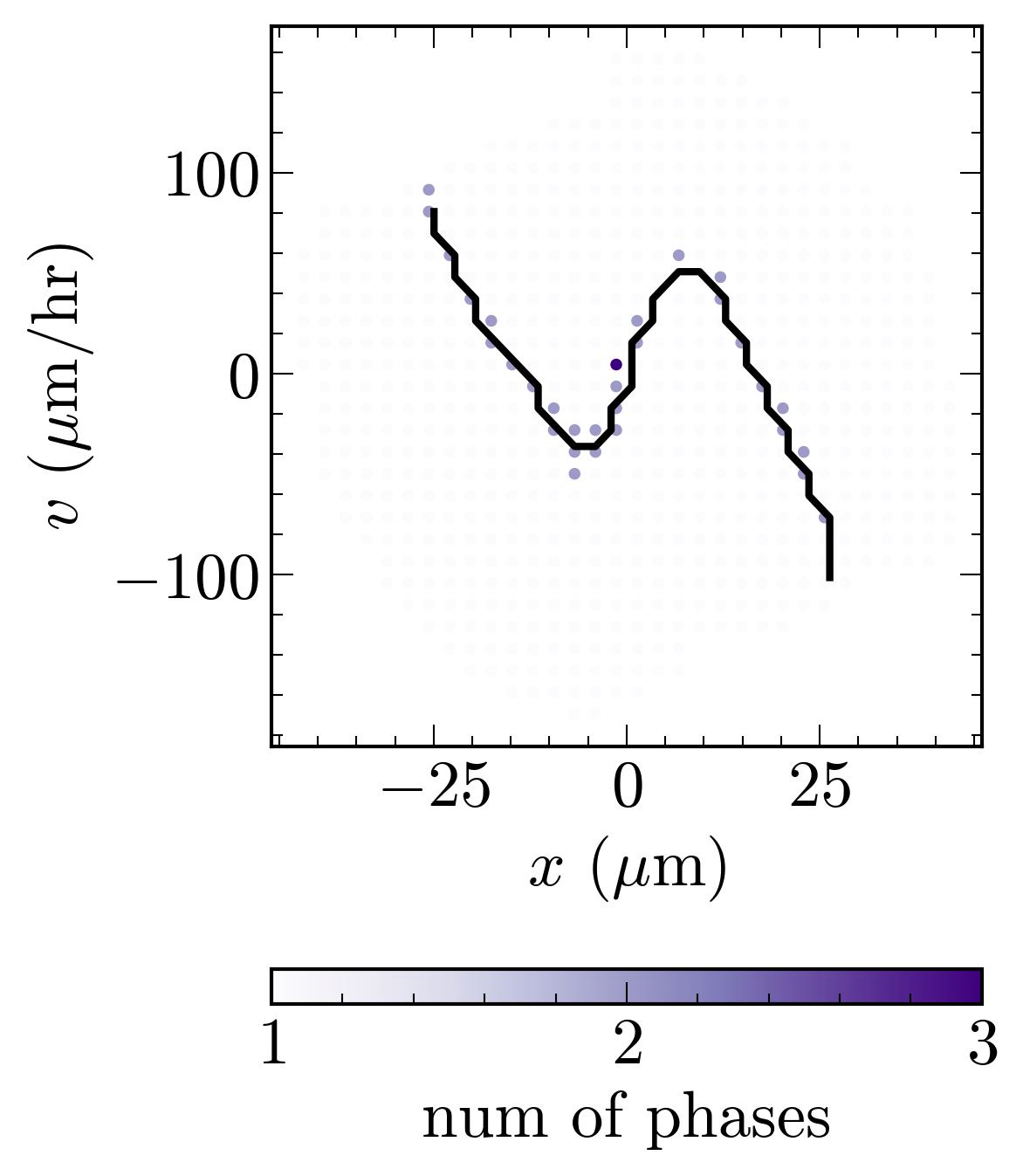}{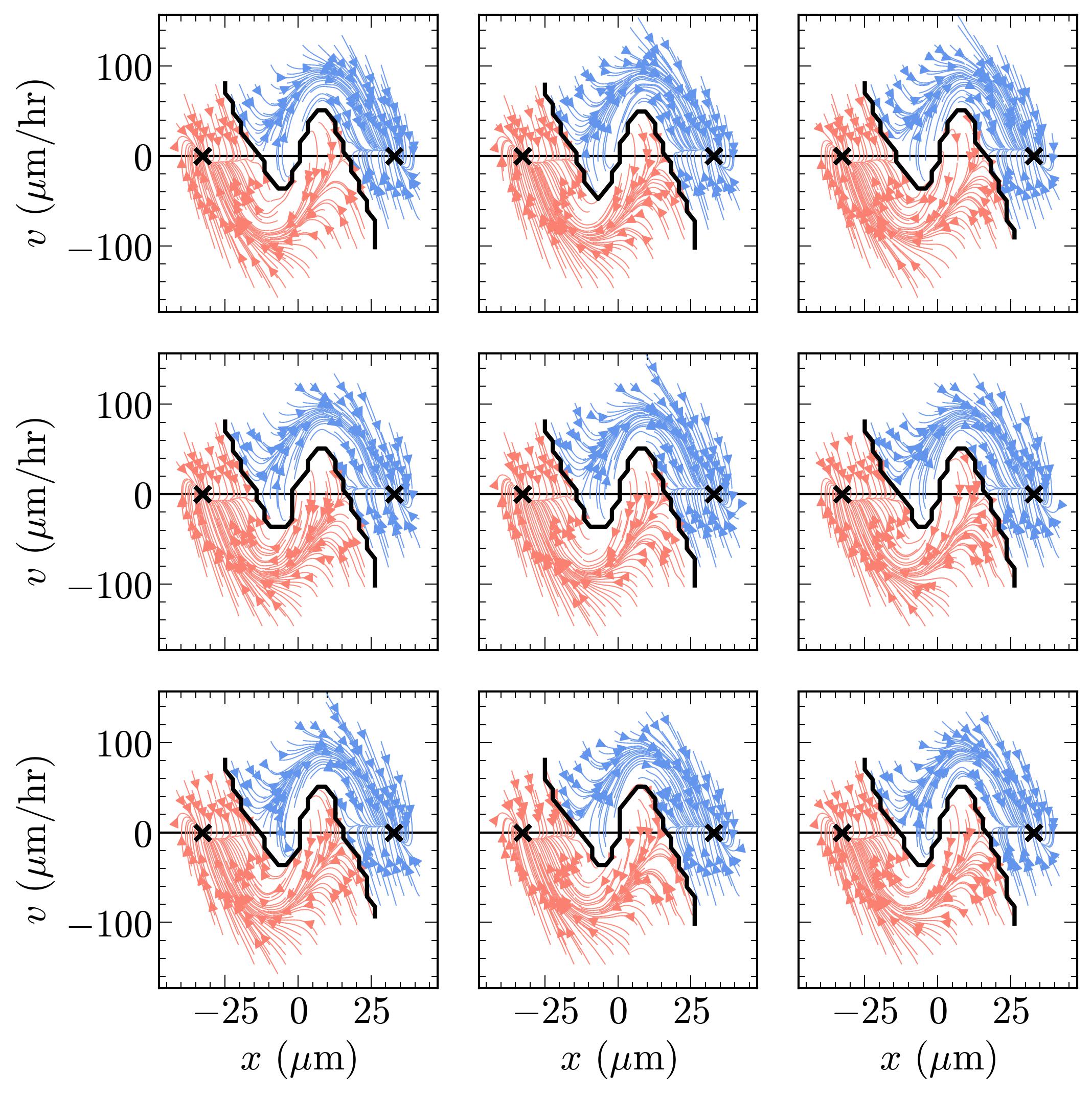}{Overview of variability in motility for the cell-micropattern pair presented in Fig. \ref{fig: bs streamline}.}{fig: bs err}
\bootview{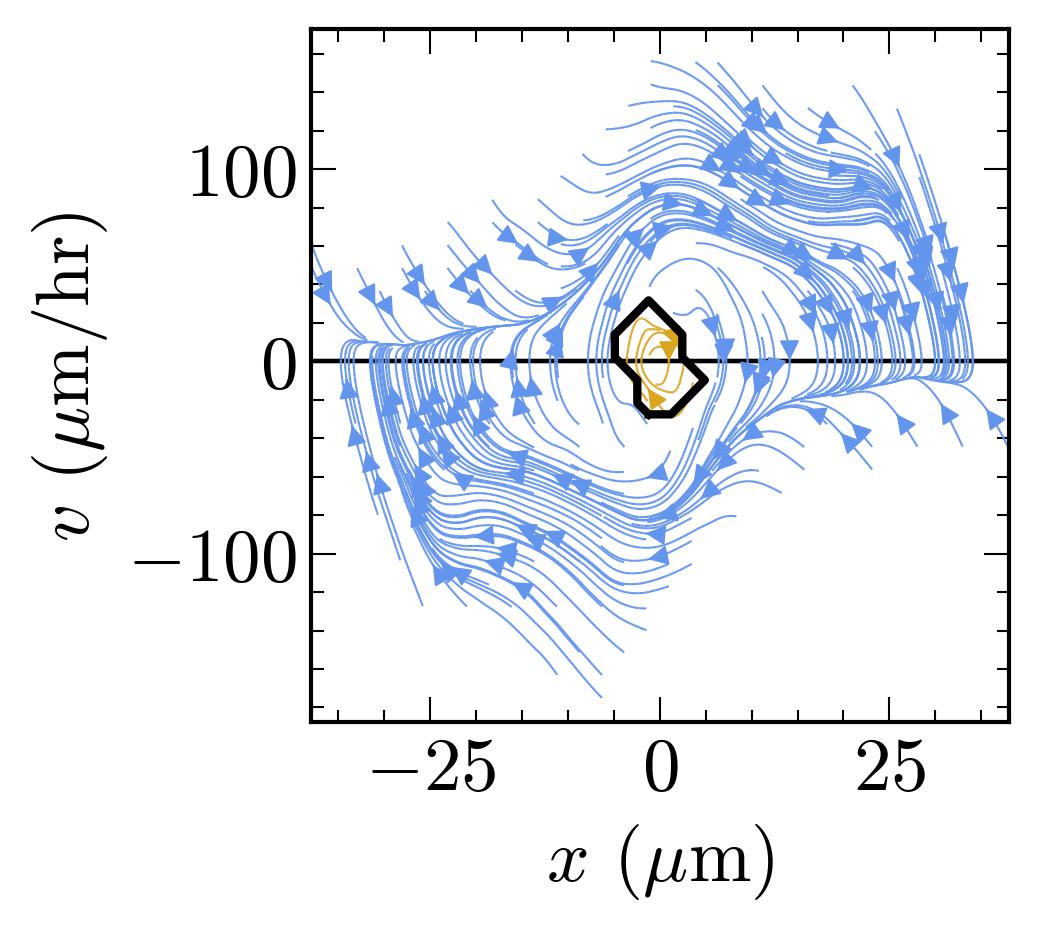}{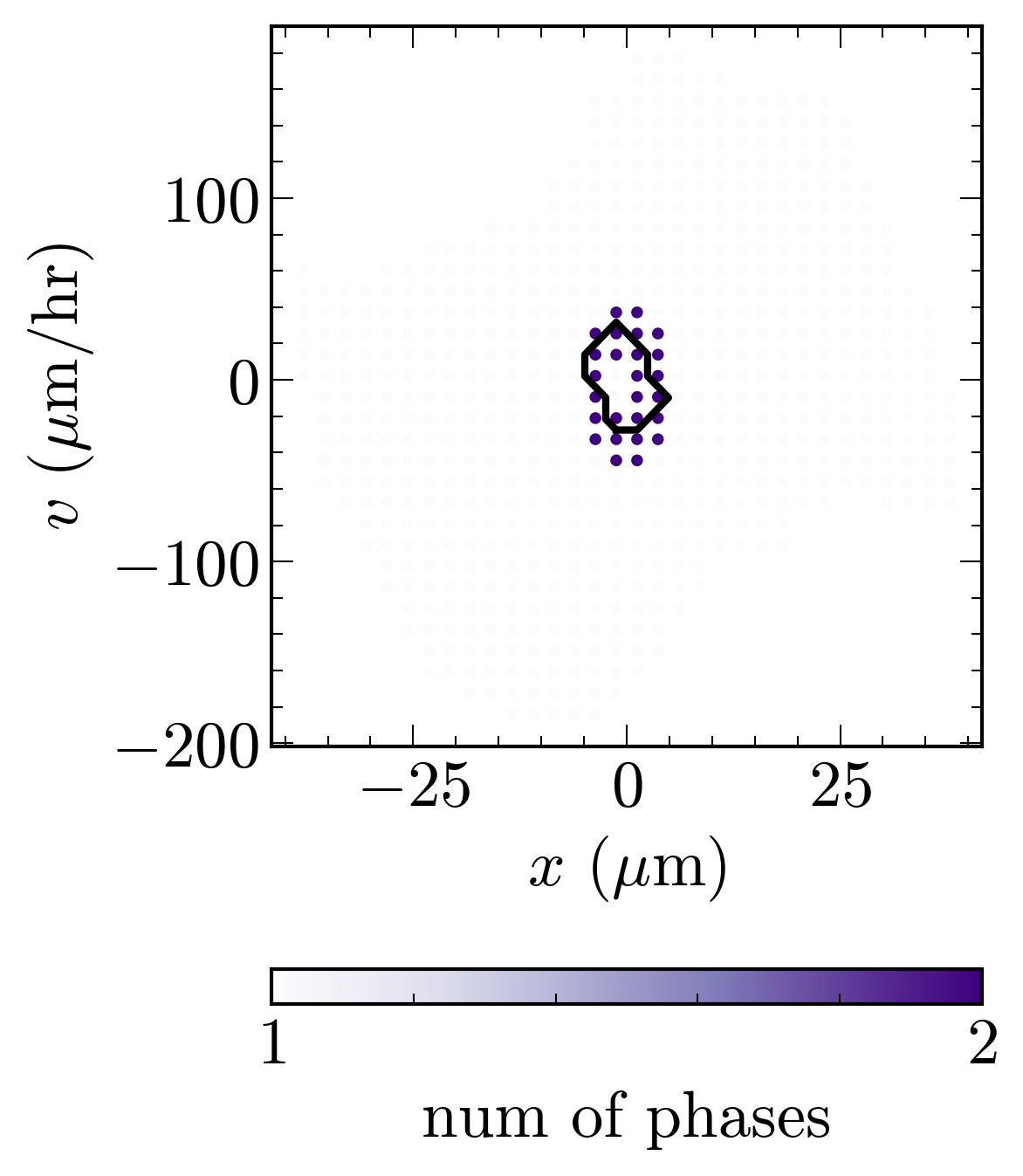}{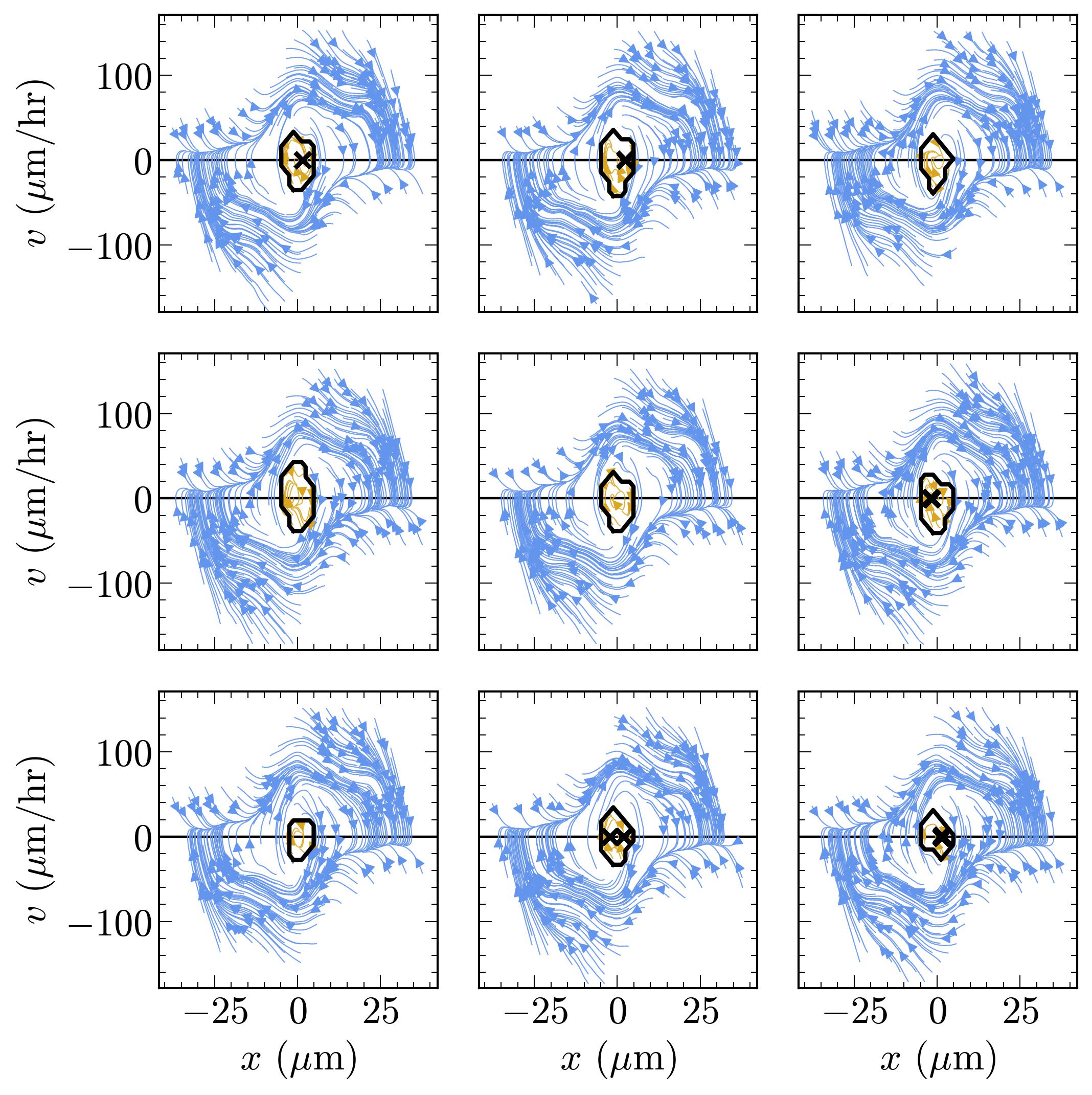}{Overview of variability in motility for the cell-micropattern pair presented in Fig. \ref{fig: tension lc}.}{fig: tension lc err}
\bootview{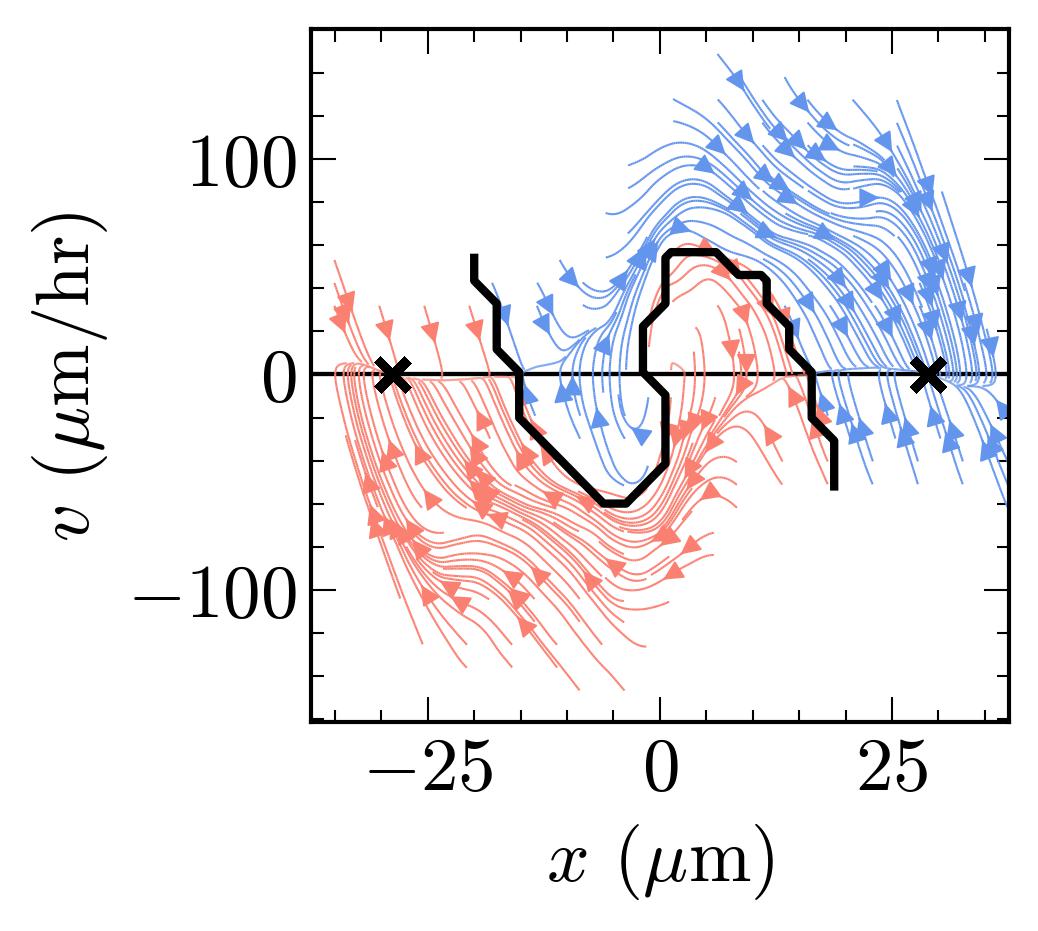}{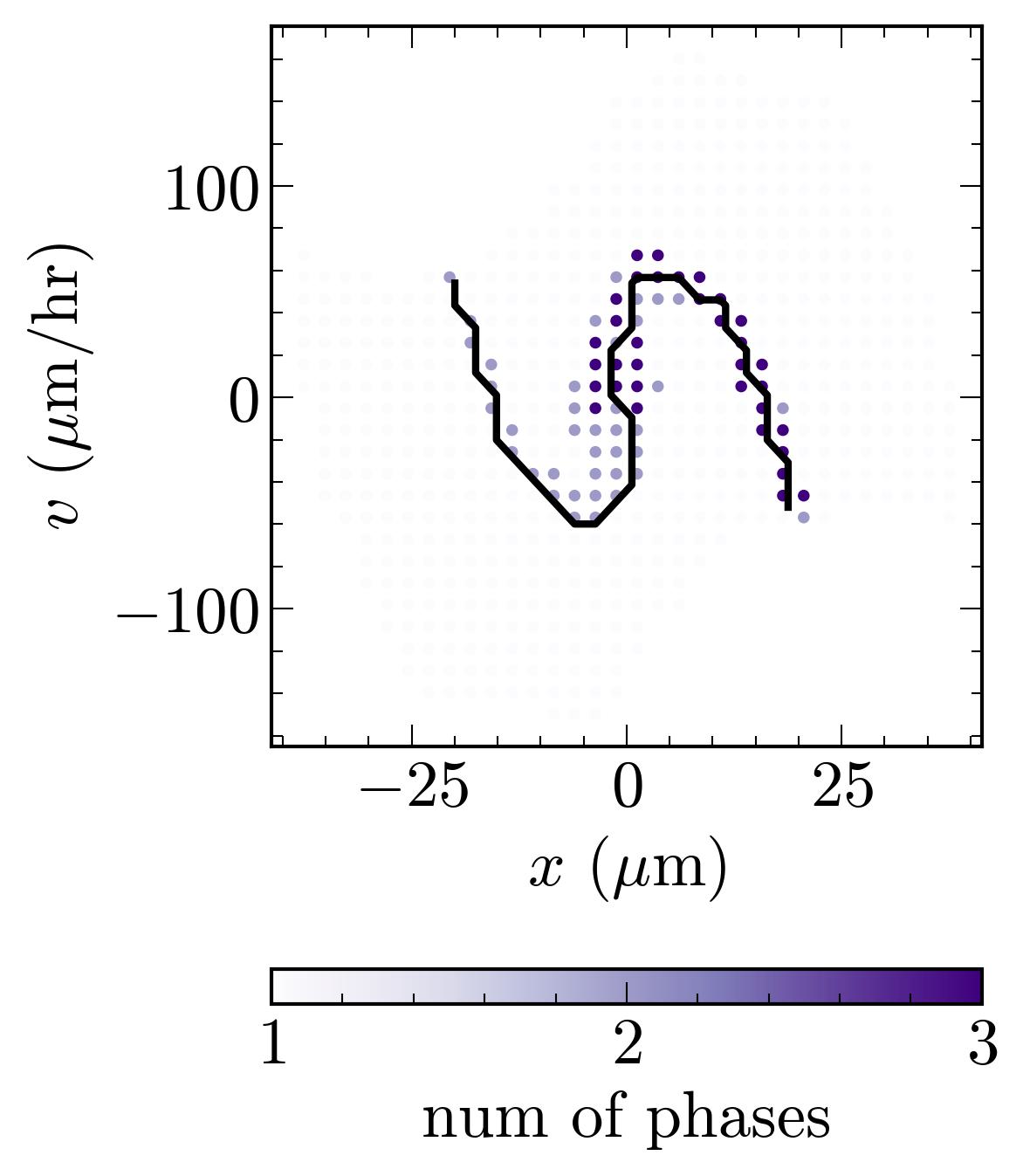}{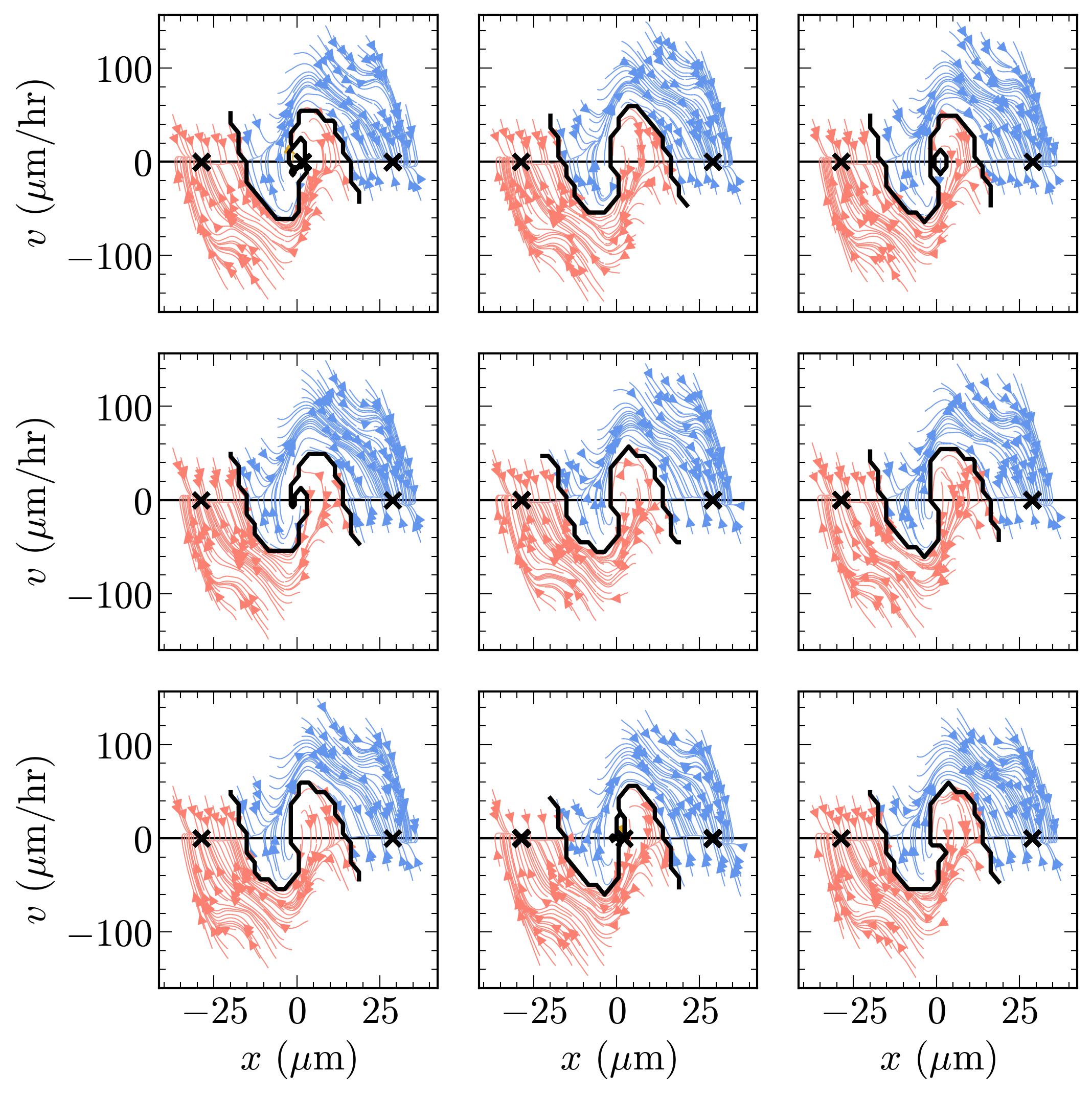}{Overview of variability in motility for the cell-micropattern pair presented in Fig. \ref{fig: tension bs}.}{fig: tension bs err}
\bootview{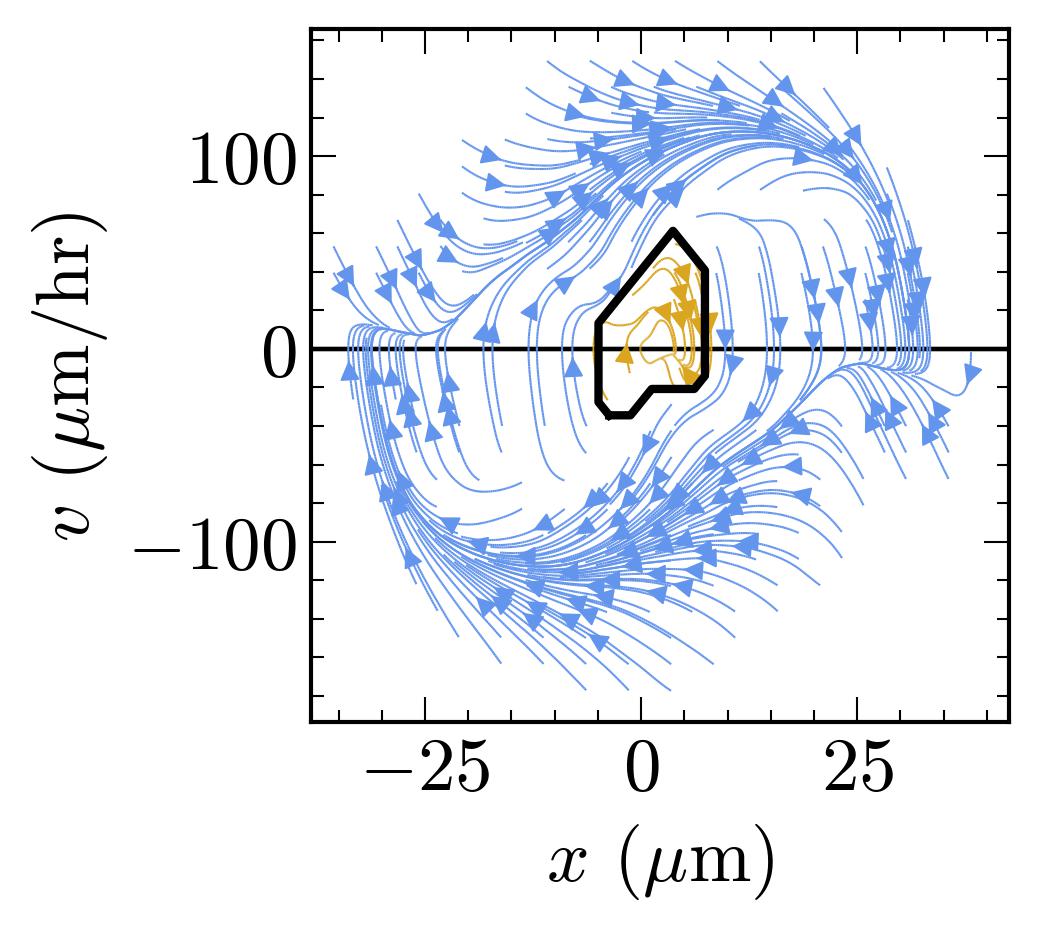}{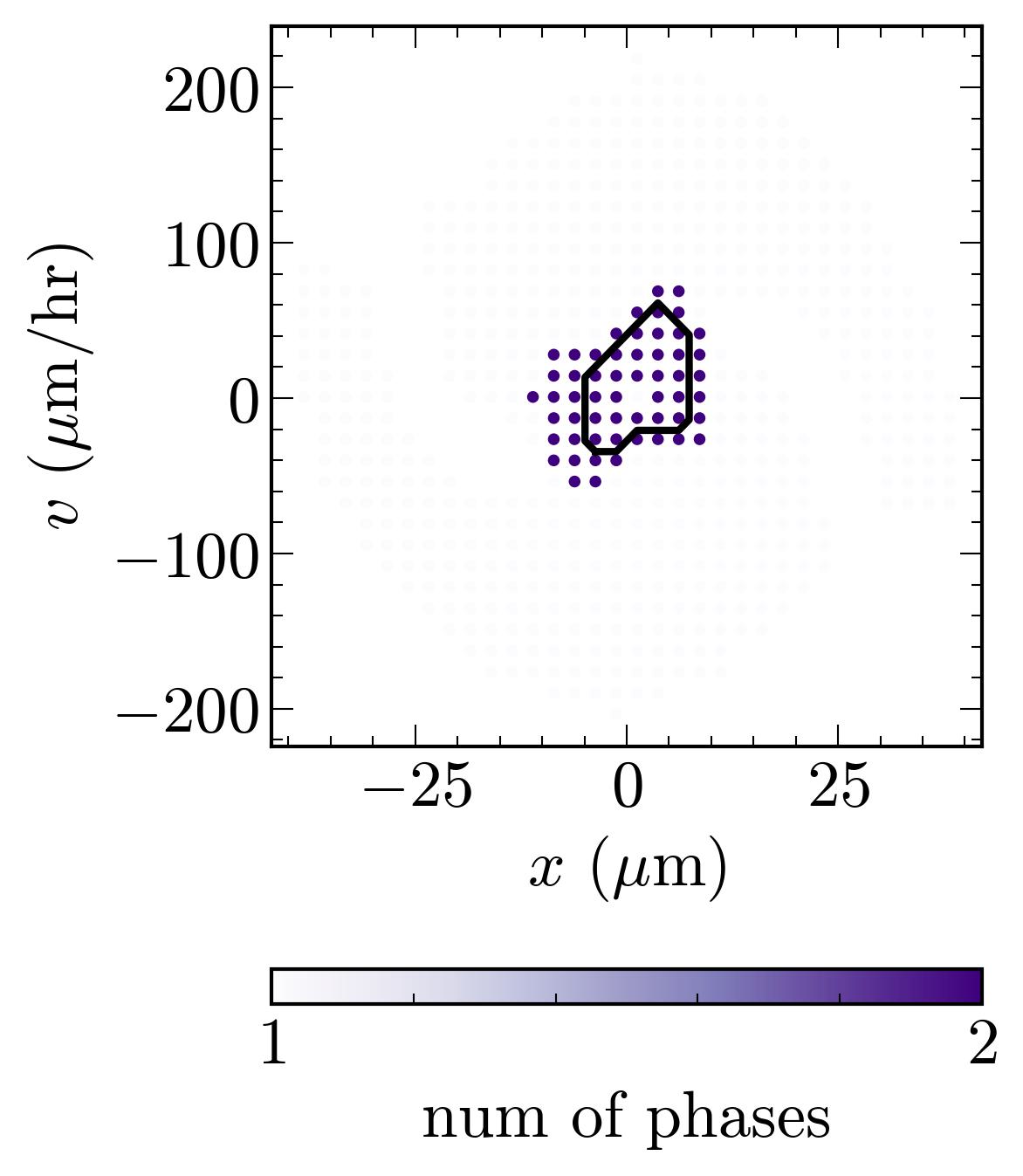}{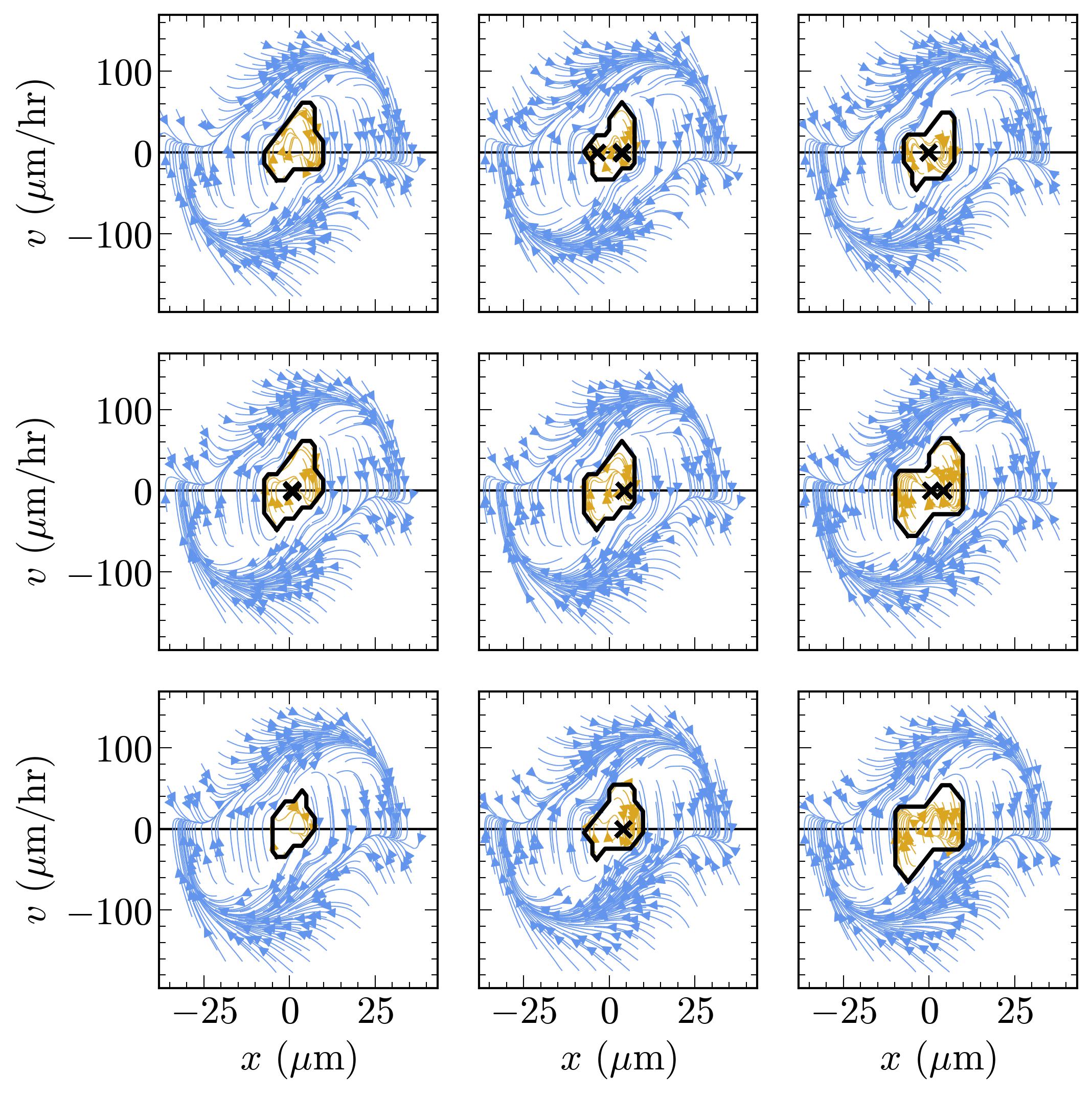}{Overview of variability in motility for $\mu_\beta=10$ perturbation on the two-state geometry presented in Fig. \ref{fig: mub perturbation}.}{fig: mub err}
\bootview{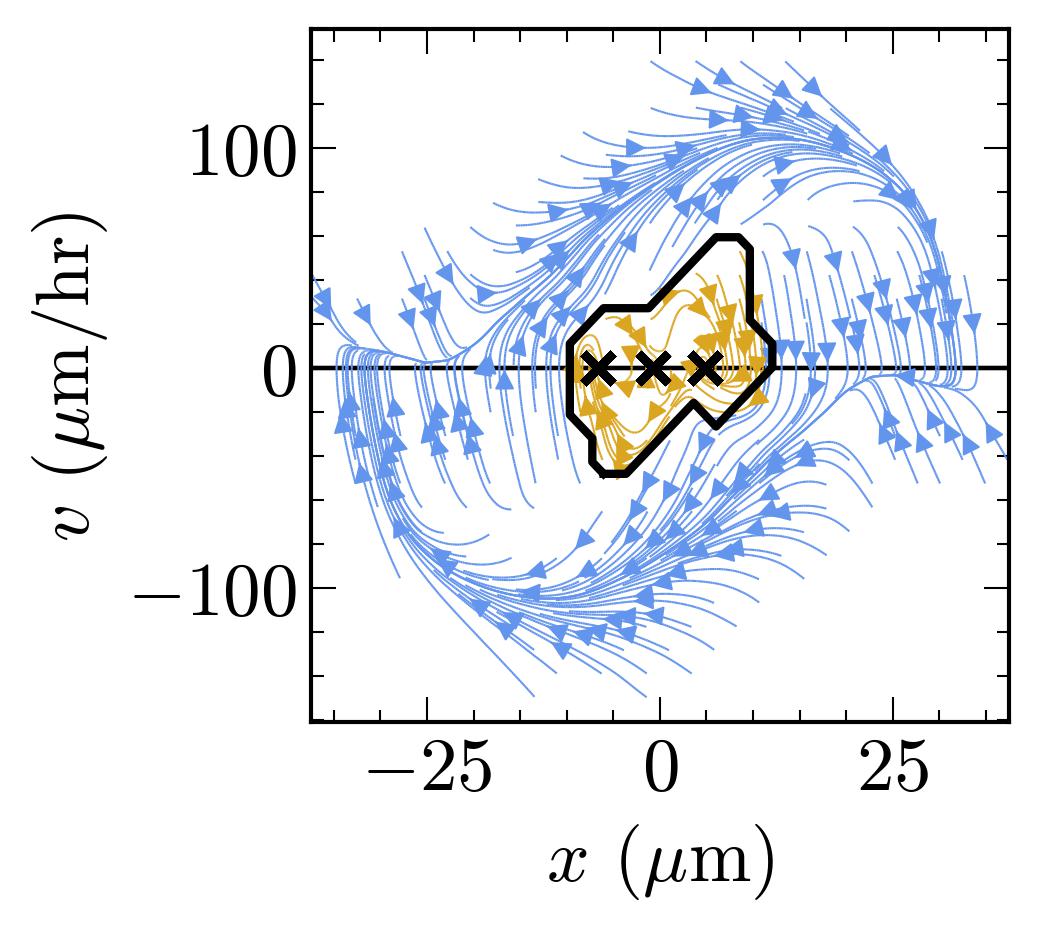}{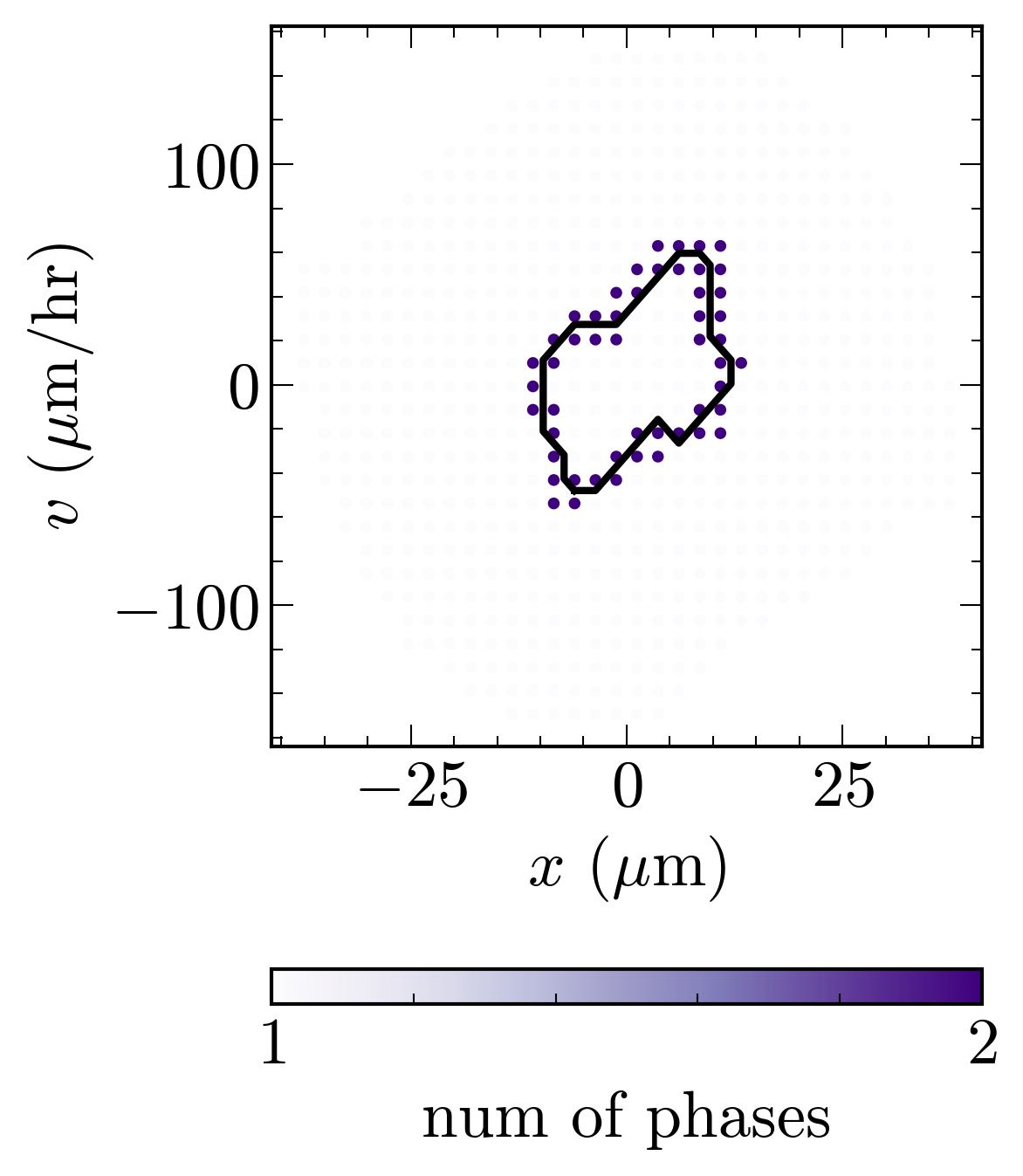}{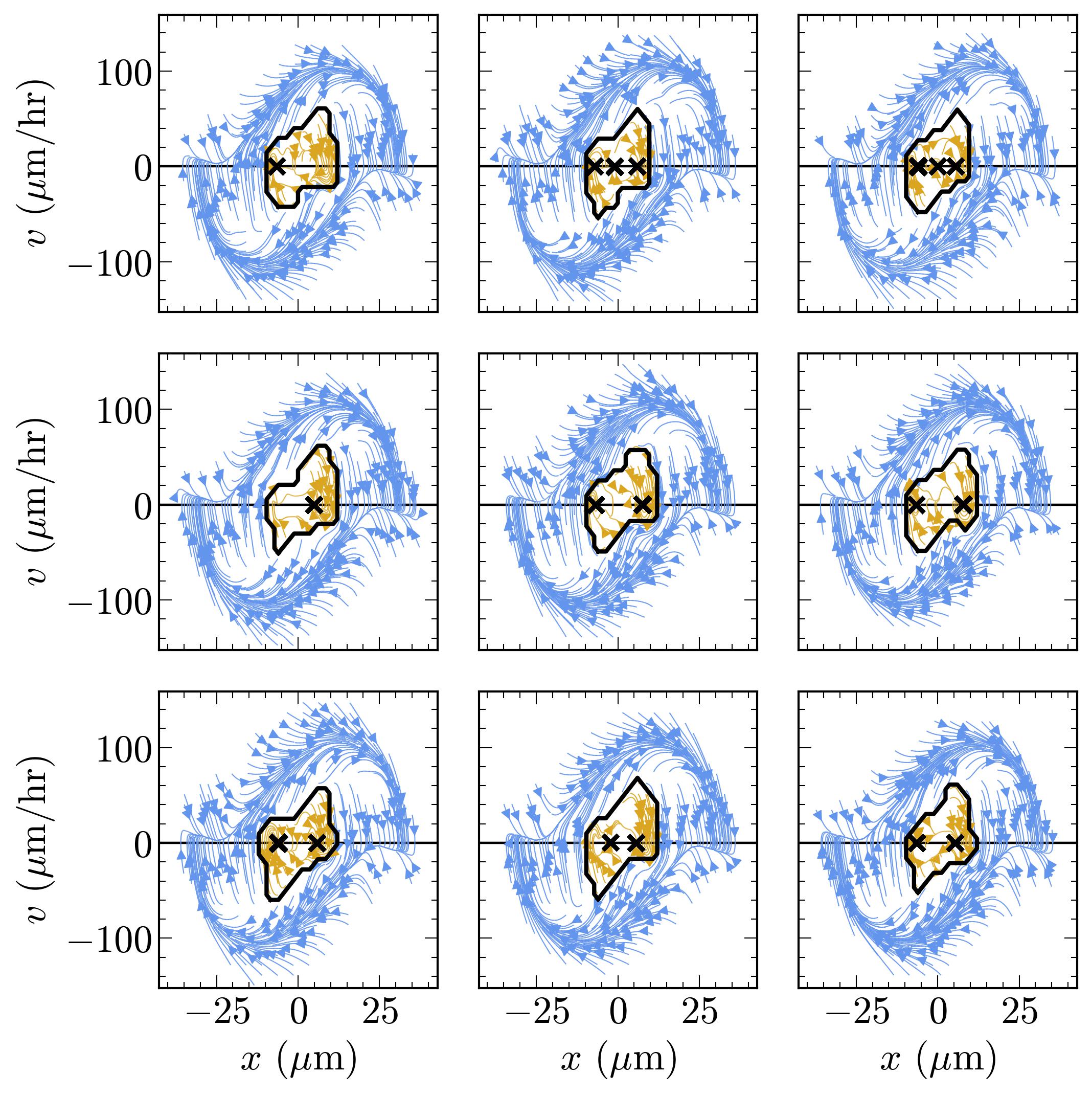}{Overview of variability in motility for $\sigma_\beta=22.5$ perturbation on the two-state geometry presented in Fig. \ref{fig: sigmab perturbation}.}{fig: sigmab err}
\bootview{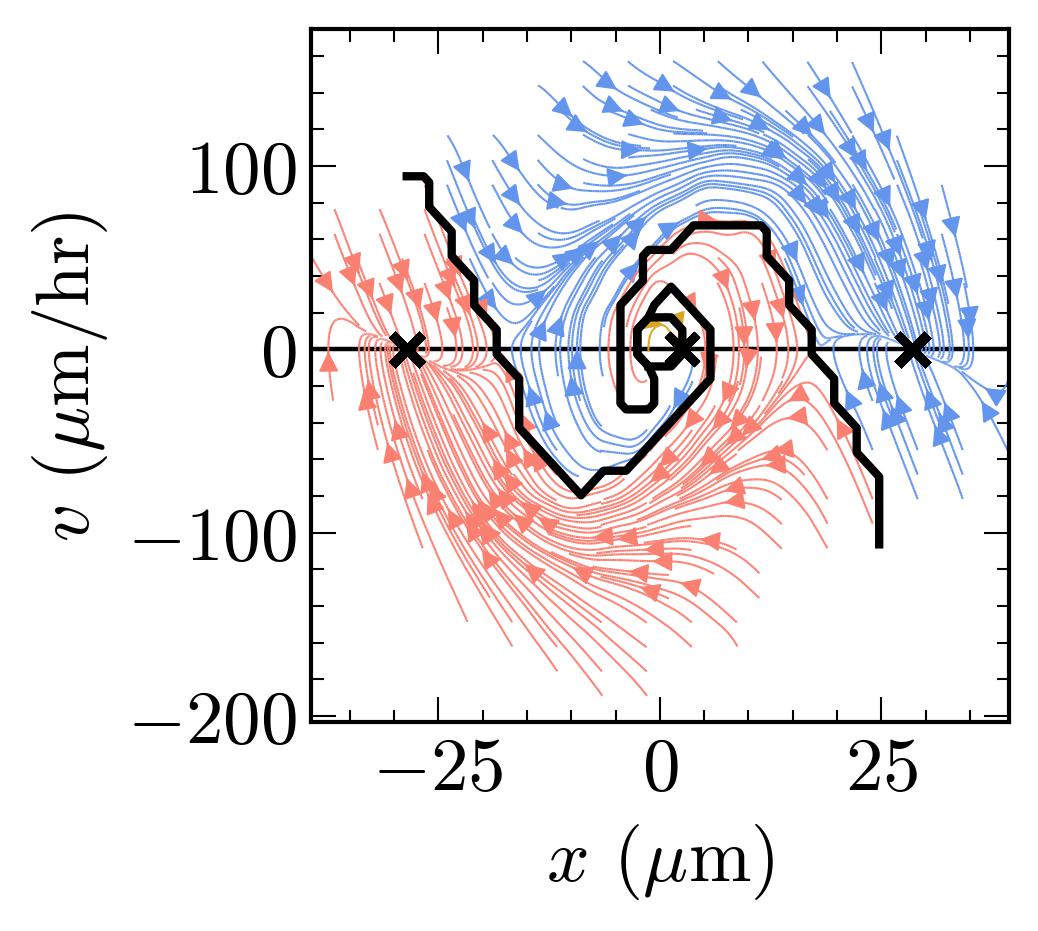}{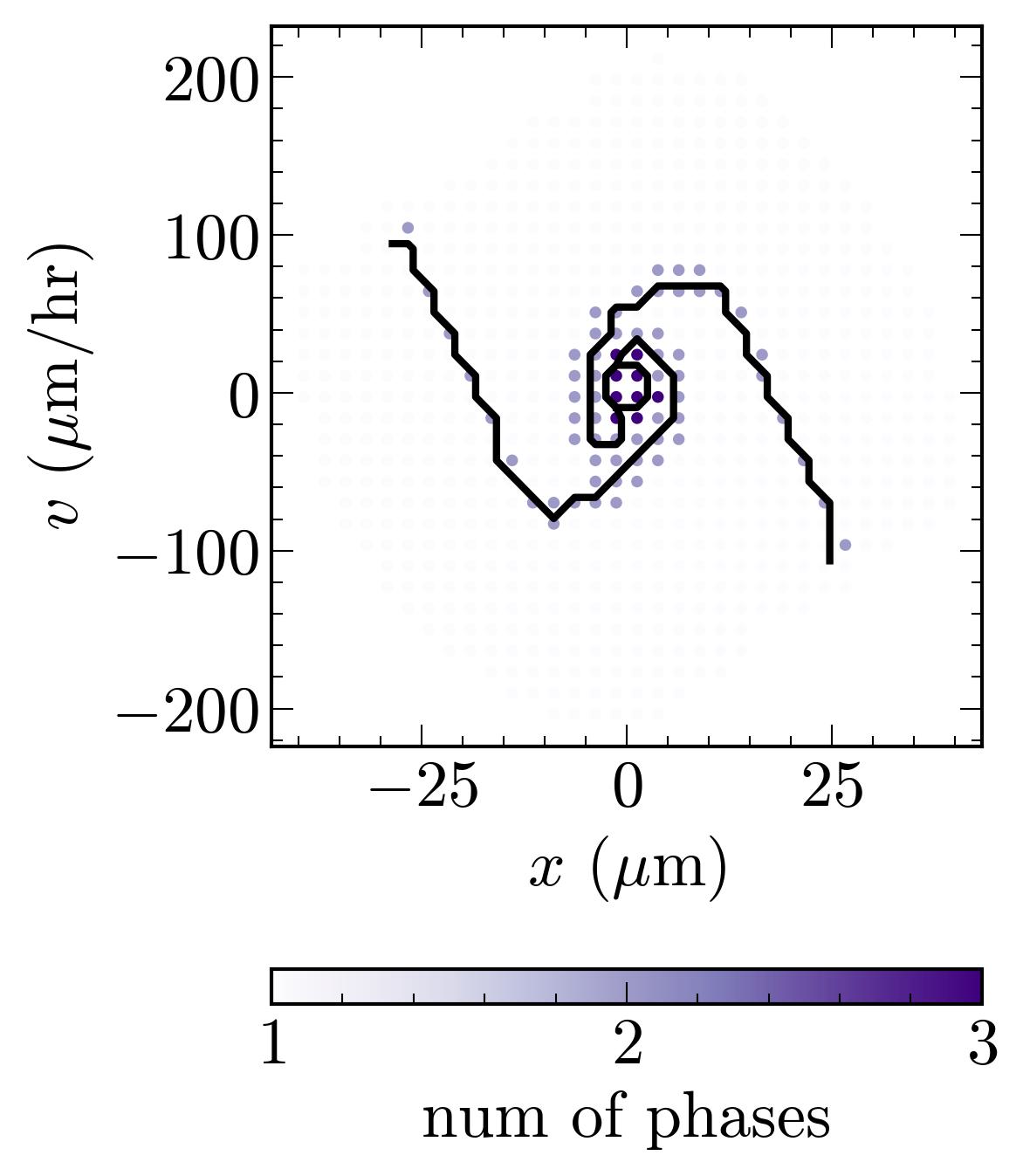}{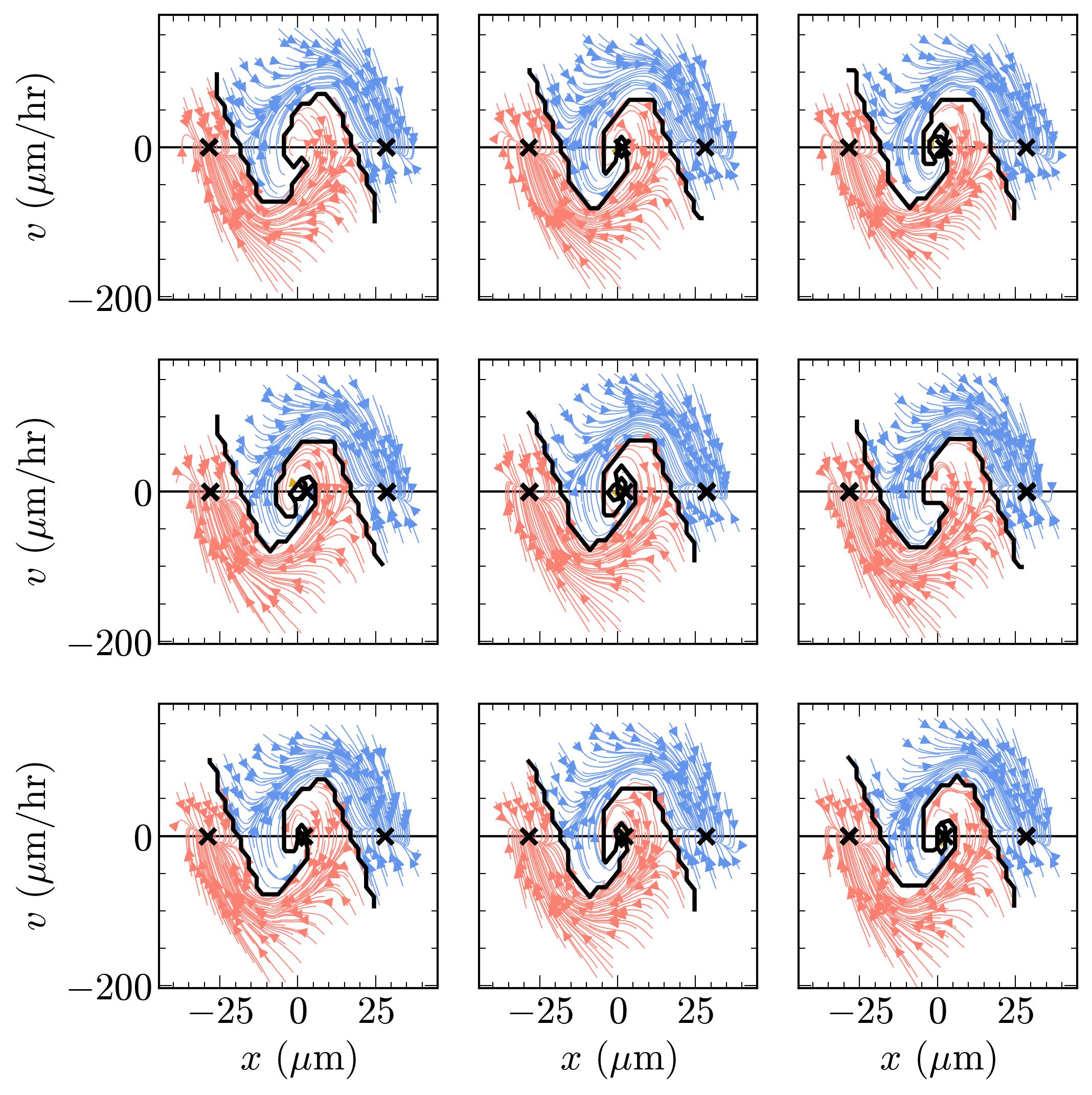}{Overview of variability in motility for $\sigma=12\micron$ perturbation on the two-state geometry presented in Fig. \ref{fig: sigma perturbation}.}{fig: sigma_ts err}
\bootview{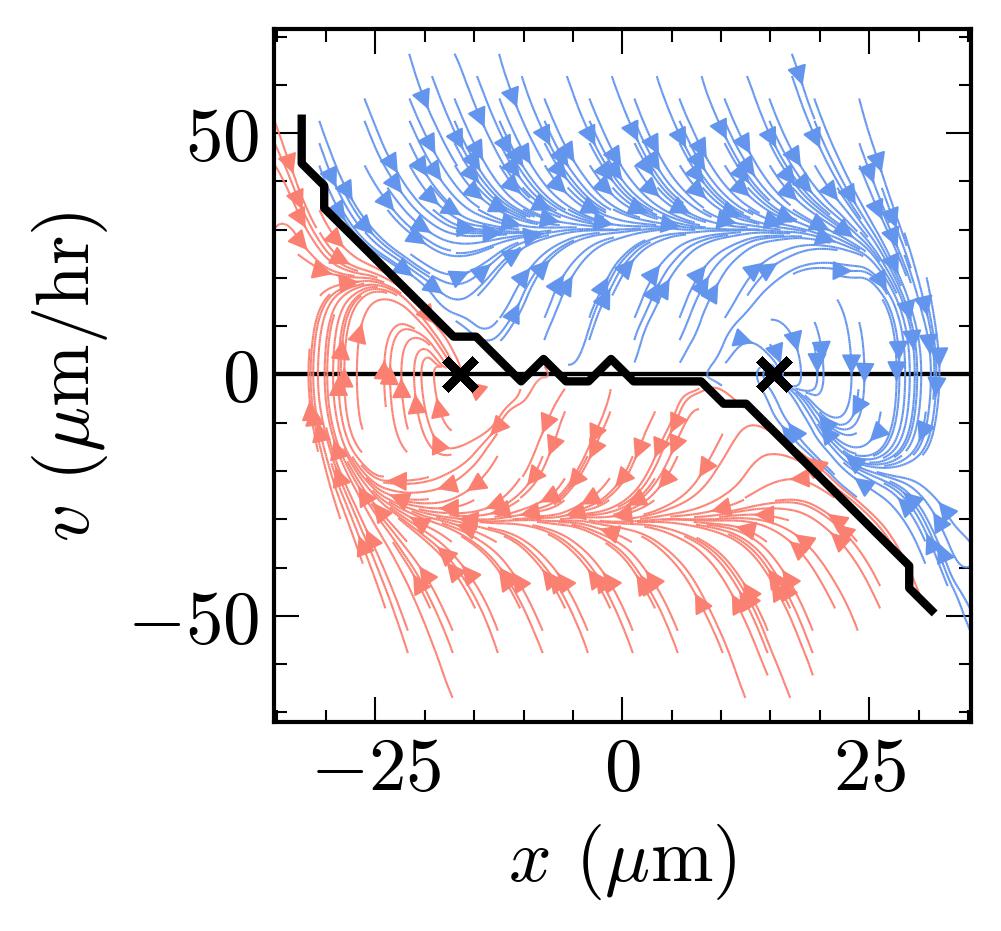}{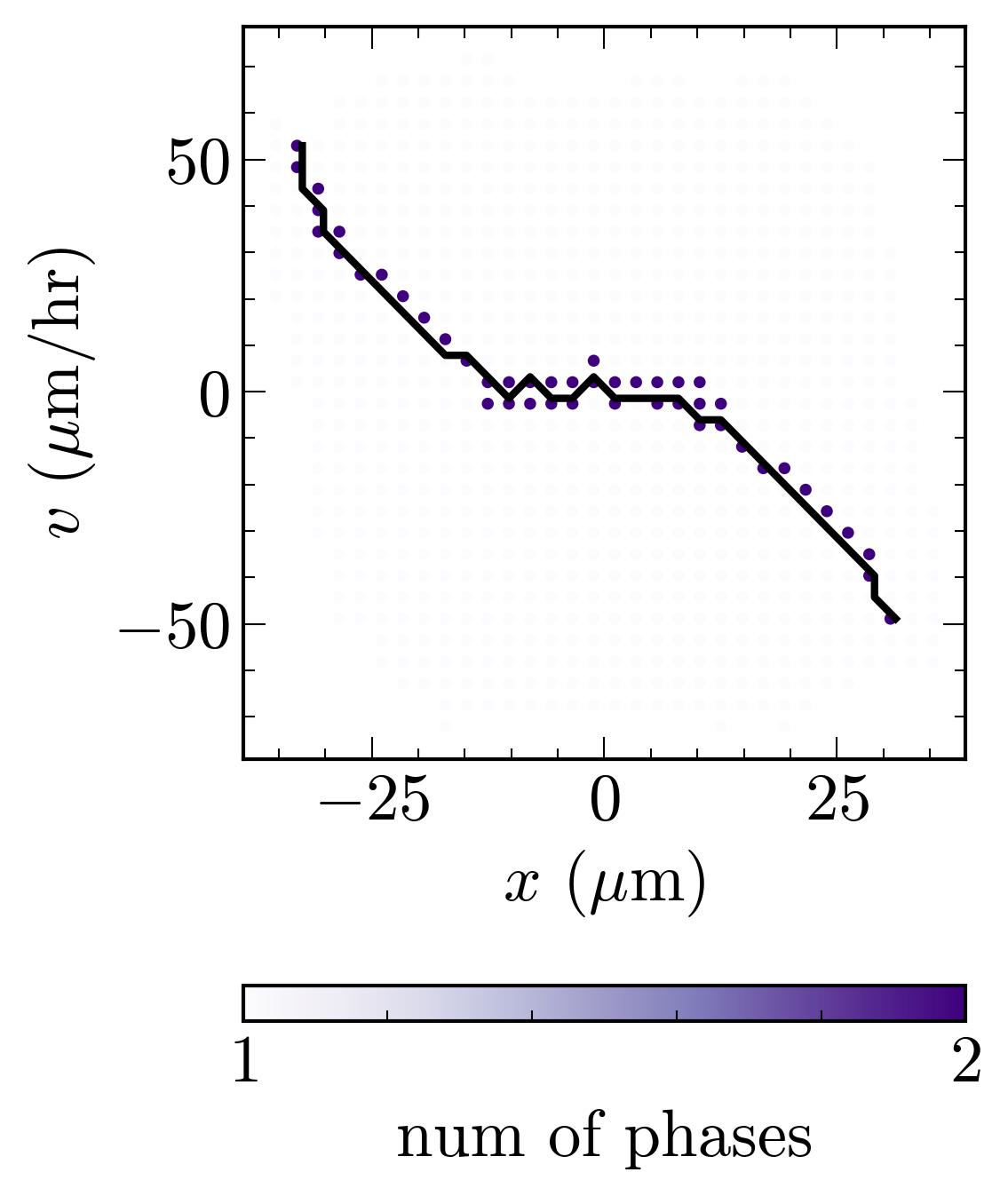}{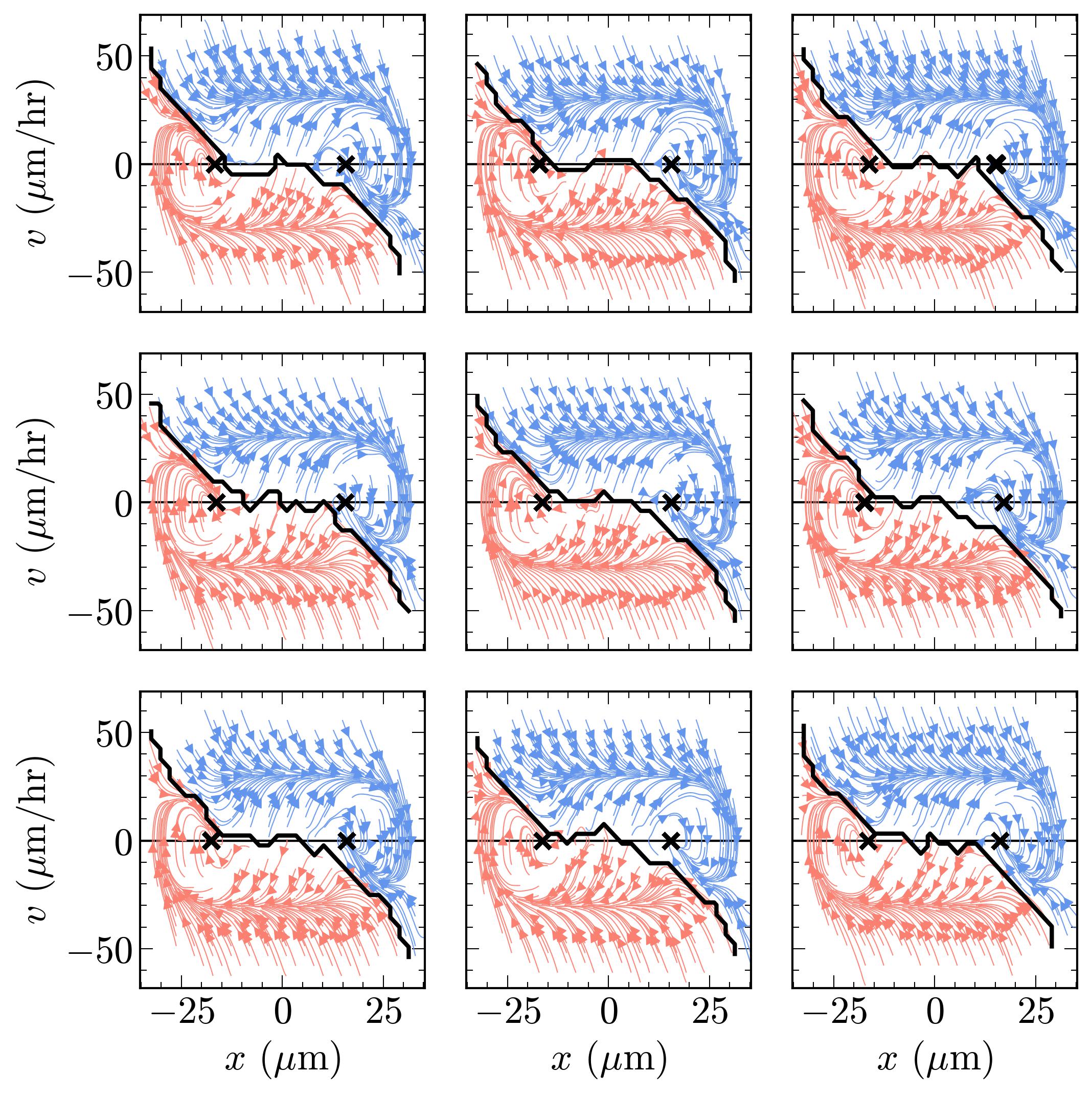}{Overview of variability in motility for $\sigma=3.4\micron$ perturbation on the rectangular geometry presented in Fig. \ref{fig: sigma perturbation}.}{fig: sigma_rect err}
\bootview{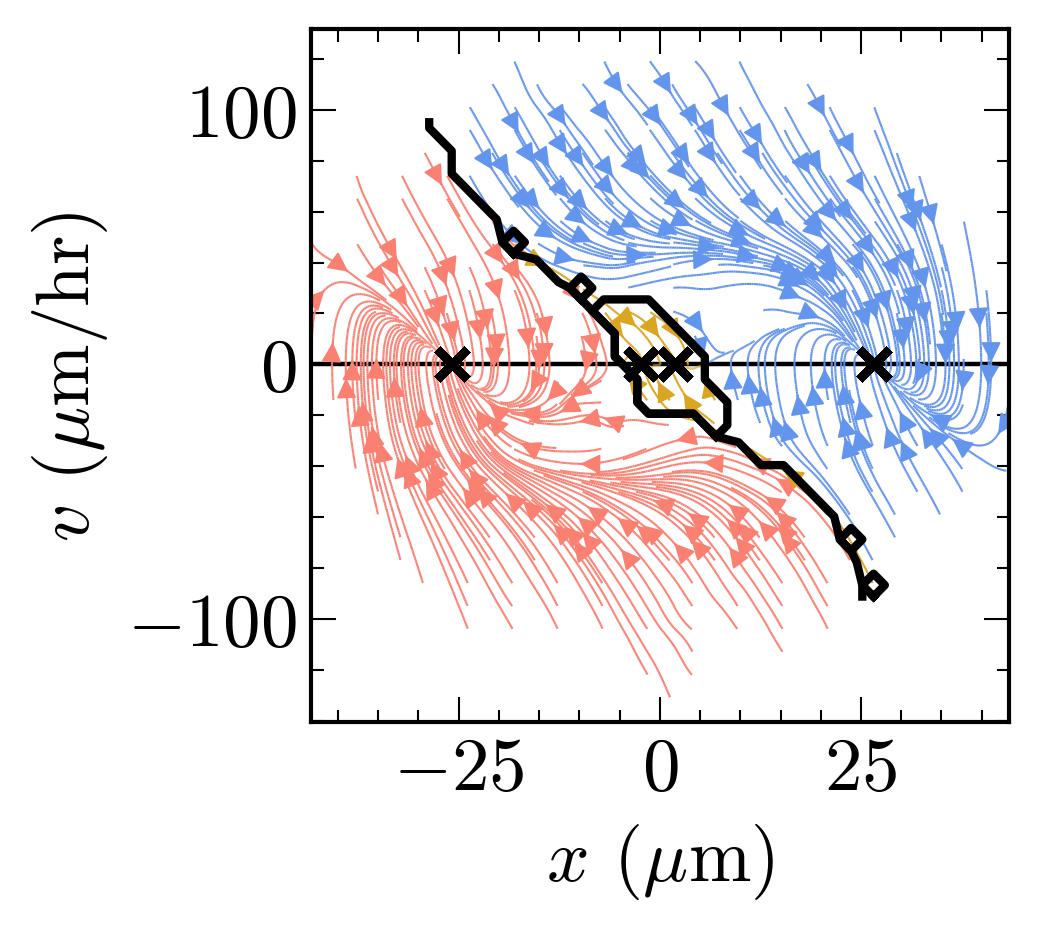}{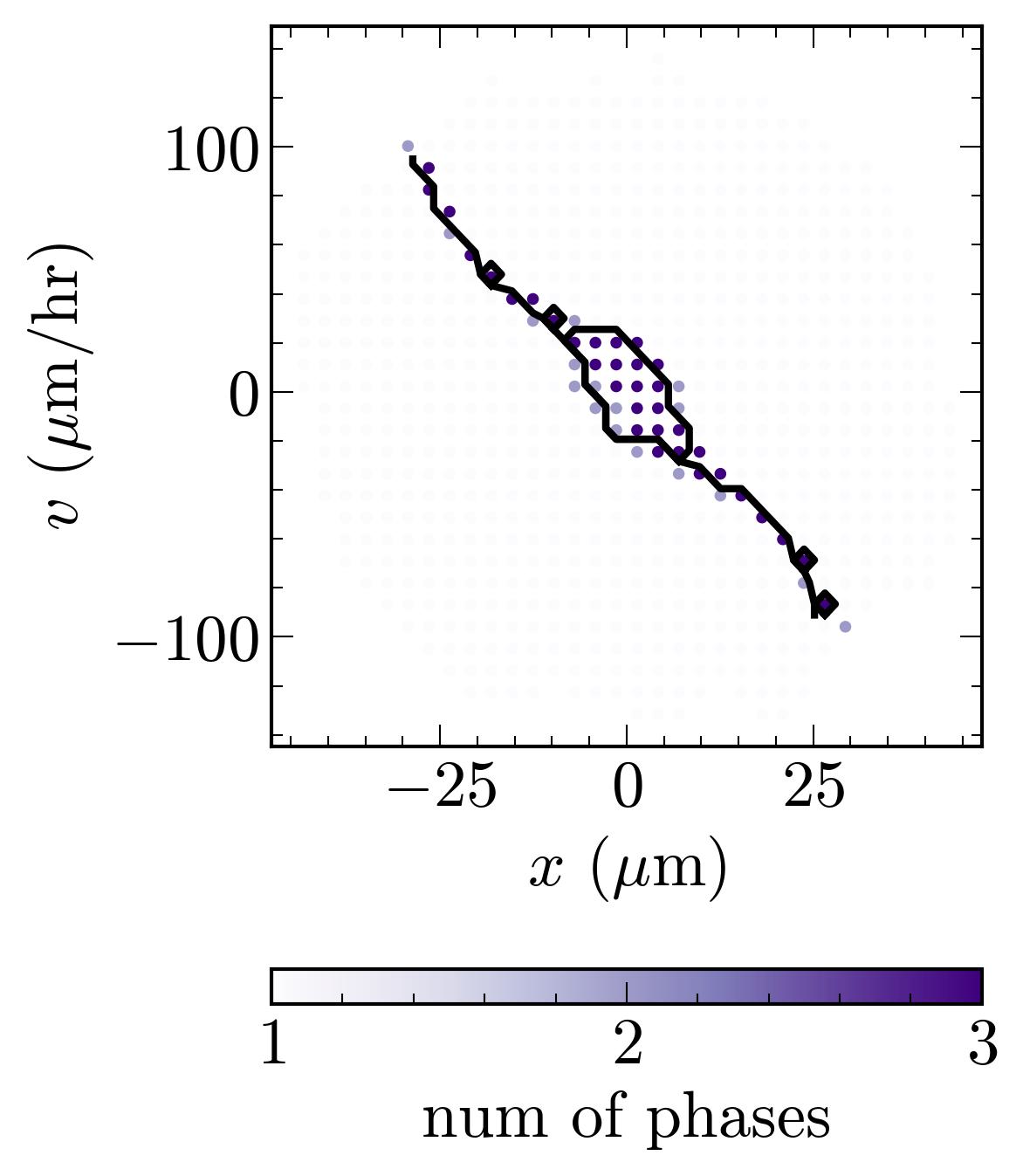}{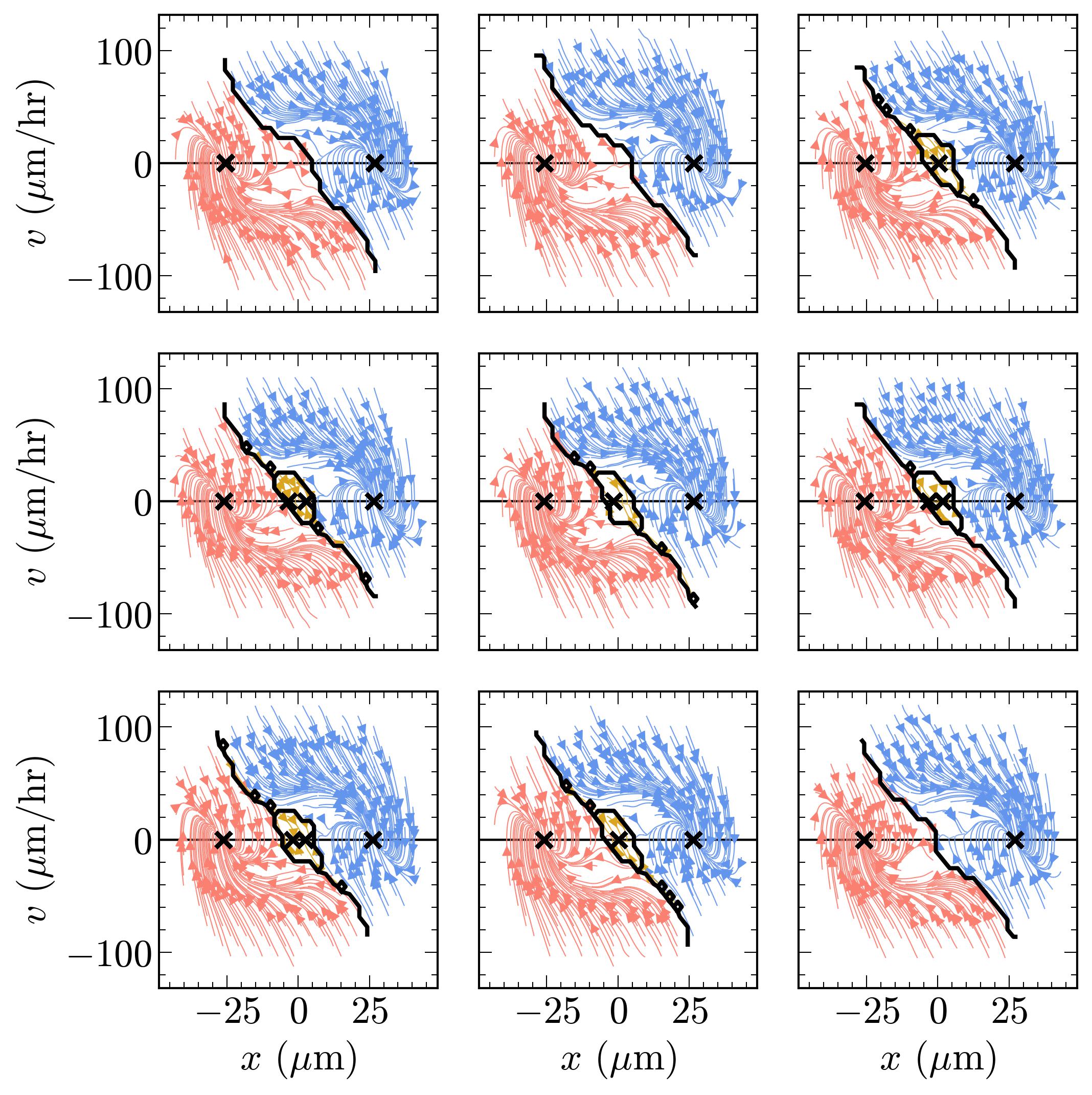}{Overview of variability in motility for the cell-micropattern pair presented in Fig. \ref{fig: tauXknx streamline}.}{fig: tauXknx err}

\section{Legend for the attached simulation movies}
All simulations are run for 24 hours.
\begin{enumerate}
    \item \label{mov: limit-cycle} \textbf{Movie 1:} Default cell limit cycling on the two-state micropattern
    \item \label{mov: stationary} \textbf{Movie 2:} Default cell migrating on the rectangular micropattern
    \item \label{mov: bistable} \textbf{Movie 3:} Default, small cell with $R_0=15\micron$ migrating on the two-state micropattern
    \item \label{mov: stiff-lc} \textbf{Movie 4:} Default, stiff cell with $\gamma=1.8\gamma_0$ migrating on the two-state micropattern
    \item \label{mov: free} \textbf{Movie 5:} Default cell freely migrating on an unconstrained micropattern (axes are in $\micron$)
    \item \label{mov: noisy-lc} \textbf{Movie 6:} Default, noisy cell with $\sigma_\beta=52.5$ migrating on the two-state micropattern
    \item \label{mov: substrate_knox} \textbf{Movie 7:} Default cell with $\tau_\chi\to\infty$ migrating on the two-state micropattern
    \item \label{mov: perim_knox} \textbf{Movie 8:} Default cell with $s_0\to\infty$ migrating on the two-state micropattern
\end{enumerate}    
\end{document}